\documentclass[12pt,preprint]{aastex}
\begin{document}

\title{Beryllium and Alpha-Element Abundances in a Large Sample of Metal-Poor
Stars}

\author{Ann Merchant Boesgaard\altaffilmark{1}, Jeffrey
A. Rich\altaffilmark{1}, Emily M. Levesque\altaffilmark{1}\altaffilmark{2} \&
Brendan P. Bowler\altaffilmark{1}}

\affil{Institute for Astronomy, University of Hawai`i at M\-anoa, \\ 2680
Woodlawn Drive, Honolulu, HI {\ \ }96822 \\ } 
\email{boes@ifa.hawaii.edu}
\email{jrich@ifa.hawaii.edu}
\email{Emily.Levesque@colorado.edu}
\email{bpbowler@ifa.hawaii.edu}
\altaffiltext{1}{Visiting Astronomer, W.~M.~Keck Observatory jointly operated
 by the California Institute of Technology and the University of California.}
\altaffiltext{2}{University of Colorado}
\begin{abstract}
The light elements, Li, Be, and B, provide tracers for many aspects of
astronomy including stellar structure, Galactic evolution, and cosmology.  We
have made observations of Be in 117 metal-poor stars ranging in metallicity
from [Fe/H] = $-$0.5 to $-$3.5 with Keck I + HIRES.  Our spectra are
high-resolution ($\sim$ 42,000) and high signal-to-noise (the median is 106
per pixel).  We have determined the stellar parameters spectroscopically from
lines of Fe I, Fe II, Ti I and Ti II.  The abundances of Be and O were derived
by spectrum synthesis techniques, while abundances of Fe, Ti, and Mg were
found from many spectral line measurements.  There is a linear relationship
between [Fe/H] and A(Be) with a slope of +0.88 $\pm$0.03 over three orders of
magnitude in [Fe/H].  We find that Be is enhanced relative to Fe; [Be/Fe] is
+0.40 near [Fe/H] $\sim$$-$3.3 and drops to 0.0 near [Fe/H] $\sim-$1.7.  For
the relationship between A(Be) and [O/H] we find a gradual change in slope
from 0.69 $\pm$0.13 for the Be-poor/O-poor stars to 1.13 $\pm$0.10 for the
Be-rich/O-rich stars.  Inasmuch as the relationship between [Fe/H] and [O/H]
seems robustly linear (slope = +0.75 $\pm$0.03), we conclude that the slope
change in Be vs.~O is due to the Be abundance.  Much of the Be would have been
formed in the vicinity of SN II in the early history of the Galaxy and by
Galactic cosmic-ray (GCR) spallation in the later eras.  Although Be is a
by-product of CNO, we have used Ti and Mg abundances as alpha-element
surrogates for O in part because O abundances are rather sensitive to both
stellar temperature and surface gravity.  We find that A(Be) tracks [Ti/H]
very well with a slope of 1.00 $\pm$0.04.  It also tracks [Mg/H] very well
with a slope of 0.88 $\pm$0.03.  We have kinematic information on 114 stars in
our sample and they divide equally into dissipative and accretive stars.
Almost the full range of [Fe/H] and [O/H] is covered in each group.  There are
distinct differences in the relationships of A(Be) and [Fe/H] and of A(Be) and
[O/H] for the dissipative and the accretive stars.  It is likely that the
formation of Be in the accretive stars was primarily in the vicinity of
SN II while the Be in the dissipative stars was primarily formed by GCR
spallation.  We find that Be is not as good a cosmochronometer as Fe.  We have
found a spread in A(Be) that is valid at the 4$\sigma$ level between [O/H] =
$-$0.5 to $-$1.0 which corresponds to $-$0.9 to $-$1.6 in [Fe/H].
\end{abstract}


\section{INTRODUCTION}

The rare light elements, Li, Be, and B, are rare relative to their neighbors
on the periodic table, light H and He and heavier C, N, and O, because their
origins are not in stellar interiors.  A method other than stellar
nucleosynthesis is needed to form LiBeB.  Although they can be formed by
nuclear fusion in stellar interiors, they are readily destroyed at
temperatures lower than the formation temperatures, i.e.~at shallower depths
in the interiors.

Beryllium can be formed outside of stars in different ways.  The ``supernovae
mechanism'' occurs in the vicinity of massive stars which accelerate plentiful
nuclei like C, N, and O during the explosion into the surrounding gas.  The
collisions break up these abundant elements into smaller units, among which
are Li, Be, and B nuclei (e.g. Duncan et al.~(1997, 1998) Lemoine et
al.~(1998).  In the general vicinity of SN II the number of Be nuclei would be
proportional to the number of CNO nuclei.  This mechanism might be the
prevalent one in the early history of the Galaxy (Boesgaard et al.~1999, Rich
\& Boesgaard 2009, Tan et al.~2009) and Smiljanic et al.~2009).  The slope
between A(Be) = log N(Be)/N(H) +12.00 and [O/H] would be $\leq$1.

A variation on this is the ``hypernovae mechanism'' (Fields et al.~2002;
Nakamura et al.~2006) which could enrich certain regions with excess light
elements.  Similarly, Parizot (2000) suggests the idea that superbubbles
containing multiple supernovae would lead to local enrichments of light
elements.  This mechanism has been suggested by Boesgaard \& Novicki (2006)
and by Smiljanic et al.~(2008) to account for some stars with strong Be
enrichments.

Another external source of formation of the rare light elements is the
classical Galactic Cosmic Ray = ``GCR spallation'' reactions outside of stars
in the general interstellar medium first proposed by (Reeves et al.~1970).
Energetic cosmic rays ($>$ 150 MeV) bombard CNO atoms in the ambient
interstellar gas and break them into smaller pieces, including Li, Be, B.
This process has been described and element ratios were predicted by Meneguzzi
et al.~(1971).  In this case the number of CNO nuclei would be proportional to
the cumulative number of SN II ($N$) and the number of cosmic rays would be
proportional to the instantaneous rate of SN II ($dN$).  The spallation
products would be $\int N dN$ = $kN^2$.  This mechanism might be the more
dominant one now producing light elements in the Galactic disk.  Then the
slope between A(Be) and [O/H] would be $\leq$2.  Processes such as mass
outflow during star formation would reduce the predicted slope to lower
values.

The early observations of Be and B did not demonstrate the expected quadratic
relationship of Be and B with metallicity as predicted by the classic GCR
spallation.  More complex models of the production of the light elements were
then made by several groups (e.g.~Ryan et al.~1992, Ramaty et al.~1997,
Prantzos et al.~(1993), Yoshii et al.~1997, Suzuki et al.~(1999), Suzuki \&
Yoshii 2001).  Ryan et al.~suggest a model with outflow from the Galactic halo
and time-dependent cosmic ray flux.  The model of Prantzos et al.~that best
fit the data at the time consists of two zones - one with gas outflow for the
halo and one with gas inflow for the disk.  Suzuki \& Yoshii produce a
self-consistent model which includes inhomogeneous conditions for the halo;
they also discuss the effect of an AGN in our Galaxy as a producer of
energetic particles as spallation ``bullets.''  Their model (their Figure 3)
does show a fit between Be and metallicity with a slope of 1.0.

Rich \& Boesgaard (2009) identified a change in slope between the abundance of
Be and O, suggesting such that the ``supernovae mechanism'' dominanted in the
early times of Galactic evolution and the ``GCR mechanism'' became the more
dominant one in the later stages of evolution.  They found a slope for the the
metal-poorer stars to be 0.74 $\pm$0.11 and for the metal-richer stars to be
1.59 $\pm$0.15.

Beryllium has some advantages in the study of light elements over Li and B.
It has only one stable isotope, $^9$Be, while Li has both $^6$Li and $^7$Li
and B has both $^{10}$B and $^{11}$B.  The small non-LTE effects tend to
cancel each other out whereas there are non-LTE effects for both Li and B.
Beryllium is apparently not produced by the $\nu$-process, but both Li and B
are predicted to be formed by that mechanism (Woosley et al.~1990).  It is
less fragile than Li to ($p,\alpha$) reactions (but more fragile than B).
Although it is less easily observed than Li, Be is observable from the ground
while B is not.  It may be that Be will be a good cosmo-chronometer as
suggested by Suzuki \& Yoshii (2001) and Pasquini et al.~(2005); this is not
the case for Li due to the Big Bang component of Li.

The study of the Be abundances in metal-poor stars has been the topic of many
papers following the detection of Be in the metal-poor star, HD 140283, by
Gilmore et al.~(1991).  Prior to that study only upper limits had been
determined for several stars with [Fe/H] $<-$1.3 (Rebolo et al.~1988, Ryan et
al.~1990).  Some recent compilations of Be abundances in metal-poor stars
include Boesgaard et al.~(1999), Boesgaard \& Novicki (2006), Tan et
al.~(2009), Smiljanic et al.~(2009), Rich \& Boesgaard (2009).

Primas et al.~(2000a, 2000b) found Be abundances in two of the three very
metal-poor stars that they studied.  They compared their results for Be and Fe
with those determined by Boesgaard et al.~(1999), finding that their Be
abundances were ``significantly'' higher than expected.  Rich \& Boesgaard
(2009), however, found no evidence for a plateau with constant Be abundance.
They did find that the four lowest metallicity stars have enhanced [Be/Fe] by
about 1 $\sigma$.

In this paper we report on an extensive collection of high-resolution, high
signal-to-noise spectra of Be in 117 metal-poor stars.  We have determined
abundances for Be, Fe, and O in these stars and have derived abundances for Ti
and Mg in the 99 stars observed with the upgraded version of HIRES on the Keck
I telescope.  Abundances of Li from the literature have been used to assess
the possibility of Be depletions.  We have found kinematic data for 114 stars
and examine the relationship between Be abundance and kinematic properties for
these stars.

\section{OBSERVATIONS AND REDUCTIONS}

Altogether 17 nights have been allocated for several programs on various
aspects of Be research.  Two of those were cancelled before they even began by
the Keck Observatory due to a snowstorm that closed the mountain access road
(2006 March 19/20) and because of damage resulting from 6.7 and 6.0 magnitude
earthquakes (2006 October 15/16).  Two other nights had some weather problems.
Altogether we obtained data on 15 nights between September 2004 and July 2010.
(None of our data was obtained in ``service'' mode.)  The median seeing for
the 15 nights was 0.7$\arcsec$.

The HIRES instrument (Vogt et al.~1994) on the Keck I was upgraded in 2004 and
was used to observe the Be II resonance doublet at 3130.421 and 3131.065 \AA.
We restricted our targets to those within declination of $-$30$\arcdeg$ to
+60$\arcdeg$ to minimize the effects of atmospheric dispersion on the
ultraviolet spectral region of the Be II lines; the latitude of Mauna Kea is
+19$\arcdeg$ 45$\arcmin$.  In addition we tried to observe all the stars near
the meridian, at the lowest possible airmass, usually within $\pm$2 hours of
crossing the meridian.  The upgraded HIRES CCD has 3 chips: blue, green, and
red.  The blue chip has a quantum efficiency of 93\% at 3130 \AA.  This is
important to counter the effects of atmospheric absorption and weak stellar
output at this wavelength in solar-temperature stars.  In the low metallicity
stars the Be II lines are weak so we tried to obtain a signal-to-noise ratio
per pixel (S/N) of at least 100.  The median S/N for our stars is 106 per
pixel.  The measured spectral resolution in the ultraviolet region is
$\sim$42,000.

The upgraded chips are each 2048 x 2048 pixels and have a pixel size of 15
$\micron$.  Our grating settings produced a spectral range on the three chips
of approximately 3000-6000 \AA.  Quartz flat-fields were obtained with
exposures tailored to the sensitivity of each chip: 50 s for the UV (blue
chip), 3-5 s for the green chip and 1 s for the red chip. We obtained 7-9
exposures for each chip.  Typically 9-11 bias frames were obtained each night
and Th-Ar comparison spectra were taken at the beginning and end of each
night.  The individual integration times for the stellar spectra were not
longer than 30-45 minutes in order to both minimize the effects of cosmic ray
hits and maximize the signal so that multiple spectra of the same star could
be co-added reliably.

The log of the observational data is presented in Table 1 where the final two
columns give the total exposure time in minutes and the S/N of the combined
spectra.  The S/N measurements are per pixel and are made near 3130 \AA.  In
this paper in addition to these new observations, we have included the data
and results for the most metal-poor stars as presented in Rich \& Boesgaard
(2009).  We have also included the stars from Boesgaard et al.~(1999) as
reanalyzed in Rich \& Boesgaard (2009) (Table 4).  We have reanalyzed the
Subaru spectra of Boesgaard \& Novicki (2006) and the four most metal-poor
stars from Boesgaard \& Hollek (2009).  The observing log for these stars are
in the original papers and do not appear in Table 1.

We have observed 18 stars in common with Smiljanic et al.~(2009), six which
they obtained from the archive and 12 which were observed for them in service
mode.  We have compared our exposure times and S/N ratios with theirs.  On
average our exposure times are 70\% of theirs while the data quality in terms
of S/N is 2.8 times better.  In addition our spectral resolution is
$\sim$42,000 compared to their $\sim$35,000.  Including the effect of the
size of the primary mirrors, we can see that Keck+HIRES is $\sim$5 times better
for UV spectroscopy than VLT+UVES.

For the data reduction we have used the IDL pipeline\footnote{
http://www.ucolick.org/~xavier/HIRedux} made for the upgraded HIRES courtesy
of J. Prochaska and standard IRAF\footnote{IRAF is distributed by the National
Optical Astronomy Observatories, which are operated by the Association of
Universities for Research in Astronomy, Inc., under cooperative agreement with
the National Science Foundation.} routines.  The pipeline performed bias
subtraction, flat field normalization (with our flat fields, not the archived
ones), spectral order extraction, and wavelength correction.  We used IRAF to
co-add multiple spectra of the same star (sometimes taken on multiple nights)
and to fit continua to the final combined, calibrated spectra.

There are five stars in Table 1 for which we did not determine Be abundances.
For two of the stars for which Li abundances have been published, our data
shows that they are double-lined spectroscopic binaries: BD +9$\arcdeg$ 352
and BD +26$\arcdeg$ 2606.  We have not tried to determine Be abundances for
these stars.  In addition, G 10-4 is Li depleted (Ryan \& Deliyannis 1998) and
too cool at 5000 K to determine a reliable Be abundance.  Two other stars, HD
106516 and HD 221377, are deficient in Li and Be and we have not included
those in this analysis.

\section{ABUNDANCES}

We have used both IRAF and
MOOG\footnote{http://www.as.utexas.edu/~chris/moog.html} (Sneden 1973) to
analyze the reduced spectra.  Equivalent widths of Fe I, Fe II, Ti I, Ti II
and Mg I were measured with the {\it splot} task in IRAF for each star.  For
each star for each line lists we removed lines weaker than 5 m\AA{} and
stronger than log W/$\lambda$ = $-$4.82.  The weak-line limit was chosen to
exclude lines with potentially poor measurements while the strong-line limit
was selected to ensure the lines we used would be on the straight-line portion
of the curve of growth.  We used IRAF ({\it listpix}) to make files of the
wavelength versus intensity in the Be II order to use in the spectrum
synthesis mode of MOOG.

\subsection{Stellar Parameter Determination}

Our stars have a range in temperatures (5500 - 6400 K) and metallicities
($-$0.4 to $-$3.5) so our original line lists were reduced slightly
differently for each star due to the limits above.  We were left with 28-86
lines of Fe I, 9-14 lines of Fe II, 5-19 lines of Ti I, 5-13 lines of Ti II
and 1-3 lines of Mg I.

We have written a collection of perl scripts that perform an iterative
determination of $T_{\rm eff}$, log g, [Fe/H], and $\xi$ using the line lists
for each star.  These scripts utilize the {\it abfind} driver in MOOG.  We
found $T_{\rm eff}$ from the agreement of the Fe abundance from Fe I lines
with a range of excitation potentials, i.e. the slope between log Fe/H and
excitation potential ($\chi$) should be zero.  We determined log g from the
ionization balance of both Fe I with Fe II and Ti I with Ti II, i.e.~the
abundances from the lines of the two ionization stages agreed at the preferred
log g value.  For almost all our stars we used the log g values from Ti I and
Ti II in part because there were more lines of Ti II than Fe II in general.
However, the differences in log g were typically $<$0.04 and the ensuing
differences in A(Be) were typically $<$0.01.  We found $\xi$ by iterating so
that the Fe I abundances were similar at all reduced equivalent widths.  Our
starting value was $\xi$ = 1.5 km s$^{-1}$; for some stars this did not
converge to a slope of zero so the iterations stopped and 1.5 km s$^{-1}$ was
used.  The median value for the 42 stars where we did determine $\xi$ is 1.42
km s$^{-1}$.  This is virtually identical to the 1.45 km s$^{-1}$ value found
by Magain (1985).

Model atmospheres were made for the specific parameters for each star by
interpolation with the Kurucz grid (1993).  All elements were reduced by the
amount that Fe is reduced relative to the sun, except for Be and O.

Table 2 contains the derived values for the stellar parameters in columns 2-5.
We show in Figure 1 the distribution of our selected stars in the $T_{\rm
eff}$ vs.~[Fe/H] plane.  We note that there are 6 stars with [Fe/H] $<$
$-$3.00.  We limited our lower temperature bound to 5500 K because below that
the metallic and molecular blending lines become too strong and the Be II
lines become too weak to obtain reliable Be abundances.

We have compared our derived parameters with those determined by Stephens \&
Boesgaard (2002) for the seven stars in common.  (We have two other stars in
common, but we adopted their parameters for G 64-12 and G 64-37.)  The average
of the temperature differences is +10 K, of the log g differences is $-$0.02
and of the [Fe/H] differences is $-$0.02 dex (in the sense of this study minus
theirs).  One star, BD +34 2476, has a $T_{\rm eff}$ difference of 187 K.  The
other stars agreed well within the nominal uncertainty of $\pm$50 K.  The
largest difference in [Fe/H] was only +0.16.  The range in the differences in
log g was $\pm$0.36.

We have also compared our parameters with those used in the study by Smiljanic
et al.~(2009) for the 16 stars in common with theirs.  (There are 18 stars in
common but they used the Stephens \& Boesgaard (2002) parameters for two of
them).  For 14 of the 16 stars they made use of the parameters derived
spectroscopically by Fulbright (2002).  For nine of the stars the agreement in
temperature is excellent, within $\pm$50 K.  However, for seven stars their
temperatures are systematically lower than ours by 250 to 600 K.  For six of
these seven stars we were able to make comparisons with effective temperatures
found by Casagrande et al.~(2010) from the infrared-flux method (IRFM).  Those
temperatures were in between our values and the ones used by Smiljanic et
al.(2009), with five being closer to ours and one closer to theirs.  On
average our spectroscopic temperatures were 138 K hotter than those from the
IRFM, while those of Smiljanic et al.~ were 285 K cooler.  The seventh star
was analyzed by Nissen et al.~(2000) who found temperatures with IRFM; that
value is also intermediate between the two spectroscopically determined
temperatures with ours again being hotter.

Thirteen of our stars were also observed by Nissen \& Schuster (2010)
in their sample of 106 stars from which they found evidence for two different
populations of halo stars in the solar neighborhood.  They also determined
their parameters spectroscopically.  On average our values for $T_{\rm eff}$
are 28 K hotter than theirs, our values for log g are smaller than theirs by
0.15 dex and we find [Fe/H] lower by -0.05 dex.  (If we exclude HD 241253 for
which we differ in temperature by 224 K, our mean difference is +12 K; for
that star we agree in [Fe/H] by 0.03 dex.)

The Be abundance is particularly sensitive to log g.  As mentioned above, we
found this parameter spectroscopically primarily from the ionization balance
between Ti I and Ti II which compared well with what we found from Fe I and Fe
II.  This study has 20 stars in common with Tomkin et al.~(1992) who found log
g values from the ionization balance between Fe I and Fe II.  The average
difference in log g from the two studies is +0.008.

Smiljanic et al.~(2009) used gravities derived from Hipparcos parallaxes or
adopted literature values.  The agreement for all but one of the 16 stars is
within $\pm$0.30 and the mean difference is $-$0.04.  For that one star, G
63-46, our log g value is higher by 0.93 dex and our Be abundance is higher by
a factor of four.  We have also tried their log g from Hipparcos in our
analysis which we present in the next section.  Neither our Be abundance nor
theirs produce an outlier in the relationships produced in section 4.

We also compared our derived values of [Fe/H] for the 16 stars.  All but one
star agree to within $\pm$0.12 with a mean difference of $-$0.03.  For BD +2
4651 we find [Fe/H] = $-$1.90 while they find $-$1.50.  This star is also not
an outlier in any of the figures in section 4 with either set of values.

\subsection{Beryllium Syntheses}

We used MOOG with the {\it synth} driver to find Be and O abundances.  We used
the appropriate model for each star to create a synthetic spectrum from 3129 -
3133 \AA{} to compare with the observed intensity-wavelength files.  For Be we
used four abundances to find the best match.  We then plotted the best fit to
the data with comparisons of A(Be) of +0.30 dex and $-$0.30 dex and one with
no Be at all.  For O we tried four abundances differing by 0.20 dex (or in
some cases 0.10 dex.)  As part of the fitting procedure we could adjust the
width of the gaussian we used for the line profile, the continuum level, and
the exact wavelength.  For some stars there was some broadening beyond the
instrumental width, but the determination of the continuum level was not a
problem in the metal-poor stars and the wavelength corrections made during the
data reduction were very accurate.

In Figure 2 we show the syntheses for two stars of very different
metallicities: [Fe/H] = $-$1.01 and $-$2.79.  This illustrates the complexity
of the spectra in the more metal-rich stars and the concomitant difficulty in
getting a good synthetic fit.  The metal-poor stars can be well fit, as
seen in the figure for BD $-$10$\arcdeg$ 388 and reliable Be abundances can be
determined.  However, the Be II lines grow weaker as the metallicity decreases
(see, e.g. Gilmore et al.~1991, 1992, Molaro et al.~(1997), Boesgaard \& King
1993, Boesgaard et al.~1999, Rich \& Boesgaard 2009).  The need for very high
S/N is clear for the lowest metallicity stars.  For example, our spectrum of
BD $-$10$\arcdeg$ 388, shown in Figure 2, has a S/N of 120.

Figure 3 shows the observed spectra and synthetic spectra for two stars with
the same temperature, but different values for log g and [Fe/H].  The lower
metallicity star has the lower Be and O abundances.  In BD +28$\arcdeg$ 2137
the value for [Fe/H] is lower by 0.41 dex, for [O/H] by 0.39 dex, for A(Be) by
0.31 dex compared to those values in BD $-$17$\arcdeg$ 484.

For stars with lower log g values the Be II lines are stronger for a given Be
abundance which can be seen in Figure 4.  For G 180-24 log g is 3.77, which is
lower than 3.98 for G 24-3 and the Be II lines can both be seen to be stronger
in G 180-24.  The Be abundance are comparable; G 180-24 does have a higher
A(Be), but not by much: 0.10 dex.  The syntheses for these two stars are shown
in Figure 5.  This implies that Be II will be more easily detected in
subgiants than in dwarfs at low metallicities.

One of our stars, G 268-32, with [Fe/H] = $-$2.51 seemed to be very difficult
to fit with an ordinary synthetic spectrum.  According to the analysis by Aoki
et al.~(2002), it is a carbon-enhanced metal-poor star (CEMP) with [C/Fe] =
2.1 and [N/Fe] = 1.2.  When we used those values in the synthesis, we derived
a much better fit.  Our parameters are very similar to those of Aoki et
al.~(2002): we derive 6230/4.60/$-$2.51/1.46, while they find
6250/4.5/$-$2.55/1.5 for $T_{\rm eff}$, log $g$, [Fe/H], and $\xi$.

Then following the example of Ito et al.~(2009), we removed the CH (and CN)
lines from the line list in order to find an upper limit for the Be abundance.
Figure 6 shows an expanded view of the region of the 3131 line of Be II.  The
feature at 3130.8 \AA{}, primarily Ti II, is well-fit.  For Be the synthetic
spectrum with no Be best matches the observed spectrum, but we adopt an upper
limit of A(Be) $<-$1.5.  This is now the second CEMP star with no Be; Ito et
al.~(2009) found A(Be) $<$ $-$2.00 in BD +44$\arcdeg$ 493.  Unlike BD
+44$\arcdeg$ 493, G 268-32 has enhanced s-process elements; Aoki et al.~(2002)
find [Ba/Fe] = +1.98.  The group of CEMP stars that are enhanced by s-process
nucleosynthesis, CEMP-${\it s}$, are thought to result from mass transfer from
an AGB companion.  This AGB connection may provide an answer to the problem of
the low (or no) Be in light of a normal Li abundance in G 268-32; Thorburn
(1994) found A(Li) = +2.09, but with a lower temperature, 5841 K, and a lower
[Fe/H] = $-$3.50.  When we use our model with her Li equivalent width, we find
A(Li) = 2.32.

Smiljanic et al.~(2008) discovered a Be-rich halo dwarf, HD 106038 which has
[Fe/H] = $-$1.26 (Nissen \& Shuster 1997).  They found a Be abundance of A(Be)
= 1.40 near the meteoritic value of 1.42 (Grevesse \& Sauval 1998) and 1.41
(Lodders 2003).  We observed this star on our 2008 Jan 16 Keck night and
obtained a S/N of 100 in a 25 min.~integration.  Our derived parameters
($T_{\rm eff}$ = 6085 K, log g = 4.63, [Fe/H] = $-$1.34, $\xi$ = 1.46) are
very similar to the ones they used and our Be abundance is similar, A(Be) =
1.47.  Tan et al.~(2009) also analyzed HD 106038 and derived A(Be) = 1.37.

The upper part of Figure 7 shows our observed spectrum of HD 106038 compared
to BD $-$8$\arcdeg$ 4501, which has similar parameters.  The difference in the
Be II lines is dramatic.  The lower half of the figure is our spectrum
synthesis for HD 106038 showing good fits for both Be and OH.  This OH line
gives [O/H] = $-$0.95; the other two OH features give $-$0.92 and $-$0.90 (see
O synthesis section below).

In section 3.1 we noted that our spectroscopic log g was quite different from
the Hipparcos one used by Smiljanic et al.~(2009) for G 63-46.  The parallax
for this star is 7.44 $\pm$1.70 mas resulting in their value of 3.77
($\pm$0.10).  We have done a Be- and O-synthesis with that log g and find A(Be)
= 0.66, lower than the +0.98 with our log g but higher than the +0.37 found by
Smiljanic et al.  The OH lines are not fit as well using the lower log g,
however.  We also tried a synthesis with all four parameters used by Smiljanic
et al.; that does not fit our data at all well, mostly because their
temperature is so low that all the blending lines near the 3130 line of Be II
are too strong.

\subsection{Oxygen Syntheses}

Beryllium is a direct by-product of O via spallation so it is useful to find
the abundance of O in our stars.  As pointed out in Rich \& Boesgaard (2009)
determining reliable O abundances in stars is not easy and the three common
features used (O I triplet, [O I], and the electronic transitions of OH in the
UV) all have drawbacks.  Our spectra do not extend beyond 6000 \AA{}, so we
could only use the OH features in the UV.  In addition to the OH feature
between the two Be II lines at 3130.6 \AA, we used two additional OH features
at 3139 and 3140 \AA.  There are seven OH lines in the 3139 region and five OH
transitions in the 3140 region as well as atomic lines.  Our line lists for
the OH regions are from B. Gustafsson (private communication).

Figure 8 shows the syntheses for these regions in two of our stars with
different values of [Fe/H].  We used the value for the Gaussian smoothing that
we found in the Be II synthesis becaused it was constrained there by many more
features.  Similarly, if needed, we used those values for the continuum and
wavelength adjustments.  In the figure we show the best fit for O and for
amounts of $\pm$0.20 dex as well as the result with no O at all.

We derive an average [O/H] abundance by weighing the O abundance derived from
3130 \AA{} region by a factor of two That extra weight was used because 1) the
3130 \AA{} line list is better determined and 2) that region is closer to the
center of the echelle order and thus has a somewhat higher S/N.  Generally
the O abundance from these two OH features agree well with the that found from
the 3130 \AA{} feature.

We used our spectrum of the Moon (see Table 1) to find a solar abundance of O
from the same three OH features.  From this we find log O/H + 12.00 = 8.63
$\pm$0.08.  This is in good agreement with the revised value for the Sun of
Asplund et al.~(2009) of 8.69.  The work of Asplund \& Garc\'\i a Perez (2001)
indicates that the use of 3-D model atmospheres reduces the O abundance found
from the OH lines, and also reduces it more at lower values of [Fe/H].

\subsection{Alpha-Element Abundances: Ti and Mg}

The O abundances are quite sensitive to the values for log g and T$_{\rm eff}$
(see $\S$3.5 below).  There is an additional uncertainty from using-D instead
of 3-D model atmospheres.  We decided to use two other alpha-elements as
surrogates for O and thus have determined the abundances of [Ti/H] and [Mg/H].

We have measured equivalent widths of 5-19 Ti I lines and 5-13 Ti II lines.
We found Ti abundances with the {\it abfind} driver in MOOG.  The standard
deviation of the Ti abundance from the agreement of the Ti I lines is 0.06 -
0.11 dex and from the Ti II lines is 0.08 - 0.11 dex.  Table 3 gives our
results for [Ti/H] for the newly observed stars and for the stars in Rich \&
Boesgaard (2009).

For Mg we have used three lines of Mg I at 4571, 4703, and 4730 \AA.  For the
most metal-poor stars ([Fe/H] $\lesssim-$2.7) only the line at 4703 \AA{} was
strong enough to use.  Conversely, in the more metal-rich stars ([Fe/H] $\sim$
$-$0.5 to $\sim$$-$1.0) that line was too strong to use.  Table 3 also gives
the Mg abundances.

\subsection{Abundance Uncertainties}

Uncertainties in the abundances are due to uncertainties in the stellar
parameters and the quality of the data, including the S/N ratio.  We have
calculated the errors in the abundances for all five elements which are due to
the uncertainties in the stellar parameters.  We have used $\pm$100 K as the
uncertainty in T$_{\rm eff}$, $\pm$0.2 dex as the uncertainty in log g,
$\pm$0.10 dex as the uncertainty in [Fe/H], and $\pm$0.2 km s$^{-1}$ as the
uncertainty in $\xi$.  We have chosen three representative stars that cover a
range in parameters from 5768 to 6222 K in temperature, 3.04 to 4.07 in log g,
$-$1.35 to $-$2.79 in [Fe/H], and 1.14 to 1.54 km s$^{-1}$ in $\xi$; the
results are in Table 4.  The adopted abundance error is the quadrature sum of
the uncertainty from each parameter.

We can estimate the uncertainty in the equivalent width measurements from the
S/N values and the spectral resolution.  Eighty-seven percent of our stars
have spectra with S/N $>$ 80 and 60\% are $>$ 100.

As can be seen from Figure 2, the more metal-rich stars have many blended
lines in the Be II region; to determine reliable Be abundances the stellar
parameters and the line list used in the synthesis must therefore be
well-determined.  For the metal-poorer stars the spectrum is less crowded,
making the Be abundances easier to determine.  However, in the most metal-poor
stars the Be lines become very weak so it is especially important to obtain
high S/N spectra.  As can be noticed in Table 4 the Be abundance is
particularly sensitive to the log g value used.

The abundance of O from the OH features is particularly sensitive to the model
temperature and gravity.  Even though the three OH features we used give very
similar abundances in each star, the uncertainties in the stellar parameters
contribute substantial error.  In addition there is the possibility of a more
systematic trend seen when the 3-D model atmospheres are used (Asplund \&
Garc\'\i a Per\'ez 2001) as mentioned in $\S$3.3.

In the figures that follow we adopt these values as mean errors: [Fe/H]
$\pm$0.09; [Ti/H] $\pm$0.09; [Mg/H] $\pm$0.07; A(Be) $\pm$0.12; and [O/H]
$\pm$0.22. For the element ratios we adopt these values: [Ti/Fe] $\pm$0.13;
[Mg/Fe] $\pm$0.11; [Be/Fe] $\pm$0.15; [O/Fe] $\pm$0.24.

\section{RESULTS}

Parameters and abundances given for 117 stars in Table 2.  Only two of these
stars have upper limits on the Be abundances: the CEMP star, G 268-32, and LP
831-70 which has [Fe/H] = $-$3.06 and for which our S/N was only 52.

\subsection{Beryllium and Iron}

Figure 9 shows the relationship between [Fe/H] and A(Be) for our stars.  Two
stars with enriched Be (HD 106038 and HD 132475) were not included in the
determination of the least-squares fit to the data.  A linear fit between
these two logarithmic quantities is a good match over three orders of
magnitude in [Fe/H].

\begin{equation}
A(Be) = 0.877 (\pm0.030) [Fe/H] + 1.207 (\pm0.060)
\end{equation}

As Rich \& Boesgaard (2009) pointed out there is no substantial evidence for a
plateau of Be at the lowest metallicities.

Smiljanic et al.~(2009) list the linear relations found between Be and Fe by
several different studies and for four subsets of their own data.  Those
slopes are all steeper than ours because none of those studies has the large
number of very metal-poor stars that we have here.  Our single slope fit is
influenced by our stars with [Fe/H] $<$$-$2.2.  When we just consider the more
metal-rich stars with [Fe/H $>-$2.2, we find a slope of 1.04 $\pm$0.06.  This
agrees, within the errors, with the slope of 1.16 $\pm$0.07 found by Smiljanic
et al.~(2009) for their thick disk star sample.

When the Be results are normalized to the Fe abundance, [Be/Fe], as seen in
Figure 10, we can see that Be is somewhat enriched over Fe from [Be/Fe] =
+0.19 at [Fe/H] $\sim$$-$3.3 and reducing to [Be/Fe] = 0 at [Fe/H] = $-$1.7.
The formation of Be can occur in the earliest generations of massive stars
during supernovae when CNO atoms accelerate out from the explosion into the
ambient gas. These ejecta strike protons and neutrons at high energies and
split into smaller atoms like Li, Be, and B.  So it is not surprising that Be
is enhanced relative to Fe in the most metal-poor stars.  According to
Tsujimoto et al.~(1995) the relative contribution of SNe Ia to the solar Fe
abundance is 57\%.

\subsection{Beryllium and Oxygen}

Oxygen is directly connected to Be as a major ``mother'' nucleus for the rare
light elements through various spallation reactions.  The relationship we
found between A(Be) and [O/H] is shown in Figure 11.  There is more scatter in
this diagram than in Figure 9 which is due in part to the larger uncertainties
in [O/H] as seen in Table 4.  [O/H] is sensitive to both T$_{\rm eff}$ and log
g; the error bar shown here is $\pm$0.22.

The linear single-slope fit shown is:

\begin{equation}
A(Be) = 1.037 (\pm0.053) [O/H] + 0.893 (\pm0.073)
\end{equation}

The scatter of the data in Figure 11 does not seem to be random, but rather
the low O points are above the best fit line as are the high O points.  We
therefore tried a two-slope fit shown in Figure 12.  This was also done by
Rich \& Boesgaard (2009).  This change could be expected if the dominant
source of Be in the O-poor and Be-poor stars is the acceleration of CNO atoms
from SN II in the early days of Galactic evolution.  The number of Be atoms
would be proportional to the number of SN II and thus the number of O atoms.
The slope would be $\leq$1 (as modified to lower values by processes like mass
outflow during star formation).  When the dominant source of Be atoms is from
the classical GCR method with energetic cosmic rays hitting CNO atoms in the
ambient interstellar gas as detailed by Meneguzzi et al.~(1971), then the
number of O atoms would depend on the cumulative rate of supernovae and the
number of energetic cosmic rays proportional to the instantaneous rate of
supernovae.  The slope for the O-rich and Be-rich stars would be $\leq$2.  The
slope we find between [O/H] and A(Be) for the O- and Be-poor stars is 0.69
$\pm$0.13 and for the O-rich and Be-rich stars is 1.30 $\pm$0.10, consistent
with the notion of a change in the dominant production mechanism.

Those expressions are:

High-O and high-Be:

\begin{equation}
A(Be) = 1.295 (\pm0.100) [O/H] + 1.127 (\pm0.104)
\end{equation}

Low-O and low-Be:

\begin{equation}
A(Be) = 0.690 (\pm0.129) [O/H] + 0.275 (\pm0.252)
\end{equation}

The slope change does seem to be caused by the Be abundances rather than the O
abundances.  In Figure 13 we show the relationship between [Fe/H] and [O/H].
There is less scatter than found in Figures 9 (Be vs.~Fe) and 11 (Be vs.~O)
and a well-defined linear fit given by:

\begin{equation}
[O/H] = 0.749 (\pm0.025) [Fe/H] + 0.126 (\pm0.050)
\end{equation}

We have done two statistical tests (chi-squared and BIC = Bayesian information
criterion) to evaluate whether the data for A(Be) and [O/H] are better fit by
a single power law or a broken power law.  Both tests indicate that the
one-slope fit is better.

Figure 14 shows the O abundances normalized to the Fe abundances as a function
of the Fe abundances.  This too can be well-represented by a linear fit with
the scatter due to the mean error in [O/Fe].

\begin{equation}
[O/Fe] = -0.252 (\pm0.025) [Fe/H] + 0.121 (\pm0.050)
\end{equation}

\subsection{Beryllium and Alpha-Elements -- Ti and Mg}

As mentioned previously, the O abundance from the OH features has a large
dependence on both T$_{\rm eff}$ and log g from the models; in addition there
is the issue of the abundance found from 3-D models vs.~1-D models as
discussed above in $\S$3.3.  Therefore we have used two alpha-elements, Ti and
Mg, as surrogates for O.

Figure 15 shows the relationship between the abundance of Ti and Fe as well as
the Ti normalized to Fe compared to Fe.  There is a remarkably tight
correlation between [Ti/H] and [Fe/H] with a slope of 0.86 $\pm$0.01.  The
closeness of this relationship, and that between [Mg/H] and [Fe/H] (shown in
Figure 17), indicates that the stellar parameters we have derived are
well-determined.  The relationship between [Ti/Fe] and [Fe/H] with a slope of
$-$0.14 $\pm$0.01 is less steep than the one between [O/H] and [Fe/H] which
has a slope of $-$0.25 $\pm$0.02.

Figure 16 shows how A(Be) tracks [Ti/H] well with a slope of 1.00 $\pm$0.04.
As in Figure 11 of A(Be) vs.~[O/H] both the lowest values of [Ti/H] and the
highest lie above the best fit but not as dramatically as for [O/H].

\begin{equation}
A(Be) = 1.002 (\pm0.042) [Ti/H] + 1.069 (\pm0.072)
\end{equation}

The relationship between [Mg/H] and [Fe/H] and the one between [Mg/Fe] and
[Fe/H] are shown in Figure 17.  There is a surprisingly close correlation
between [Mg/H] and [Fe/H] with a slope of 0.94 $\pm$0.01.  This is impressive
in part because we have only 1-3 Mg I lines and only the strongest one could
be used in the Fe-poor stars.  The relationship between [Mg/Fe] and [Fe/H]
(slope = $-$0.07 $\pm$0.01) is even flatter than those of [Ti/Fe] with [Fe/H]
and [O/H] with [Fe/H].

The plot of [Mg/H] (as a surrogate for [O/H]) with A(Be) is shown in Figure
18.  The slope of this relationship is 0.87 $\pm$0.03 which is very similar
to the slope between A(Be) and [Fe/H] of 0.87 $\pm$0.03.  The relation between
A(Be) and [Mg/H] is

\begin{equation}
A(Be) = 0.870 (\pm0.033) [Mg/H] + 0.815 (\pm0.056)
\end{equation}

\subsection{Lithium}

Table 3 also gives the abundance of Li from the literature for most of our
stars along with the reference for each.  This is not meant to be a
comprehensive compilation where multiple studies of a given star are combined
in some way.  We did make use of the compilation done by Charbonnel \& Primas
(2005) when the Li abundances were available for our stars to give some
consistency.  Our purpose here is only to compare the Li and Be abundances in
a general way and to check for potential Be depletions in Li-depleted stars.

In Figure 19 we show our Be abundances compared to Li abundances found in the
literature.  Most of our sample have values of A(Li) near 2.2, corresponding
to the halo star Li plateau reflecting the Big Bang production of Li as first
found by Spite \& Spite (1982).  There are seven stars in our sample that are
Li-deficient: BD +37$\arcdeg$ 1458 at A(Li) = 1.37, HD 64090 at 1.21, HD
109303 at $<$1.65, HD 188510 at 1.61, HD 201889 at 1.04, G 74-5 at 1.48, and
Ross 390 at 1.15.  Although Li is depleted in those seven stars, the Be
abundances are apparently normal for their [Fe/H] values.  Four of the seven
have T$_{\rm eff}$ $<$ 5600 K (HD 64090, HD 188510, HD 201889, and BD
+37$\arcdeg$ 1458); these temperatures are cool enough for Li depletion to
have occurred (e.g.~Boesgaard et al.~2005).  Only the coolest, BD +37$\arcdeg$
1458 at 5492 K, might be mildly Be-deficient as judged by its position in
Figure 9 where it is 0.35 dex below the fit.  The other three stars have
[Fe/H] values $<-$0.9, which are too metal-rich to be part of the halo star Li
plateau; they may have suffered Li depletion like that in Pop I stars.

We could not find Li abundances in the literature for 21 of our 117 stars.
Only one, HD 184499, may be cool enough to have had Li depletion, but it has
normal Be.  Our two stars with enhanced Be are also enhanced in Li.  HD 106038
with A(Be) = 1.47 has A(Li) = 2.48 (Asplund et al.~2006) and 2.55 (Tan et
al.~2009).  HD 132475 was found by Novicki (2005) to have enhanced Li with
A(Li) = 2.39.  Boesgaard \& Novicki (2006) determined a Be abundance of A(Be)
= +0.57 and Tan et al.~(2009) found A(Be) = +0.62, i.e.~it is Be-rich for its
[Fe/H] of $-$1.50, as seen in Figure 9.

\subsection{Kinematics}

Our kinematic classification of these stars is drawn from Gratton et al.\
(2003), who use Galactic orbit calculations to determine criteria for
distinguishing between stellar populations corresponding to different Galactic
components. Stars with a galactic rotation velocity larger than 40 km s$^{-1}$
and an apogalactic distance ($R_{apo}$) less than 15 kpc comprise a kinematic
class associated with the dissipative collapse population (Eggen et al.\
1962), including stars from the classical thick disk and halo. The remainder
of the stars in our sample can be associated with the accretion-process
population first proposed by Searle \& Zinn (1978). These are mainly halo
stars, a subset of which can be further distinguished due to their retrograde
orbits.  Retrograde stars have a V velocity of $<-$220 km s$^{-1}$.  The $UVW$
velocities have positive $U$ values away from the Galactic center, positive
$V$ values in the direction of the solar motion, and positive $W$ values
paralleling the direction of the north Galactic pole.  To convert these values
relative to the local standard of rest we used solar values of $U_{\odot}$ =
$-$9 km s$^{-1}$, $V_{\odot}$ = +12 km s$^{-1}$, and $W_{\odot}$ = +7 km
s$^{-1}$.  Here we define a star's Galactic rotation velocity as $v_{rot} =
V + 220$ km s$^{-1}$, where 220 km s$^{-1}$ is the rotational velocity of the
local standard of rest with respect to the Galaxy (Fulbright 2002).  The
Galactic rest-frame velocities, $v_{RF}$, are {[$U^2$ + ($V$ + 220)$^2$ +
$W^2$]$^{1/2}$.

Kinematic properties for our full sample are given in Table 5, which lists the
stellar radial velocity, the $U_{lsr}$, $V_{lsr}$, and $W_{lsr}$ velocities
relative to the local standard of rest, all in km s$^{-1}$, the distance to
the apogalaticon of the orbit ($R_{apo}$), and the distance above the Galactic
plane $Z_{max}$, both in kpc.  The reference for the orbital parameters is
included in Table 5, with the majority from the compilation of Carney et
al.~(1994).  For each star we also list $v_{RF}$ in km s$^{-1}$.  We include
our final classification of each star as dissipative (D) or accretive (A) or
accretive/retrograde (A,R).  In total our sample consists of 57 dissipative
stars and 57 accretive stars (33 of which are also classified as retrograde).
For two stars in our sample (HD 24289 and BD $-$13 3442), we were unable to
make a conclusive kinematic classification as no data was available in the
literature.

The distribution of A(Be) with [Fe/H] for the four groups -- all stars,
dissipative, accretive, and retrograde -- is shown in Figure 20.  The stars in
the dissipative group span nearly the full range in [Fe/H].  The slope we find
for them is 0.94 $\pm$0.04.  However, the slopes for the accretive and
retrograde stars are both lower at 0.68 $\pm$0.04.  These slopes are different
at the 4.6 $\sigma$ level.  For comparison, in Figure 21 we have plotted the
accretive stars and the dissipative stars with the best fit linear relations
overplotted.  They separate well into two distinctive groups.  The slopes
intersect at [Fe/H] near $-$2.2 which somewhat blurs the distinction in A(Be)
for low metallicity stars in the two populations.  We suggest that the
accretive stars may have originated in environments where less Be was formed.
They may have had their Be formed in the vicinity of SN II through the
acceleration of CNO atoms into the surrounding gas as discussed in $\S$4.2.

Similar plots between A(Be) and [O/H] are shown in Figures 22 and 23.  In
Figure 22 we see that the stars in the dissipative group span nearly the full
range in [O/H].  The slope between A(Be) and [O/H] for this group is 1.13
$\pm$0.08.  As with Be and Fe, the accretive and retrograde stars have a
shallower slope for Be and O of 0.76 $\pm$0.06.  These slopes are different at
the 3.6 $\sigma$ level.  The two-slope fit that we see in Figure 12 of 1.30
$\pm$0.10 and 0.69 $\pm$0.13 may be partly due to the fact that there are two
populations of stars.  And this in turn may result from the formation of a
larger fraction of Be made by the CNO cosmic rays produced by SN II for the
accretive stars while most of the dissipative stars inherited Be from the
galactic cosmis rays.  A direct comparison between those two groups in Figure
23 shows that they follow distinct relations, but again the intersection of
the two fits occurs at [O/H] $\sim$ $-$1.9 and the differences at Low [O/H]
are cloudy.

In Figure 24 we show the distribution of the Be abundances with the rest-frame
velocity ($v_{RF}$).  There are 4 panels showing the distribution for all
stars, for just the dissipative stars, for just the accretive stars, and for
just the retrograde stars.  For the dissipative stars the full range of Be
abundances is present, but there are no stars with $v_{RF}$ $>$ 280 km
s$^{-1}$ (by definition).  For the accretive and retrograde stars there are no
stars with A(Be) $>$ +0.35 with the exception of the two with enriched Be (HD
106038 and HD 132475).  Based on the A(Li) values and the positions in the
A(Be)-[Fe/H] plane, there is no evidence that there has been Be depletion
(with the possible exception of BD +37$\arcdeg$ 1458).  Therefore it seems
plausible that the accretive and retrograde stars come from an environment
that has {\it not} been as enriched in Be as the dissipative sample was.  This
is in qualitative agreement with Tan \& Zhao (2011) who find that low-$\alpha$
(accretive) stars have generally lower Be abundances than high-$\alpha$
although their sample is only from [Fe/H] = $-$0.49 to $-$1.55.

Figure 25 shows the distribution of Be normalized to Fe ([Be/Fe]) with the
rest-frame velocity ($v_{RF}$).  The accretive stars and their subset of
retrograde stars show a flat distribution of [Be/Fe] with $v_{RF}$; the
star with the highest velocity at 409 km s$^{-1}$ is G 64-12 which has low
values of A(Be) = $-$1.43 and [Fe/H] = $-$3.45 but a comparatively high value
for [Be/Fe] = +0.60.  The dissipative stars may be connected to $v_{RF}$ with
the lower velocity stars showing higher values of [Be/Fe].

\subsection{Beryllium as a Cosmochronometer}

Pasquini et al.~(2005) have suggested that A(Be) in stars can track early star
formation in the Galaxy.  They use [O/Fe] to reveal the variation in the star
formation rate and A(Be) as a measure of time, a cosmochronometer, from the
beginning of star formation.  They use the Galactic evolution model of
Travaglio et al.~(1999).  Model predictions have been made by Valle et
al.~(2002) for Li, Be, and B.  They take the stellar production of Fe and O
from Thielemann et al.~(1996) for SNe I and from Woosley \& Weaver (1995) for
SN II.

From observations of Be, Fe, and O from Boesgaard et al.~(1999) they compare
[O/Fe] with A(Be) and find good agreement between the data for the dissipative
stars and the model for the thick disk (see their Figure 2).  That sample was
only eight stars, however.  Our sample size here is 116, excluding the CEMP
star, G 268-32.  

Figure 26 shows our data for [O/Fe] and A(Be) for all 116 stars.  We show a
linear fit and the 1$\sigma$ error bars for [O/Fe].  The relationship is:

\begin{equation}
[O/Fe] = -0.269 (\pm0.029) A(Be) + 0.476 (\pm0.023)
\end{equation}

We have also separated the data into dissipative, accretive, and retrograde
sub-samples of 57, 57, and 33 stars.  There is virtually no difference between
the fits for the full sample and these sub-samples.  The slopes are $-$0.26
$\pm$0.04, $-$0.25 $\pm$0.05, and $-$0.29 $\pm$0.08, respectively.  The curve
fit from Pasquini et al.~(2005) does not match the data as well as the
straight line fit.  Note that in Figure 14 we have plotted [O/Fe] with [Fe/H]
as the abscissa (as opposed to A(Be) as the abscissa in Figure 26).  In both
plots there is a larger spread at higher values of the abscissa: [Fe/H]
$>-$1.4 and A(Be) $>$ 0.0.  The vast majority of these stars belong to the
dissipative population.  We conclude that Be is no better as a
cosmochronometer than Fe, and may in fact be worse.

\subsection{Spread in Beryllium}

Our Figure 7, top panel, shows two stars with similar metallicity and log g
values, but very different Be abundances.  Smiljanic et al.~(2009) show four
pairs of stars from their paper in Figure 14 with similar parameters and very
different values for A(Be).  Another compelling example can be seen in Figure
14 of Boesgaard \& Hollek (2009) of two stars differing by 35 K in
temperature, 0.07 in log $g$, 0.06 in [Fe/H], but differing by a factor of 2
in both A(Li) and A(Be); their masses are similar at 0.95 and 0.96 $M_{\odot}$
as are their ages at 9.17 and 9.35 Gyr.  These examples seem to indicate that
there is a real spread in the Be abundance in stars of similar stellar
parameters.

We have examined the possibility that there is a real spread in Be at a given
interval of [Fe/H], [O/H], [Ti/H], and [Mg/H].  Our approach is to determine
whether the data show unusually large scatter within a given range of [X/H]
compared to the rest of the observations outside that interval.  This is done
using a prediction interval, which is the expected range of an unobserved
random variable $Y$ (here abundance of Be) at a given confidence (defined by
100(1-$\alpha$)\%) based on a set of observed data (Casella \& Berger 2002,
Section 11.3).  We assume the data are independent, obey a linear
relationship, and are normally distributed about the line in $y$.

For each interval of interest $\Delta$[X/H] and confidence limit $\alpha$ we
use observations outside that region to derive the best-fit slope, $y$-offset,
and prediction interval in that region of interest.  We determine how many
points within $\Delta$[X/H] lie inside and outside the prediction interval and
then use the binomial distribution to calculate the probability of this
occurring by chance.  If the probability is $<$0.0063\% ($>$4$\sigma$) and is
not highly sensitive to either the width of $\Delta$[X/H] nor the chosen
confidence limit $\alpha$, then we interpret that as evidence for a spread in
A(Be) for that particular $\Delta$[X/H].

We ran our analysis for [Fe/H], [O/H], [Ti/H], and [Mg/H] with a running
interval in steps of 0.01 in [X/H] for $\Delta$[X/H] = 0.50, 0.30, and 0.20
($\pm$0.25, $\pm$0.15, $\pm$0.10) and confidence levels $\alpha$ = 0.01, 0.05,
and 0.10 (99\%, 95\%, 90\%).  No significant spread in A(Be) was observed in
[Fe/H], [Ti/H], or [Mg/H].  We did, however, find evidence of a spread for
[O/H] from $\sim$$-$0.5 to $-$1.1 dex.  In Figure 27 we show an example of the
analysis for A(Be) vs.~[O/H].  At the [O/H] value of $-$0.8 dex, an interval
of $\Delta$[O/H] = 0.50, and a confidence limit of 95\% for the prediction
interval, 10 out of 30 stars fall outside the prediction interval (top panel).
The probability of this occurring by chance given the 5\% expectation rate is
1.1$\times$10$^{-6}$ (middle panel).  Our results are independent of whether
we include or exclude the two anomalously Be-rich stars HD~106038 and
HD~132475 (bottom panel).  The spread is real at the 4$^+$$\sigma$ level of
confidence.  It is also real at the 3 $\sigma$ level of confidence when our
$\Delta$[O/H] = 0.30.

The interval where there is the most spread in Be is between [O/H] = $-$0.5 to
$-$1.0, which corresponds to $-$0.9 to $-$1.6 in [Fe/H].

\section{SUMMARY AND CONCLUSIONS}

The study of the light elements - Li, Be, and B - is important in advancing
our knowledge of cosmology, Galactic evolution, and stellar structure and
evolution.  We have made observations to determine Be abundances in 117 stars,
most of which were obtained from the upgraded HIRES spectrometer on the Keck
10-m telescope.  Our spectral resolution is $\sim$42,000 and our median S/N
ratio is 106 per pixel.  We have also found abundances of Fe, O, Mg, and Ti.
The Be and O abundances were found from spectrum synthesis from the Be II
resonance lines and the OH electronic transitions in the ultraviolet spectral
region near 3100 \AA.  Equivalent widths were measured for Fe I, Fe II, Ti I,
Ti II, and Mg I lines.

We have determined the stellar parameters spectroscopically from the Fe I
lines for $T_{\rm eff}$ and the ionization balance of Ti I and Ti II or
occasionally from Fe I and Fe II for log g.  The value for [Fe/H] comes from
the Fe I lines.  We have made error estimates from the uncertainties in the
stellar parameters for three stars which cover the range in our stellar
parameters in $T_{\rm eff}$, log g, [Fe/H], and $\xi$.

One of our stars, G 268-32, is a carbon-enhanced metal-poor star (CEMP)
according to Aoki et al.~(2002).  It has [Fe/H] = $-$2.51.  Like BD +44 493, a
CEMP star studied by Ito et al.~(2009), G 268-32 seems to be devoid of Be and
we set an upper limit of A(Be) = $-$1.5.  Although the compilation by
Charbonnel \& Primas (2005) gives A(Li) = 1.99 for G 268-32 (HIP 3446), when
we use our model and Thorburn's (1994) equivalent width for Li, we find A(Li)
= 2.32.

We found a linear fit between the [Fe/H] and A(Be) over three orders of
magnitude in [Fe/H] with a slope of 0.88 $\pm$0.03.  Although we found no
support for the idea of a plateau in Be at low metallicities, we did find that
Be is enriched relative to Fe changing from [Be/Fe] near +0.2 at [Fe/H] near
$-$3.3 down to $\sim$0.0 near [Fe/H] $\sim-$1.7.  This supports the idea that
much of the Be was formed by the ``supernovae'' mechanism in the early
history of the Galaxy.  The Fe is formed primarily from intermediate mass
stars from SNe Ia.

We examined the relationship between Be and O because they are directly
related as the light elements are formed by spallation with CNO atoms.  This
relationship can also be fit by a straight line with a slope of 1.04 $\pm$0.05
over two and a half orders of magnitude in [O/H].  There is more scatter in
the Be-O data than in the Be-Fe data and the scatter is not uniformly
distributed; both the lowest Be points and the highest Be points lie above the
best fit line.  We tried fitting the results with two lines dividing the
sample at [O/H] = $-$1.6.  The Be-poor and O-poor stars are fit with a slope
of 0.69 $\pm$0.13 while the Be-rich/O-rich stars have a steeper slope of 1.30
$\pm$0.10.  A change of slope between A(Be) and [O/H] would be expected due to
the change in the dominant mechanism of Be formation going from the
``supernovae'' mechanism to the classical GCR mechanism.  It is clear that
the change in slope is caused by the Be abundance rather than by the O
abundance as indicated by the smaller scatter and clear linear fit between
[Fe/H] and [O/H].

The O abundance found from the UV lines of OH is sensitive to both the
temperature and the gravity.  Therefore, we decided to use the alpha-elements,
Ti and Mg, as surrogates for O.  We found a remarkably tight correlation
between [Ti/H] and [Fe/H] and between [Mg/H] and [Fe/H].  This indicates that
our stellar parameters are well-determined.  The slope between [Mg/H] and
[Fe/H] is 0.94 $\pm$0.01.

We found that A(Be) tracks [Ti/H] very well with a slope of 1.00 $\pm$0.04
and that A(Be) tracks [Mg/H] very well with a slope 0.88 $\pm$0.03.  As with
the Be vs.~O relation, both the lowest Be stars and the highest Be stars lie
above the linear fit which may indicate that two slopes would be a better
representation of the data.

We have searched the literature for Li abundances in our stars and found
results for 96 of our 117 stars.  Most of our stars have Li abundances near
the observed Li plateau of A(Li) $\sim$2.2.  Seven stars were below A(Li) =
1.8.  With the possible exception of our coolest star, BD +37$\arcdeg$
1458, none of the Li-depleted stars are Be-deficient for their {fe/H] values.
If BD +37$\arcdeg$ 1458 is depleted in Be it is only a mild depletion.  We
note that our two stars with enhanced Be, HD 106038 and HD 132475, also have
enhanced Li.

Gratton et al.~(2003) have established kinematic criteria for distinguishing
between a dissipative collapse population (mainly thick disk and halo stars)
and an accretive population (mainly halo stars).  We have kinematic
information on 114 stars and our sample is equally divided between the two
groups.  Most of our stars have prograde motion, but 33 are on retrograde
orbits with V $<$ $-$220 km s$^{-1}$.  The stars in all three groups cover
nearly the full range in both [Fe/H] and [O/H].  

The dependence of A(Be) on [Fe/H] and on [O/H] shows distinct differences
between the dissipative and the accretive groups.  The dissipative stars have
a steeper slope of A(Be) with both [Fe/H] and [O/H] than the accretive and
retrograde stars.  For the Be relationship with [Fe/H] the slope for the
accretive stars is 0.68 $\pm$.04 and for the dissipative stars is 0.94
$\pm$.04.  The slopes of A(Be) with respect to [O/H] are both steeper than
with [Fe/H]: 0.76 $\pm$0.06 for the accretive stars and for 1.13 $\pm$0.08 the
dissipative stars.  The Be in the accretive stars may have made in the
vicinity of SN II stars in the early days of the Galaxy formation and
accretion of small stellar systems.  The Be in the dissipative stars may have
been made preferentially by GCR spallation reaction.

We have found that there are no stars in the accretive and retrograde groups
with A(Be) $>$0.35, except for the two stars with enhanced Be.  Those two
stars are thought to have originated in the vicinity of a hypernova or in
entrained superbubbles with multiple supernovae (Boesgaard \& Novicki 2006,
Smiljanic et al.~2009).  In general, the accretive and retrograde stars were
not formed in an environment rich in Be.

The distribution of [Be/Fe] with the rest frame velocity is flat for the
accretive and retrograde stars with the mean over the range of 400 km s$^{-1}$
near [Be/Fe] = 0.0.  The dissipative stars have higher [Be/Fe] at low values
of $v_{RF}$.

It was suggested by Suzuki \& Yoshii (2001) and Pasquini et al.~(2005) and
then further explored by Smiljanic et al.~(2009) that the stellar abundance of
Be can be used as a chronometer.  In our large sample of 116 stars we find
that Be is not as good as Fe as a chronometer.  In the usual plot of [O/Fe]
vs.~A(Be) there is more scatter in the data than in the plot of [O/Fe]
vs.~[Fe/H].  The Be abundance is less well-determined than the Fe abundance.
If we divide the stars into the dissipative and accretive (and retrograde)
subgroups, the fit to the data is no different from the fit with the full
sample.  For all plots there is more scatter in A(Be) for data with A(Be)
$>$0.0 and with [Fe/H] $>$ $-$1.4, i.e.~for the more recently formed stars.

Related to the use of Be as a chronometer is the issue of whether there is a
spread in Be at a given metallicity, as measured by Fe, O, Ti, Mg.  There is
empirical evidence for a spread through comparisons of Be in pairs of stars
with the same stellar parameters.  Smiljanic et al.~(2009) show four such
pairs.  Boesgaard et al.~(2010) have done statistical tests which show a real
dispersion in A(Be) at a given [O/H] at the 4$\sigma$ level and at a given
[Fe/H] at the 3$\sigma$ level.  With a similar analysis done here with the
larger sample, we find more support for the reality of the spread in A(Be)
with [O/H].  There is evidence at the 4$\sigma$ level for a spread in A(Be) at
[O/H] from $-$0.5 to $-$1.0 which corresponds to [Fe/H] = $-$0.9 to $-$1.6

\acknowledgements We are grateful to the various Support Astronomers and
Observing Assistants who helped us over the many observing runs for this
project.  We are indebted to Hai Fu for his IRAF modification that enabled us
to measure equivalent widths quickly and easily.  We thank Gabriel Dima for
determining Li abundances in six of our stars.  We acknowledge support from
NSF through grant AST 05-05899 to A.M.B.

\clearpage
\tightenlines
\singlespace
\centering
\begin{deluxetable}{lrlrllclcl} 
\footnotesize
\tablewidth{0pc}
\tablecolumns{10}
\tablenum{1} 
\tablecaption{Log of the Beryllium Observations} 
\tablehead{ \colhead{Star}  & HIP & Other &  \colhead{Code\tablenotemark{a}} &
\colhead{R.A.}  
& \colhead{Dec.}  & \colhead{V} & \colhead{Date U.T.} & \colhead{Exp.} & 
\colhead{S/N} 
}

\startdata
G 130-65     & \nodata & LP 349-6 & 1 & 00 22 &  +23 54 & 11.64  & 2004 Nov 18 & \phn 60 &\phn 59 \\
G 268-32     &  3446 & LP 706-7 & 2  & 00 44 & $-$13 55 & 12.10 & 2004 Sep 2 & & \\
	     &  &    &       & &         &       & 2004 Nov 18 & & \\
	     &  &	  &	&  &	     &       & 2006 Jan 2 & 270 & 113 \\
BD +4 302    & \nodata & G 3-16 & 3  & 01 43 & +04 51   & 10.47 & 2007 Nov 7 & & \\
             &  &    &   &    &          &       & 2005 Sep 27 & \phn 90 & 151 \\
BD +2 263    & \nodata & G 71-33 & 4  & 01 45 & +03 30   & 10.63 & 2005 Sep 27 & \phn 50 & 108 \\
BD $-$10 388 & 8572 & G 271-162 & 5  & 01 50 & $-$09 21 & 10.37 & 2004 Nov 7  & \phn 55 & 123 \\
G 74-5       & 10140 & BD +29 366  & 6  & 02 10 & +29 48   & \phn 8.77 & 2008 Jan 16 & \phn 10 & 137\\ 
BD $-$1 306  & 10449 & G 159-50 & 7 & 02 14 & $-$01 12 & \phn 9.09 & 2008 Jan 16 & \phn 15 & 135 \\
BD $-$9 466  & \nodata & NLTT 8103 & 8  & 02 28 & $-$08 59 & 11.19 & 2006 Jan 2 & \phn 30 & \phn 89\\
BD $-$17 484 & 11729 & LTT 1242 & 9  & 02 31 & $-$16 59 & 10.43 & 2008 Jan 16 & \phn 45 & 108 \\
HD 16031     & 11952 & BD $-$13 482 & 10 & 02 34 & $-$12 23 & \phn 9.78 & 2006 Jan 2 & \phn 20 & 135 \\
BD +9 352    & 12529 & G 76-21 & \nodata   & 02 41 & +09 46   & 10.17  & 2004 Nov 18 & \phn 50 & 115 \\
G 4-37       & 12807 & G 76-26 & 11 & 02 44 & +08 28   & 11.43  & 2005 Sep 27 & \phn 90 & 104 \\
BD +22 396   & 13111 & G 5-1 & 12 & 02 48 & +22 35   & 10.10  & 2007 Nov 19 & \phn 20 & \phn 69 \\
G 5-19       & \nodata & LHS 6057 & 13 & 03 11 & +12 37   & 11.12 & 2005 Sep 27 & \phn 60 & \phn 90 \\
LTT 1566     & 15396 & Ross 570 &14 & 03 18 & $-$07 08 & 11.22 & 2004 Nov 18 & \phn 90 & \phn 86 \\ 
CD $-$24 1656 & 16063  & NLTT 10967 & 15 & 03 26 & $-$23 43 & 10.89 & 2005 Jan 31 & & \\
	     &  &	&  &  &	    &       & 2006 Jan 2 & \phn 79 & \phn 82 \\
HD 24289     & 18082 & BD $-$4 680 & 16 & 03 51 & $-$03 49 & \phn 9.96 & 2007 Nov 7 & \phn50 & 126 \\
BD +21 607   & 19797 & HD 284248 & 17 & 04 14 & +22 21   & \phn 9.22 & 2004 Nov 18 & \phn 15 &\phn 98 \\
HD 31128     & 22632 & CD $-$27 1935 & 18 & 04 52 & $-$27 03 & \phn 9.14 & 2008 Jan 16 & \phn 30 &\phn 76 \\ 
HD 241253    & 24030 & G 97-22 & 19 & 05 09 & +05 33   & \phn 9.72 & 2004 Nov 18 & \phn 25 & \phn 95\\
HD 247168    & 27111 & BD +9 946 & 20 & 05 44 & +09 14   & 11.82 & 2004 Nov 18 & \phn 90 & \phn 68 \\ 
HD 247297    & 27182 & BD +14 1018 & 21 & 05 45 & +14 41   & \phn 9.11 & 2005 Sep 27 & \phn 12 &\phn 99 \\
Ross 797     & 27880 & LTT 2402 & 22 & 05 53 & $-$14 22 & 11.47 & 2006 Jan 2 & \phn 60 & \phn 89 \\
G 191-55     & \nodata & BD +58 876 & 23 & 05 57 & +58 40   & 10.47 &2005 Jan 31 & \phn 90 & \phn 98 \\ 
BD +19 1185  & 28671 & HD 250792 & 24 & 06 03 & +19 21   & \phn 9.32 & 2005 Sep 27 & \phn 14 & 108 \\
BD +37 1458  & 29759 & G 101-29 & 25 & 06 16 & +37 43   & \phn 8.92 & 2005 Jan 31 & \phn 30 & 145 \\
G 192-43     & 32567 & G 193-4 & 26 & 06 47 & +58 38   & 10.32  & 2007 Nov 19 & \phn 60 & 106 \\
G 88-10      & 34630 & LTT 11991 & 27 & 07 10 & +24 20  &  11.86  & 2007 Nov 19 & & \\
             &  &  &	&       &     &         & 2008 Jan 16  & 150 & 121 \\ 
G 108-58     & \nodata & LP 708-12 & 28  & 07 10 & $-$01 17 & 11.82  & 2008 Jan 16 & \phn 90 & \phn 91 \\
\\
\\
\\
\\
G 90-3       & 36430 & LTT 17974 & 29 & 07 29 & +32 51   & 10.50  & 2004 Nov 7 & \phn 56 & 123 \\
BD +24 1676  & 36513 & G 88-32 & 30 & 07 30 & +24 05   & 10.80  & 2004 Nov 18 & 125 & 140 \\
Ross 390     & 36878 & LTT 2886 & 31 & 07 34 & $-$10 23 & 11.10 & 2006 Jan 2 & \phn 60 & \phn 96 \\
G 113-9      & \nodata & NLTT 18799 & 32 & 08 00 & $-$04 05 & 11.00 & 2008 Jan 16 &  \phn 60 & 100 \\
G 113-22     & \nodata & BD +0 2245 & 33 & 08 16 & +00 01 & \phn 9.69 & 2006 Jan 2 &  \phn 30 &140 \\
HD 233511    & 40778 & BD +54 1216 & 34 & 08 19 & +54 05 & \phn  9.71 & 2005 Jan 31 & \phn 30 & 108 \\
BD +39 2173  & \nodata & G 115-34 & 35 & 08 55 & +38 39 & 11.22 & 2007 Nov 19 & \phn 60 & \phn 89 \\
BD $-$3 2525 & 44124 & G 114-26 & 36 & 08 59 & $-$04 01 & \phn 9.67 & 2004 Nov 7 & \phn 30 & 130 \\
G 115-49     & 44605 & LTT 12383 & 37 & 09 05 & +38 47 & 11.60 & 2007 Nov 19 & \phn 60 & \phn 81 \\
G 10-4       & 54639 & G 45-38 & 38 & 11 11 & +06 25 & 11.41 & 2005 May 15 & & \\
             &  &  &	&       &     &         & 2006 Jan 2 & 120 & \phn 99 \\
BD +36 2165  & 54772 & LTT 13019 & 39 & 11 12 & +35 43 & \phn 9.75 & 2005 Jan 31 & \phn 35 & 110 \\
BD +51 1696  & 57450 & LTT 13244 & 40 & 11 46 & +50 52 & \phn 9.90 & 2005 Jan 31 & & \\
             &  &  &	&          &  &         & 2005 Apr 1 & \phn 43 & 117 \\
HD 104056    & 58443 & BD $-$3 3216 & 41 & 11 59 & $-$04 46 & \phn 9.01 & 2006 Jan 2 & \phn 12 & 139 \\
BD $-$4 3208 & 59109 & G 13-9 & 42 & 12 07 & $-$05 44 & \phn 9.99 & 2005 Apr 1 & \phn 60 & 109 \\ 
HD 106038    & 59490 & BD +14 2481 & 43 & 12 12 & +13 15 &  10.16 & 2008 Jan 16 & \phn 25 & 107 \\
HD 106516    & 59750 & BD $-$9 3468 & \nodata & 12 15 & $-$10 18 & \phn 6.11 & 2007 Jan 12 & \phn 17 & 276 \\
HD 108177    & 60632 & BD +2 2538 & 44 & 12 25 & +01 17 & \phn 9.66 & 2005 Jan 31 & \phn 30 & \phn93 \\ 
HD 109303    & 61264 & BD +50 1929 & 45 & 12 33 & +49 18 & \phn 8.15 & 2010 Jul 4 & \phn\phn 6 & 110 \\ 
BD +28 2137  & 61545 & G 59-27 & 46 & 12 36 & +27 28 & 10.86 & 2007 Jun 10 & \phn 30 & 100 \\
G 63-46      & 66665 & BD +13 2698 & 47 & 13 39 & +12 35 & \phn 9.37 & 2006 Jun 19 & \phn 20 & \phn 93 \\
BD +34 2476  & 68321 & G 165-39 & 48 & 13 59 & +33 51 & 10.06 & 2006 Jun 19 & \phn 55 & 123 \\
G 64-37      & 68592 & LTT 5476 & 49 & 14 02 & $-$05 39 & 11.14 & 2010 Jul 4 & 150 & 139 \\
BD +26 2606  & 72461 & G 166-45 & \nodata & 14 49 & +25 42 & \phn 9.72 & 2006 Jun 19 & \phn 25 & 104 \\
BD +26 2621  & 72920 & G 166-54 & 50 & 14 54 & +25 34 & 10.99 & 2005 Jan 31 & 60 & 125 \\
G 153-21     & 78620 & BD $-$6 4346 & 51 & 16 03 & $-$06 27 & 10.20 & 2007 Jun 10 & \phn 30 & \phn 94 \\
G 180-24     & 78640 & BD +42 2667 & 52 & 16 03 & +42 14 & \phn 9.85 & 2006 Jun 19 & \phn 30 & \phn86 \\
G 181-28     & \nodata & LTT 15067 & 53 & 17 07 & +34 21 & 12.02 & 2010 Jul 4 & 150 & 107 \\
BD +2 3375   & 86443 & G 20-8 & 54 & 17 39 & +02 24 & \phn 9.93 & 2006 Jun 19 & \phn 45 & 115\\
BD $-$8 4501 & 87062 & G 20-15 & 55 & 17 47 & $-$08 46 & 10.59 & 2004 Sep 7 & \phn 55 & \phn 77 \\
HD 161770    & 87101 & BD $-$9 4604 & 56 & 17 47 & $-$09 36 & \phn 9.66 & 2004 Sep 7 & \phn 25 & \phn78 \\
BD +36 2964  & 87467 & G 182-31 & 57 & 17 52 & +36 24  &10.37 & 2004 Sep 7 & \phn 40 & \phn87 \\
BD +20 3603  & 87693 & G 183-11 & 58 & 17 54 & +20 16  & \phn 9.69 & 2006 Jun 19 & \phn 30 & 102 \\
G 20-24      & 88827 & BD +1 3579 & 59 & 18 07 & +01 52  & 11.09 & 2007 Jun 10 & \phn 60 & 110 \\
\\
\\
\\
BD +13 3683  & 90957 & G 141-19 & 60 & 18 33 & +13 09  & 10.55 & 2005 Jul 5 & & \\
             &  &    &	  &       &   &      & 2006 Jun 19 & \phn 80 & \phn 93 \\  
HD 179626    & 94449 & BD $-$0 3676 & 61 & 19 13 &  $-$00 35 & \phn 9.14 & 2005 Jul 6 & \phn 25 & \phn 70 \\
HD 188510    & 98020 & BD +10 4091 & 62 & 19 55 & +10 44 & \phn 8.82 & 2010 Jul 4 & \phn 15 & 132 \\
G 24-3       & 98989 & G 92-52 & 63 & 20 05 & +04 02 & 10.44 & 2005 Jul 5 & \phn 50 & \phn 95 \\
BD +42 3607  & 99267 & G 125-64 & 64 & 20 09 & +42 51 & 10.11 & 2007 Jun 10 & \phn 30 & 120 \\
BD +23 3912  & \nodata & HD 345957 & 65 & 20 10 & +23 57 & \phn 8.93 & 2004 Nov 7 & \phn 25 & 113 \\
HD 194598    & 100792 & BD +9 4529 & \nodata & 20 26 & +09 27 & \phn  8.36 & 2004 Nov 7 & \phn 10 & 132 \\
G 24-25	     & \nodata & LTT 16036 & 66 & 20 40 & +00 33 & 10.61 & 2010 Jul 4 & \phn 15 & \phn 56 \\  
BD $-$14 5850 & 102602 & Ross 190 & 67 & 20 47 & $-$14 25 & 10.96   & 2004 Nov 18 & & \\
             & &  &	 &          &  &         & 2005 Sep 27 & 140 & 103 \\
BD +4 4551   & 102718 & SAO 126242 & 68 & 20 48 & +05 11 & \phn 9.69  & 2004 Nov 7 & \phn 20 & \phn90 \\
G 26-12	     & 106447 & LP 638-7   & 69 & 21 33 & +00 23 & 12.15 & 2010 Jul 4 & 150 & \phn96 \\
G 188-22     & 107294 & BD +26 4251 & 70 & 21 43 & +27 23 & 10.05 & 2006 Jun 19 & \phn 30 & 121 \\
BD +19 4788  & \nodata & G 126-36 & 71 & 21 48 & +19 58 & \phn 9.93 & 2010 Jul 4 & \phn30 & 106 \\
G 126-52     & \nodata & LTT 16447 & 72 & 22 04 & +19 32 & 11.02 & 2004 Sep 7 & & \\ 
             &  &   &	 &        &    &         & 2006 Jun 19 &  110  & 121 \\
BD +17 4708  & 109558 & LTT 16493 & 73 & 22 11 & +18 05 & \phn 9.47 & 2004 Nov 7 & & \\
             &  &   &	 &         &   &         & 2005 Jul 5 & \phn 40 & 122\\
BD +7 4841  & 110140 & G 18-39 & 74 & 22 18 & +08 26 & 10.38 & 2004 Nov 7 & \phn 35 & \phn 75 \\
HD 218502    & 114271 & BD $-$15 6355 & 75 & 23 08 & $-$15 03 & \phn 8.50 & 2010 Jul 4 & \phn12 & 125 \\ 
BD +2 4651  & 115167 & G29-23 & 76 & 23 19 & +03 22 & 10.19 & 2004 Nov 7 & & \\
             &  &   &	 &      &      &         & 2005 Sep 27 &  \phn 75 & 134\\
BD +59 2723  & 115704 & Ross 233 & 77 & 23 26 & +60 37 & 10.47 & 2004 Nov 7 & \phn 75 & \phn 61 \\
HD 221377    & 116082 & BD +51 3630 & \nodata & 23 31 & +52 24 & \phn 7.57 & 2005 Jul 6  & \phn 9 & 116 \\
Moon	     &  &    &       &     &   &           & 2005 May 15 & \phn 10 & 470 \\
\enddata
\tablenotetext{a}{Code is an ID number for the star in the Appendix Table of
equivalent widths.}

\end{deluxetable}

\clearpage
\tightenlines
\singlespace
\centering
\begin{deluxetable}{lccclcccc} 
\footnotesize
\tablewidth{0pc}
\tablecolumns{9}
\tablenum{2} 
\tablecaption{Stellar Parameters and Abundances}
\tablehead{ \colhead{Star}  &  \colhead{$T_{\rm eff}$} &
\colhead{[Fe/H]}  
& \colhead{log g}  & \colhead{$\xi$} & \colhead{A(Be)} & \colhead{[O/H]} & 
\colhead{[Be/Fe]} & \colhead{[O/Fe]}
}
\startdata
HD 16031 & 6162 & $-$1.81 & 3.89 & 1.55 & $-$0.38 & $-$0.97 & $+$0.01 & +0.84 \\
HD 19445 & 5853 & $-$2.10 & 4.41 & 1.5  & $-$0.48 & $-$1.53 & $+$0.20 & +0.57 \\
HD 24289 & 5700 & $-$2.22 & 3.50 & 1.5  & $-$0.83 & $-$1.67 & $-$0.03 & +0.55 \\
HD 30743 & 6222 & $-$0.62 & 4.15 & 1.88 & $+$0.78 & $-$0.73 & $-$0.02 & +0.11 \\
HD 31128 & 5882 & $-$1.56 & 3.94 & 1.27 & $-$0.13 & $-$0.83 & $+$0.01 & +0.74 \\
HD 64090 & 5500 & $-$1.77 & 4.73 & 1.5  & $-$0.09 & $-$1.36 & $+$0.26 & +0.41 \\
HD 74000 & 6134 & $-$2.05 & 4.26 & 1.5  & $-$0.49 & $-$1.56 & $+$0.14 & +0.49 \\
HD 76932 & 5807 & $-$0.95 & 4.00 & 1.5  & $+$0.77 & $-$0.65 & $+$0.30 & +0.30 \\
HD 84937 & 6206 & $-$2.20 & 3.89 & 1.5  & $-$0.83 & $-$1.49 & $-$0.05 & +0.71 \\
HD 94028 & 5907 & $-$1.54 & 4.44 & 1.5  & $+$0.45 & $-$1.15 & $+$0.57 & +0.39 \\
HD 104056 & 6085 & $-$0.66 & 4.43 & 1.50 & $+$0.44 & $-$0.34 & $-$0.32 & +0.32 \\
HD 106038 & 6085 & $-$1.34 & 4.63 & 1.46 & $+$1.47 & $-$0.93 & $+$1.39 & +0.41 \\
HD 108177 & 6105 & $-$1.72 & 3.91 & 1.47 & $-$0.41 & $-$1.06 & $-$0.11 & +0.66 \\
HD 109303 & 6230 & $-$0.47 & 4.04 & 1.5  & +0.80 & $-$0.20 &$-$0.15 & +0.27 \\
HD 118244 & 6234 & $-$0.53 & 4.13 & 1.92 & $+$0.76 & $-$0.65 & $-$0.13 & $-$0.12 \\
HD 132475 & 5765 & $-$1.50 & 3.60 & 1.5  & $+$0.78 & $-$0.78 & $+$0.86 & +0.72 \\
HD 134169 & 5759 & $-$0.94 & 3.68 & 1.5  & $+$0.55 & $-$0.66 & $+$0.07 & +0.28 \\
HD 140283 & 5692 & $-$2.56 & 3.47 & 1.5  & $-$1.18 & $-$1.72 & $-$0.04 & +0.84 \\
HD 161770 & 5708 & $-$1.50 & 3.32 & 1.32 & $-$0.20 & $-$0.62 & $-$0.12 & +0.66 \\
HD 179626 & 5882 & $-$1.01 & 3.74 & 1.08 & $+$0.47 & $-$0.59 & $+$0.06 & +0.42 \\
HD 184499 & 5670 & $-$0.51 & 4.00 & 1.5  & $+$1.12 & $-$0.41 & $+$0.21 & +0.10 \\
HD 188510 & 5600 & $-$1.49 & 4.43 & 1.46 & $-$0.41 & $-$0.93 & $-$0.34 & 0.56 \\
HD 194598 & 5875 & $-$1.23 & 4.20 & 1.5  & $+$0.12 & $-$1.00 & $-$0.07 & +0.23 \\
HD 195633 & 5986 & $-$0.88 & 3.89 & 1.5  & $+$0.66 & $-$0.76 & $+$0.12 & +0.12 \\
HD 200580 & 5853 & $-$0.54 & 4.04 & 1.72 & $+$0.62 & $-$0.77 & $-$0.26 & $-$0.23 \\
HD 201889 & 5553 & $-$0.95 & 3.74 & 1.5  & $+$0.58 & $-$1.02 & $+$0.11 & $-$0.07 \\
HD 201891 & 5806 & $-$1.07 & 4.42 & 1.5  & $+$0.56 & $-$0.87 & $+$0.21 & +0.20 \\
HD 208906 & 5929 & $-$0.73 & 4.39 & 1.34 & $+$0.72 & $-$0.79 & $+$0.03 & $-$0.06 \\
HD 218502 & 6155 & $-$1.86 & 3.73 & 1.37 & $-$0.46 & $-$1.17 & $-$0.02 & +0.69 \\
HD 219617 & 5872 & $-$1.58 & 4.52 & 1.5  & $-$0.23 & $-$1.27 & $-$0.07 & +0.31 \\
HD 233511 & 6075 & $-$1.62 & 4.17 & 1.5  & $-$0.16 & $-$0.86 & $+$0.04 & +0.76 \\
HD 241253 & 6055 & $-$1.07 & 4.13 & 1.42 & $+$0.60 & $-$0.38 & $+$0.25 & +0.69 \\
HD 247168 & 5620 & $-$1.74 & 4.33 & 1.14 & $-$0.84 & $-$1.47 & $-$0.52 & +0.27 \\
HD 247297 & 5758 & $-$0.55 & 4.07 & 1.5  & $-$0.08 & $-$0.52 & $-$0.95 & +0.03 \\
\\
\\
\\
\\
CD $-$24 1656 & 6172 & $-$2.03 & 3.85 & 1.46 & $-$0.56 & $-$1.21 & +0.05 & +0.82 \\
BD $-$17 484 & 6110 & $-$1.56 & 3.63 & 1.32 & $-$0.37 & $-$0.95 & $-$0.23 & +0.61 \\
BD $-$14 5850 & 5777 & $-$2.18 & 4.12 & 1.5  & $-$0.79 & $-$1.38 & $-$0.03 & +0.80 \\
BD $-$13 3442 & 6090 & $-$2.91 & 4.11 & 1.5  & $-$1.12 & $-$2.15 & $+$0.37 & +0.76 \\
BD $-$10 388 & 5768 & $-$2.79 & 3.04 & 1.54 & $-$1.30 & $-$1.98 & $+$0.07 & +0.81 \\
BD $-$9 466  & 5990 & $-$1.89 & 3.79 & 1.54 & $-$0.61 & $-$1.12 & $-$0.14 & +0.77 \\
BD $-$8 4501  & 6392 & $-$1.28 & 4.39 & 1.46 & $-$0.18 & $-$0.40 & $-$0.32 & +0.88 \\
BD $-$4 3208 & 6210 & $-$2.35 & 4.03 & 1.5 & $-$0.73 & $-$1.68 & $+$0.20 & +0.67 \\
BD $-$3 2525 & 6115 & $-$1.91 & 4.04 & 1.5 & $-$0.19 & $-$1.07 & $+$0.30 & +0.84 \\
BD $-$1 306 & 6060 & $-$0.78 & 4.77 & 1.5 & $+$0.87 & $-$0.35 & $+$0.23 & +0.43 \\
BD +1 2341p & 6402 & $-$2.67 & 4.24 & 1.5  & $-$1.00 & $-$1.76 & $+$0.25 & +0.91 \\
BD +2 263 & 5842 & $-$1.92 & 3.95 & 1.10 & $-$0.73 & $-$1.56 & $-$0.23 & +0.36 \\
BD +2 3375 & 6008 & $-$2.22 & 4.04 & 1.28 & $-$0.68 & $-$1.48 & $+$0.12 & +0.74 \\
BD +2 4651 & 5992 & $-$1.90 & 3.33 & 1.44 & $-$0.58 & $-$1.18 & $-$0.10 & +0.72 \\
BD +3 740 & 6030 & $-$2.95 & 3.83 & 1.5  & $-$1.40 & $-$2.26 & $+$0.13 & +0.69 \\
BD +4 302 & 6095 & $-$2.17 & 3.71 & 1.48 & $-$0.76 & $-$1.34 & $-$0.01 & +0.83 \\
BD +4 4551 & 5990 & $-$1.43 & 3.85 & 1.41 & $+$0.22 & $-$0.89 & $+$0.23 & +0.54 \\
BD +7 4841 & 6018 & $-$1.56 & 3.60 & 1.50 & $-$0.18 & $-$0.93 & $-$0.04 & +0.63 \\
BD +9 2190 & 6008 & $-$3.00 & 3.85 & 1.5  & $-$1.22 & $-$2.38 & $+$0.36 & +0.62 \\
BD +13 3683 & 5502 & $-$2.38 & 3.06 & 1.47 & $-$1.23 & $-$1.46 & $-$0.27 & +0.92 \\
BD +17 4708 & 5992 & $-$1.70 & 3.54 & 1.36 & $-$0.45 & $-$1.09 & $-$0.17 & +0.61 \\
BD +19 1185 & 5835 & $-$0.92 & 4.62 & 1.50 & $+$0.17 & $-$0.78 & $-$0.33 & +0.14 \\
BD +19 4788 & 6020 & $-$0.78 & 4.92 & 1.5  & $+$0.72 & $-$0.23 & $+$0.08 & +0.55 \\
BD +20 2030 & 5978 & $-$2.77 & 3.61 & 1.5  & $-$1.23 & $-$2.06 & $+$0.12 & +0.71 \\
BD +20 3603 & 5908 & $-$2.18 & 3.61 & 1.01 & $-$0.93 & $-$1.85 & $-$0.17 & +0.33 \\
BD +21 607 & 6097 & $-$1.72 & 4.11 & 1.5  & $-$0.35 & $-$1.17 & $-$0.05 & +0.55 \\
BD +22 396 & 6050 & $-$0.88 & 4.89 & 1.05 & $+$0.65 & $-$0.48 & $+$0.11 & +0.40 \\
BD +23 3912 & 5815 & $-$1.46 & 3.36 & 1.44 & $-$0.16 & $-$1.09 & $-$0.12 & +0.37 \\
BD +24 1676 & 6125 & $-$2.55 & 3.74 & 1.45 & $-$1.28 & $-$1.94 & $-$0.15 & +0.61 \\
BD +26 2621 & 6266 & $-$2.69 & 4.50 & 1.5  & $-$0.94 & $-$1.96 & $+$0.33 & +0.73 \\
BD +26 3578 & 6158 & $-$2.32 & 3.94 & 1.5  & $-$0.90 & $-$1.69 & $+$0.00 & +0.63 \\
BD +28 2137 & 6110 & $-$1.97 & 3.83 & 1.07 & $-$0.68 & $-$1.33 & $-$0.13 & +0.64 \\
BD +34 2476 & 6248 & $-$1.94 & 3.72 & 1.23 & $-$0.76 & $-$1.25 & $-$0.24 & +0.69 \\
BD +36 2165 & 6052 & $-$1.71 & 3.78 & 1.5 & $-$0.45 & $-$1.28 & $-$0.16 & +0.43 \\
\\
\\
\\
\\
BD +36 2964 & 6152 & $-$2.20 & 3.85 & 1.39 & $-$0.47 & $-$1.18 & $+$0.31 & +1.02 \\
BD +37 1458 & 5492 & $-$2.02 & 3.85 & 1.5  & $-$0.95 & $-$1.37 & $-$0.35 & +0.65 \\
BD +39 2173 & 6200 & $-$1.99 & 3.76 & 1.13 & $-$0.58 & $-$1.33 & $-$0.01 & +0.66 \\
BD +42 3607 & 5655 & $-$2.29 & 3.81 & 1.43 & $-$1.10 & $-$1.48 & $-$0.23 & +0.81 \\
BD +44 1910 & 5878 & $-$2.64 & 3.56 & 1.5  & $-$1.11 & $-$1.96 & $+$0.11 & +0.68 \\
BD +51 1696 & 5852 & $-$1.21 & 4.19 & 1.5  & $-$0.33 & $-$0.53 & $-$0.54 & +0.68 \\
BD +59 2723 & 5945 & $-$2.20 & 4.50 & 1.02 & $-$0.35 & $-$1.70 & $+$0.43 & +0.50 \\
G 4-37	& 6120 & $-$2.50 & 3.82 & 1.33 & $-$0.75 & $-$1.62 & $+$0.33 & +0.88 \\
G 5-19	& 5975 & $-$1.13 & 3.93 & 1.34 & $-$0.08 & $-$0.62 & $-$0.37 & +0.51 \\
G 11-44	& 5820 & $-$2.29 & 3.58 & 1.5  & $-$1.04 & $-$1.63 & $-$0.17 & +0.66 \\
G 20-24	& 6222 & $-$1.89 & 4.07 & 1.14 & $-$0.57 & $-$1.41 & $-$0.10 & +0.48 \\
G 21-22	& 5916 & $-$1.02 & 4.59 & 1.5  & $+$0.31 & $-$1.02 & $-$0.09 & +0.00 \\
G 24-3	& 6000 & $-$1.62 & 3.98 & 1.47 & $-$0.28 & $-$1.21 & $-$0.08 & +0.41 \\
G 24-25 & 5752 & $-$1.56 & 3.69 & 1.43 & $-$0.73 & $-$0.98 & $-$0.59 & +0.58 \\
G 26-12 & 6135 & $-$2.33 & 3.84 & 1.39 & $-$0.88 & $-$1.35 & $+$0.03 & +0.98 \\
G 59-24	& 6112 & $-$2.32 & 4.10 & 1.5  & $-$0.69 & $-$1.32 & $+$0.21 & +1.00 \\
G 63-46	& 6125 & $-$0.60 & 4.70 & 1.50 & $+$0.98 & $-$0.04 & $+$0.16 & +0.56 \\
G 64-12	& 6074 & $-$3.45 & 3.72 & 1.5  & $-$1.43 & $-$2.24 & $+$0.60 & +1.21 \\
G 64-37	& 6122 & $-$3.28 & 3.87 & 1.5  & $-$1.40 & $-$2.32 & $+$0.46 & +0.96 \\
G 74-5	& 6025 & $-$0.84 & 4.77 & 1.50 & $+$0.81 & $-$0.21 & $+$0.23 & +0.63 \\
G 75-56	& 5890 & $-$2.38 & 3.83 & 1.5  & $-$0.84 & $-$1.74 & $+$0.12 & +0.64 \\
G 88-10	& 5945 & $-$2.61 & 4.00 & 1.5  & $-$1.08 & $-$1.86 & $+$0.11 & +0.75 \\
G 90-3	& 5710 & $-$2.28 & 3.19 & 1.48 & $-$0.88 & $-$1.79 & $-$0.02 & +0.49 \\
G 92-6	& 6115 & $-$2.70 & 4.79 & 1.5  & $-$0.91 & $-$2.28 & $+$0.37 & +0.42 \\
G 108-58 & 5865 & $-$2.37 & 4.03 & 1.5  & $-$1.12 & $-$1.33 & $-$0.17 & +1.04 \\
G 113-9	& 5998 & $-$1.75 & 3.68 & 1.27 & $-$0.26 & $-$1.01 & $+$0.07 & +0.74 \\
G 113-22 & 5802 & $-$1.01 & 3.95 & 1.50 & $+$0.70 & $-$0.40 & $+$0.29 & +0.61 \\
G 115-49 & 5605 & $-$2.23 & 3.78 & 1.16 & $-$1.03 & $-$1.47 & $-$0.22 & +0.76 \\
G 126-52 & 6182 & $-$2.36 & 3.95 & 1.47 & $-$0.81 & $-$1.67 & $+$0.13 & +0.69 \\
G 130-65 & 6018 & $-$2.21 & 3.65 & 1.54 & $-$0.70 & $-$1.24 & $+$0.09 & +0.97 \\
G 153-21 & 6142 & $-$0.39 & 4.55 & 1.50 & $+$1.03 & $-$0.05 & $+$0.00 & +0.34 \\
G 180-24 & 6008 & $-$1.44 & 3.77 & 1.31 & $-$0.18 & $-$0.88 & $-$0.16 & +0.56 \\
G 181-28 & 5965 & $-$2.42 & 3.98 & 1.33 & $-$1.20 & $-$1.62 & $-$0.20 & +0.80 \\ 
G 188-22 & 5975 & $-$1.35 & 3.72 & 1.31 & $+$0.22 & $-$0.78 & $+$0.15 & +0.57 \\
\\
\\
\\
\\
G 191-55 & 5828 & $-$1.81 & 4.11 & 1.06 & $-$0.80 & $-$1.37 & $-$0.41 & +0.44 \\
G 192-43 & 6140 & $-$1.42 & 3.78 & 1.28 & $-$0.38 & $-$0.93 & $-$0.38 & +0.49 \\
G 201-5	& 5950 & $-$2.54 & 4.00 & 1.5  & $-$1.27 & $-$1.97 & $-$0.15 & +0.57 \\
G 206-34 & 5825 & $-$3.12 & 3.99 & 1.5  & $-$1.20 & $-$2.37 & $+$0.50 & +0.75 \\
G 268-32 & 6230 & $-$2.51 & 4.60 & 1.46 & $<-$1.50 & \nodata & $<-$0.41 & \nodata\\
G 275-4	& 5942 & $-$3.42 & 4.05 & 1.5	& $-$1.53 & $-$2.48 & $+$0.47 & +0.94 \\
LP 553-62 & 6128 & $-$2.73 & 3.93 & 1.5 & $-$1.00 & $-$2.02 & $+$0.31 & +0.71 \\
LP 635-14 & 5932 & $-$2.71 & 3.57 & 1.5 & $-$1.17 & $-$2.00 & $+$0.12 & +0.71 \\
LP 651-4 & 6030 & $-$2.89 & 4.26 & 1.5 & $-$1.12 & $-$2.04 & $+$0.35 & +0.85 \\
LP 752-17 & 5738 & $-$2.38 & 3.20 & 1.5 & $-$0.86 & $-$1.80 & $+$0.10 & +0.58 \\
LP 815-43 & 6405 & $-$2.76 & 4.37 & 1.5 & $-$0.95 & $-$1.86 & $+$0.39 & +0.90 \\
LP 831-70 & 6005 & $-$3.05 & 3.40 & 1.47 & $<-$1.10 &$-$1.85  & $<$0.53 & +1.20 \\
LTT 1566 & 6025 & $-$2.36 & 3.93 & 1.49 & $-$0.90 & $-$1.71 & $+$0.04 & +0.65 \\
Ross 390 & 5920 & $-$0.78 & 4.71 & 1.50	& $+$0.07 & $-$0.13 & $-$0.57 & +0.75 \\
Ross 797 & 5838 & $-$1.17 & 4.17 & 1.21	& $-$0.24 & $-$0.71 & $-$0.49 & +0.46 \\

\enddata

\end{deluxetable}

\clearpage
\tightenlines
\singlespace
\centering
\begin{deluxetable}{lcccccl} 
\footnotesize
\tablewidth{0pc}
\tablecolumns{7}
\tablenum{3} 
\tablecaption{Stellar Abundances of Ti, Mg, and Li}
\tablehead{ \colhead{Star}  &  \colhead{$T_{\rm eff}$} &
\colhead{[Fe/H]}  
& \colhead{[Ti/H]}  & \colhead{[Mg/H]} & \colhead{A(Li)} 
& \colhead{Li ref.\tablenotemark{1}}  
}
\startdata
HD 16031 & 6162 & $-$1.81 & $-$1.39    & $-$1.32  & 2.18   & CP05 \\
HD 19445 & 5853 & $-$2.10 & \nodata    & \nodata  & 2.18   & CP05 \\
HD 24289 & 5700 & $-$2.22 & $-$1.94    & $-$1.91  & 2.38   & CP05 \\
HD 30743 & 6222 & $-$0.62 & \nodata    & \nodata  & 2.35   & B90 \\
HD 31128 & 5882 & $-$1.56 & $-$1.30    & $-$1.14  & 2.16   & CP05 \\
HD 64090 & 5500 & $-$1.77 & \nodata    & \nodata  & 1.21   & CP05 \\
HD 74000 & 6134 & $-$2.05 & \nodata    & \nodata  & 2.14   & CP05 \\
HD 76932 & 5807 & $-$0.95 & \nodata    & \nodata  & 2.03   & C01 \\  
HD 84937 & 6206 & $-$2.20 & \nodata    & \nodata  & 2.28   & CP05 \\
HD 94028 & 5907 & $-$1.54 & \nodata    & \nodata  & 2.21   & CP05 \\
HD 104056 & 6085 & $-$0.66 & $-$0.37   & $-$0.14  & \nodata & \nodata \\
HD 106038 & 6085 & $-$1.34 & $-$1.06   & $-$0.82  & 2.48   & A06 \\
HD 108177 & 6105 & $-$1.72 & $-$1.42   & $-$1.29  & 2.20   & CP05 \\
HD 109303 & 6230 & $-$0.47 & $-$0.33   & $-$0.22  & $<$1.65 & C01  \\
HD 118244 & 6234 & $-$0.53 & \nodata   & \nodata  & 2.07   & C01 \\  
HD 132475 & 5765 & $-$1.50 & \nodata   & \nodata  & 2.39   & MN05 \\
HD 134169 & 5759 & $-$0.94 & \nodata   & \nodata  & 2.22   & CP05 \\
HD 140283 & 5692 & $-$2.56 & \nodata   & \nodata  & 2.26   & CP05 \\
HD 161770 & 5708 & $-$1.50 & $-$1.27   & $-$1.05  & 2.12   & CP05 \\
HD 179626 & 5882 & $-$1.01 & $-$0.76   & $-$0.67  & 1.81   & CP05 \\
HD 184499 & 5670 & $-$0.51 & \nodata   & \nodata  & \nodata & \nodata \\
HD 188510 & 5600 & $-$1.49 & $-$1.29   & $-$1.09  & 1.61   & S07 \\    
HD 194598 & 5875 & $-$1.23 & \nodata   & \nodata  & 2.00   & CP05 \\
HD 195633 & 5986 & $-$0.88 & \nodata   & \nodata  & 2.15 & R88 \\
HD 200580 & 5853 & $-$0.54 & \nodata   & \nodata  & 2.08   & C01 \\    
HD 201889 & 5553 & $-$0.95 & \nodata   & \nodata  & 1.04   & CP05 \\
HD 201891 & 5806 & $-$1.07 & \nodata   & \nodata  & 1.98   & MN05 \\
HD 208906 & 5929 & $-$0.73 & \nodata   & \nodata  & 2.31   & C01 \\    
HD 218502 & 6155 & $-$1.86 & $-$1.46   & $-$1.58  & 2.31   & M10 \\    
HD 219617 & 5872 & $-$1.58 & \nodata   & \nodata  & 2.23   & CP05 \\
HD 233511 & 6075 & $-$1.62 & $-$1.30   & $-$1.16  & 2.12   & CP05 \\
HD 241253 & 6055 & $-$1.07 & $-$0.81   & $-$0.66  & 1.98   & M10\\    
HD 247168 & 5620 & $-$1.74 & $-$1.60   & $-$1.58  & 2.18   & GD \\   
HD 247297 & 5758 & $-$0.55 & $-$0.21   & $-$0.24  & \nodata & \nodata \\
\\
\\
\\
\\
CD $-$24 1656 & 6172 & $-$2.03 & $-$1.61 & $-$1.65 & \nodata & \nodata \\
BD $-$17 484 & 6110 & $-$1.57 & $-$1.29 & $-$1.19 & \nodata & \nodata \\
BD $-$14 5850 & 5777 & $-$2.18 & $-$1.95 & $-$1.74  & \nodata & \nodata \\
BD $-$13 3442 & 6090 & $-$2.91 & $-$2.16 & $-$2.31  & 2.18   & R99 \\
BD $-$10 388 & 5768 & $-$2.79 & $-$2.33  & $-$2.16  & 2.26   & CP05 \\
BD $-$9 466  & 5990 & $-$1.89 & $-$1.55  & $-$1.47  & 2.20   & GD \\
BD $-$8 4501  & 6392 & $-$1.28 & $-$1.05 & $-$1.05  & 2.11   & CP05 \\
BD $-$4 3208 & 6210 & $-$2.35 & $-$1.92  & $-$1.89  & 2.30   & CP05 \\
BD $-$3 2525 & 6115 & $-$1.91 & $-$1.59  & $-$1.55  & \nodata & \nodata \\
BD $-$1 306 & 6060 & $-$0.78 & $-$0.45   & $-$0.33  & \nodata & \nodata \\    
BD +1 2341p & 6402 & $-$2.67 & $-$2.09   & $-$2.30  & 2.19   & MN05\\  
BD +2 263 & 5842 & $-$1.92 & $-$1.62     & $-$1.64  & 2.12   & R96 \\ 
BD +2 3375 & 6008 & $-$2.22 & $-$1.94    & $-$1.85  & 2.06   & CP05 \\  
BD +2 4651 & 5992 & $-$1.90 & $-$1.57    & $-$1.48  & 2.18   & CP05  \\
BD +3 740 & 6030 & $-$2.95 & $-$2.27     & $-$2.45  & 2.16   & CP05 \\
BD +4 302 & 6095 & $-$2.17 & $-$1.71     & $-$1.73  & 2.33   & GD \\
BD +4 4551 & 5990 & $-$1.43 & $-$1.17    & $-$0.97  & 1.97   & CP05 \\
BD +7 4841 & 6018 & $-$1.56 & $-$1.24    & $-$1.06  & 2.22   & CP05 \\
BD +9 2190 & 6008 & $-$3.00 & $-$2.41    & $-$2.41  & 2.18   & CP05 \\
BD +13 3683 & 5502 & $-$2.38 & $-$2.14   & $-$1.94  & 1.94   & T94 \\
BD +17 4708 & 5992 & $-$1.70 & $-$1.42   & $-$1.31  & 2.10   & CP05 \\
BD +19 1185 & 5835 & $-$0.92 & $-$0.68   & $-$0.62  & \nodata & \nodata \\
BD +19 4788 & 6020 & $-$0.78 & $-$0.43   & $-$0.33  & \nodata & \nodata \\
BD +20 2030 & 5978 & $-$2.77 & $-$2.24   & $-$2.29  & 2.16   & R96 \\
BD +20 3603 & 5908 & $-$2.18 & $-$1.72   & $-$1.72  & 2.22   & CP05 \\
BD +21 607 & 6097 & $-$1.72 & $-$1.44    & $-$1.40  & 2.14   & CP05 \\
BD +22 396 & 6050 & $-$0.88 & $-$0.60    & $-$0.42  & \nodata & \nodata \\
BD +23 3912 & 5815 & $-$1.46 & $-$1.18   & $-$0.97  & 2.51   & MN05 \\
BD +24 1676 & 6125 & $-$2.55 & $-$2.06   & $-$2.11  & 2.16   & CP05 \\
BD +26 2621 & 6266 & $-$2.69 & $-$2.15   & $-$2.31  & 2.21   & CP05 \\
BD +26 3578 & 6158 & $-$2.32 & \nodata   & \nodata  & 2.25   & CP05 \\
BD +28 2137 & 6110 & $-$1.97 & $-$1.59   & $-$1.67  & 2.16   & CP05 \\
BD +34 2476 & 6248 & $-$1.94 & $-$1.45   & $-$1.71  & 2.17   & CP05 \\
BD +36 2165 & 6052 & $-$1.71 & $-$1.39   & $-$1.39  & 2.42 & M10 \\
\\
\\
\\
\\
BD +36 2964 & 6152 & $-$2.20 & $-$1.82  & $-$1.90  & 2.27   & R96 \\
BD +37 1458 & 5492 & $-$2.02 & $-$1.72  & $-$1.54  & 1.37   & CP05 \\
BD +39 2173 & 6200 & $-$1.99 & $-$1.56  & $-$1.70  & 2.21   & R96 \\
BD +42 3607 & 5655 & $-$2.29 & $-$2.04  & $-$1.83  & 2.33   & R96 \\
BD +44 1910 & 5878 & $-$2.64 & $-$2.14  & $-$2.25  & \nodata & \nodata \\
BD +51 1696 & 5852 & $-$1.21 & $-$1.02  & $-$0.87  & 1.80 & M10 \\
BD +59 2723 & 5945 & $-$2.20 & $-$1.65  & $-$1.89  & 2.16   & CP05 \\
G 4-37	& 6120 & $-$2.50     & $-$2.03  & $-$2.11  & 1.92   & CP05 \\
G 5-19	& 5975 & $-$1.13     & $-$1.02  & $-$0.88  & 2.26   & B05\\
G 11-44	& 5820 & $-$2.29     & $-$1.97  & $-$1.90  & 2.12   & CP05 \\
G 20-24	& 6222 & $-$1.89     & $-$1.42  & $-$1.61  & 2.35   & CP05 \\
G 21-22	& 5916 & $-$1.02     & \nodata  & \nodata  & 2.48   & MN05 \\
G 24-3	& 6000 & $-$1.62     & $-$1.31  & $-$1.24  & 2.09   & CP05  \\
G 24-25 & 5752 & $-$1.56     & $-$1.33  & $-$1.22  & \nodata & \nodata \\
G 26-12 & 6135 & $-$2.33     & $-$1.99  & $-$1.95  & 2.24   & B05  \\
G 59-24	& 6112 & $-$2.32     & $-$1.97  & $-$1.89  & 2.25   & CP05  \\
G 63-46	& 6125 & $-$0.60     & $-$0.24  & $-$0.15  & \nodata & \nodata \\
G 64-12	& 6074 & $-$3.45     & $-$2.80  & $-$2.76  & 2.35   & CP05  \\
G 64-37	& 6122 & $-$3.28     & $-$2.83  & $-$2.71  & 2.06   & MN05  \\
G 74-5	& 6025 & $-$0.84     & $-$0.53  & $-$0.45  & 1.48   & CP05  \\
G 75-56	& 5890 & $-$2.38     & $-$1.89  & $-$1.96  & 2.06   & R96  \\
G 88-10	& 5945 & $-$2.61     & $-$2.09  & $-$2.08  & 2.13   & CP05  \\
G 90-3	& 5710 & $-$2.28     & $-$2.00  & $-$1.96  & 2.36   & CP05  \\
G 92-6	& 6115 & $-$2.70     & $-$1.99  & $-$2.36  & 2.43   & MN05  \\
G 108-58 & 5865 & $-$2.37    & $-$2.10  & $-$2.04  & \nodata & \nodata \\
G 113-9	& 5998 & $-$1.75     & $-$1.43  & $-$1.26  & \nodata & \nodata \\
G 113-22 & 5802 & $-$1.01    & $-$0.71  & $-$0.52  & \nodata & \nodata \\
G 115-49 & 5605 & $-$2.23    & $-$1.99  & $-$1.79  & 2.09   & CP05  \\
G 126-52 & 6182 & $-$2.36    & $-$1.91  & $-$1.95  & 2.14   & R99 \\
G 130-65 & 6018 & $-$2.21    & $-$1.86  & $-$1.69  & 2.15   & GD  \\
G 153-21 & 6142 & $-$0.39    & $-$0.14  & $+$0.02  & \nodata & \nodata \\
G 180-24 & 6008 & $-$1.44    & $-$1.16  & $-$1.06  & 2.08   & CP05  \\
G 181-28 & 5965 & $-$2.42    & $-$2.16  & $-$2.10  & 2.22   & R96  \\ 
G 188-22 & 5975 & $-$1.35    & $-$1.11  & $-$0.97  & \nodata & \nodata \\
\\
\\
\\
\\
G 191-55 & 5828 & $-$1.81    & $-$1.54  & $-$1.49  & \nodata & \nodata \\
G 192-43 & 6140 & $-$1.42    & $-$1.15  & $-$1.11  & 2.32   &  CP05 \\
G 201-5	& 5950 & $-$2.54   & $-$2.09  & $-$2.13   & 2.27   & MN05  \\
G 206-34 & 5825 & $-$3.12  & $-$2.65  & $-$2.55   & 2.27   & R96  \\
G 268-32 & 6230 & $-$2.51  & $-$2.38  & $-$2.18   & 1.97   & CP05  \\
G 275-4	& 5942 & $-$3.42   & $-$2.84  & $-$2.82   & 2.21   & T94 \\ 
LP 553-62 & 6128 & $-$2.73 & $-$2.05  & $-$2.17   & 1.97   & R96  \\
LP 635-14 & 5932 & $-$2.71 & $-$2.20  & $-$2.21   & 2.35   & M10  \\
LP 651-4 & 6030 & $-$2.89  & $-$2.24  & $-$2.33   & 2.18   & R99  \\
LP 752-17 & 5738 & $-$2.38 & $-$1.98  & $-$1.93   & \nodata & \nodata \\
LP 815-43 & 6405 & $-$2.76 & $-$2.06  & $-$2.45   & 2.26   & M10  \\
LP 831-70 & 6005 & $-$3.05 & $-$2.53  & $-$2.50   & 2.28   & M10  \\
LTT 1566 & 6025 & $-$2.36  & $-$1.98  & $-$2.00   & 2.26   & GD  \\
Ross 390 & 5920 & $-$0.78  & $-$0.55  & $-$0.53   & 1.15   & RD98 \\
Ross 797 & 5838 & $-$1.17  & $-$1.05  & $-$0.84   & 2.32   & GD  \\

\enddata

\tablenotetext{1}{
A06 = Asplund et al.~(2006),
B90 = Balachandran (1990),
B05 = Boesgaard et al.~(2005),
C01 = Chen et al.~(2001),
CP05 = Charbonnel \& Primas (2005) compilation,
GD = private communication from G. Dima,
M10 = Melendez et al.~(2010),
MN05 = Novicki (2005),
T94 = Thorburn (1994),
R88 = Rebolo et al.(1988),
R96 = Ryan et al.~(1996),
R99 = Ryan et al.~(1999),
RD98 = Ryan \& Deliyannis (1998)
S07 = Shi et al.~(2007).
}

\end{deluxetable}

\clearpage
\tightenlines
\singlespace
\centering
\begin{deluxetable}{lccccc} 
\footnotesize
\tablewidth{0pc}
\tablecolumns{6}
\tablenum{4} 
\tablecaption{Abundance Uncertainties for Three Representative Stars}
\tablehead{ 
\multicolumn{1}{l}{Element} & 
\multicolumn{1}{c}{Abundance} & 
\multicolumn{1}{c}{T: $\pm$100 K} & 
\multicolumn{1}{c}{log g: $\pm$0.2} & 
\multicolumn{1}{l}{[Fe/H]: $\pm$0.1} & 
\multicolumn{1}{l}{$\xi$: $\pm$0.2}
} 
\startdata 
\multicolumn{6}{l}{BD $-$10 388: T = 5768 K, log g = 3.04, [Fe/H] = $-$2.79, 
$\xi$ = 1.54} \\
\hline
$[$Fe/H$]$ &$-$2.79 & $\pm$0.07 & $\mp$0.02 & $\mp$0.01 & $\mp$0.02 \\
$[$Ti/H$]$ &$-$2.23 & $\pm$0.09 & $\mp$0.01 & \phn0.00 & \phn0.00 \\
$[$Mg/H$]$ &$-$2.16 & $\pm$0.04 & $\mp$0.01 & \phn0.00 & $\mp$0.01 \\
A(Be)      &$-$1.30 & $\pm$0.03 & $\pm$0.06 & $\mp$0.01 & \phn0.00 \\
$[$O/H$]$  &$-$2.00 & $\pm$0.20 & $\mp$0.10 & $\pm$0.01 & \phn0.00 \\
\hline
\multicolumn{6}{l}{G 188-22:  T = 5975 K, log g = 3.72, [Fe/H] = $-$1.35, 
$\xi$ = 1.31} \\
\hline
$[$Fe/H$]$ &  $-$1.35 & $\pm$0.06 & $\mp$0.02 & $\mp$0.01 & $\mp$0.05 \\
$[$Ti/H$]$ &  $-$1.11 & $\pm$0.09 & \phn0.00 & \phn0.00 & $\mp$0.02 \\
$[$Mg/H$]$ &  $-$0.97 & $\pm$0.07 & $\mp$0.03 & \phn0.00 & $\mp$0.02 \\
A(Be)      &  $+$0.22 & $\pm$0.05 & $\pm$0.08 & $\pm$0.01 & $\pm$0.01 \\
$[$O/H$]$  &  $-$0.80 & $\pm$0.19 & $\mp$0.05 & $\pm$0.01 & $\mp$0.01 \\
\hline
\multicolumn{6}{l}{G 20-24: T = 6222 K, log g = 4.07, [Fe/H] = $-$1.89, 
$\xi$ = 1.14} \\
\hline
$[$Fe/H$]$ &  $-$1.89 & $\pm$0.06 & $\mp$0.02 & \phn0.00 & $\mp$0.03 \\
$[$Ti/H$]$ &  $-$1.42 & $\pm$0.07 & $\mp$0.01 & \phn0.00 & $\mp$0.01 \\
$[$Mg/H$]$ &  $-$1.61 & $\pm$0.07 & $\mp$0.02 & \phn0.00 & $\mp$0.01 \\
A(Be)      &  $-$0.57 & $\pm$0.06 & $\pm$0.08 & $\mp$0.01 & \phn0.00 \\     
$[$O/H$]$  &  $-$1.46 & $\pm$0.19 & $\mp$0.08 & \phn0.00 & \phn0.00 \\

\enddata

\end{deluxetable}

\clearpage
\tightenlines
\singlespace
\centering
\begin{deluxetable}{lrrrrrrlrc}
\footnotesize
\tablewidth{0pc}
\tablecolumns{10}
\tablenum{5}
\tablecaption{Kinematic Properties}
\tablehead{ \colhead{Star}  &  \colhead{Rad.Vel.} & \colhead{U$_{lsr}$}
& \colhead{V$_{lsr}$}  & \colhead{W$_{lsr}$} & \colhead{R$_{apo}$}
& \colhead{Z$_{max}$} &
\colhead{ref.\tablenotemark{1}} & \colhead{$v_{\rm RF}$} & \colhead{D or A}
}
\startdata
HD 16031 &23.6 &$-$31 &$-$49 &$-$27 &8.2 &0.3 &C &175.9 &D       \\
HD 19445 &$-$139.3 &$-$159 &$-$94 &$-$35 &11.1 &0.4 &C &205.9 &D        \\
HD 24289 &143 &140  &$-$90  &$-$9   &\nodata &\nodata &RN, Be &190.7 &? \\
HD 30743 &$-$3.0    &$-$25.8 &$-$5.4  &$-$23.6 &9.9 &\nodata &Bk05 &217.4 &D \\
HD 31128 &105    &$-$63   &$-$100  &$-$31   &8.4  &0.4 &GC &139.0 &D       \\
HD 64090 &$-$240    &$-$265  &$-$178  &$-$84   &17.3 &2.9 &C &281.1 &A    \\
HD 74000 &204.2  &$-$112  &$-$267   &69   &9.2  &0.9 &C &139.7 &A,R      \\
HD 76932 &120.8   &40   &$-$80    &77   &8.8  &0.8  &F &164.7 &D     \\
HD 84937 &$-$16.7   &$-$139  &$-$117  &$-$3    &10.0 &0.0 &C &173.0 &D     \\
HD 94028 &61.9    &23   &$-$96    &29   &8.1  &0.3 &C &129.4 &D      \\
HD 104056 &$-$22.8   &$-$68   &$-$1    &$-$37   &9.8  &0.5 &C &232.3 &D    \\
HD 106038 &95      &13   &$-$270   &19   &8.6  &0.6 &G &55.0 &A,R       \\
HD 108177 &159    &$-$111  &$-$184   &70   &8.8  &1.9 &C &136.1 &D         \\
HD 109303 &23.8 &20.7 &$-$22.7 &34.7 &8.79 &0.48 &N, Bk05,Bo &246.0 &D \\
HD 118244 &$-$11.1    &48.8 &$-$5.7   &6.8  &9.8  &\nodata &Ch,So &219.9 & D \\
HD 132475 &167    &$-$51   &$-$354   &62   &8.8  &0.6 &F &156.2 &A,R \\
HD 134169 &18.8   &$-$24    &10    &20   &9.2  &0.2 &F &232.1 &D       \\
HD 140283 &7.2     &240  &$-$239   &48   &14.7 &0.6 &F &245.5 &A,R       \\
HD 161770 &$-$129    &$-$54   &$-$273   &$-$1   &9.2  &0.2 &GC &75.7 &A,R  \\
HD 179626 &$-$70.8   &$-$58   &$-$162   &45   &8.0  &0.9 &C &93.6 &D     \\
HD 184499 &$-$163.3   &53   &$-$144   &39   &8.2  &0.4 &C &100.5 &D    \\
HD 188510 &$-$192.2 &143 &$-$102 &69 &6.77 &0.87 &V,Bo,Re06 &128.6 &D \\
HD 194598 &$-$246.3   &68   &$-$264  &$-$22   &8.9  &0.2 &F &83.9 &A,R    \\
HD 195633 &$-$46.1    &63   &$-$26    &3    &9.4  &0.0  &F &204.0 &D     \\
HD 200580 &$-$6.4    &$-$106  &$-$69.6  &16.4 &10.5 &\nodata &V,So &184.7 &D \\
HD 201889 &$-$102.5   &119  &$-$69   &$-$30   &10.5 &0.3 &F &194.6 &D    \\
HD 201891 &$-$45.1   &$-$100  &$-$102  &$-$51   &9.6  &0.5 &F &162.9 &D   \\
HD 208906 &8.4    &$-$63.5 &$-$0.8  &$-$11.4 &11.6 &\nodata &V,M &228.5 &D \\
HD 218502 &$-$32 &$-$12 &$-$92 &1 &8.5 &0.06 &V,G03, Bk05 &128.6 &D \\
HD 219617 &10.1   &$-$183  &$-$125  &$-$23   &11.6 &0.3 &C &207.5 &D     \\
HD 233511 &59      &108  &$-$191   &41   &8.8  &0.9 &C &119.1 &A       \\
HD 241253 &$-$16.0   &$-$19   &$-$61    &88   &8.2  &1.8 &C &182.7 &D     \\
HD 247168 &$-$4      &$-$196  &$-$433  &$-$175  &24.1 &7.8 &C &338.2 &A,R  \\
HD 247297 &38.3    &18   &$-$31   &$-$5    &8.1  &0.1 &C &189.9 &D       \\
\\
\\
\\
\\
CD $-$24 1656 &66  &$-$73   &$-$188  &$-$27 &\nodata &\nodata &RN,Sc &84.2 &A\\
BD $-$17 484 &234.7   &226  &$-$189  &$-$134  &\nodata &\nodata &Eg &264.6 &A\\
BD $-$14 5850 &0.0  &$-$140  &$-$136   &71.8 &14.6 &\nodata &Bo,IDL &178.4 &D\\
BD $-$13 3442 &159  &$-$246  &$-$102   &21   &\nodata &\nodata &RN &273.6 &?\\
BD $-$10 388 &36.2    &100  &$-$24    &8    &10.2  &0.1 &C &220.2 &D \\
BD $-$9 466  &$-$164 &$-$193  &$-$226  &90 &\nodata &\nodata &RN &213.0 &A,R \\
BD $-$8 4501  &91     &$-$136  &$-$16   &$-$136  &14.2 &4.2 &C &280.4 &D \\
BD $-$4 3208 &56      &38   &$-$146  &$-$23   &8.1  &0.2 &C &86.3 &D \\
BD $-$3 2525 &25     &$-$246  &$-$192   &48   &14.6 &2.6 &C &252.2 &A \\
BD $-$1 306   &19.2    &166  &$-$157   &53   &10.8  &0.7  &C &185.3 &D  \\
BD +1 2341p   &$-$59     &$-$234  &$-$168  &$-$112  &15.5 &3.6 &C &264.6 &A  \\
BD +2 263 &$-$12.5    &67   &$-$36    &45   &9.0  &0.6 &C &200.9 &D       \\
BD +2 3375      &$-$397.8   &375  &$-$208   &14   &21.7 &3.2 &C &375.5 &A  \\
BD +2 4651  &$-$253   &140  &$-$264 &98 &\nodata &\nodata &RN,Bo &176.5 &A,R \\
BD +3 740 &173.9   &131  &$-$110  &$-$36   &9.9  &0.4  &C &174.8 &D     \\
BD +4 302 &$-$70    &65   &$-$201   &53   &\nodata &\nodata &RN &86.0 &A\\
BD +4 4551 &$-$119.2  &$-$75  &$-$46 &76 &10.3 &\nodata &Sc,IDL,Ka &204.1 &D \\
BD +7 4841     &$-$234.8   &115  &$-$225   &49   &8.6  &2.1 &C &125.1 &A,R   \\
BD +9 2190      &266.1  &$-$67   &$-$319   &193  &9.5  &6.5  &C &227.0 &A,R  \\
BD +13 3683      &101.2   &136   &7    &$-$10   &13.9 &0.1  &GC &264.8 &D  \\
BD +17 4708     &$-$291.3   &178  &$-$248   &80   &11.1 &2.2 &C &197.2 &A,R  \\
BD +19 1185     &$-$190    &$-$227  &$-$115   &64   &14.8 &1.2 &C &258.2 &D  \\
BD +19 4788 &$-$87 &$-$83 &$-$102 &12 &8.7 &0.1 &C &144.8 &D \\
BD +20 2030  &$-$65.7   &$-$169  &$-$229  &$-$42   &10.4 &1.5 &C &174.4 &A,R \\
BD +20 3603    &$-$242.7   &37   &$-$273  &$-$49   &8.0  &1.0 &C &81.1 &A,R  \\
BD +21 607      &339     &349  &$-$130  &$-$76   &30.9 &1.8 &F &368.3 &A  \\
BD +22 396    &$-$22.4   &$-$55   &$-$93   &$-$84   &8.5  &1.6 &C &161.9 &D  \\
BD +23 3912    &$-$115.2   &21   &$-$92    &107  &8.6  &1.4 &F,Ka &168.1 &D \\
BD +24 1676   &$-$238.6  &$-$283  &$-$92   &$-$39   &21.4 &1.3 &C &313.0 &A  \\
BD +26 2621   &$-$63     &$-$63   &$-$201  &$-$18   &7.7  &1.3 &C &68.2 &A  \\
BD +26 3578   &$-$129.1  &$-$40   &$-$143  &$-$57   &8.6  &0.5 &F &103.8 &D \\
BD +28 2137 &$-$132.5  &$-$132  &$-$426  &$-$148  &8.6  &7.2 &GC &285.9 &A,R \\
BD +34 2476   &$-$162    &$-$146  &$-$130  &$-$153  &11.7 &4.8 &C &229.8 &D  \\
BD +36 2165   &$-$198    &$-$184  &$-$156  &$-$139  &12.8 &5.2 &C &239.3 &D  \\
\\
\\
\\
\\
BD +36 2964    &$-$60.5   &$-$81   &$-$96    &1    &8.7  &0.0 &C &148.1 &D  \\
BD +37 1458      &242.2   &264  &$-$251  &$-$28   &17.0 &0.3 &F &267.3 &A,R  \\
BD +39 2173   &$-$79.5   &$-$117  &$-$183  &$-$6    &9.3  &0.1 &C &122.9 &D  \\
BD +42 3607      &$-$196.1   &147  &$-$156   &19   &10.0 &0.2 &C &161.5 &D  \\
BD +44 1910 &$-$86  &74  &$-$211  &$-$40   & 8.8 & 0.6 &APS &84.6 &A  \\
BD +51 1696      &64.3    &202  &$-$234   &61   &11.8 &2.6 &C &211.5 &A \\
BD +59 2723    &$-$105.8   &157  &$-$168  &$-$42   &10.1 &1.4 &C &170.6 &D  \\
G 4-37  &$-$108.4  &$-$170  &$-$87   &$-$21   &11.8 &0.3  &C &216.9 &D     \\
G 5-19  &$-$216.7  &$-$254  &$-$169   &4    &15.6 &0.1 &C &259.1 &A       \\
G 11-44 &98.4   &$-$135  &$-$266  &$-$42   &9.7  &0.3  &C &148.7 &A,R     \\
G 20-24 &34.4   &$-$122  &$-$129   &56   &9.4  &0.8  &C &162.2 &D     \\
G 21-22 &59.4   &$-$232  &$-$228  &$-$26   &12.4 &4.8  &C &233.6 &A,R      \\
G 24-3  &$-$208.9   &53   &$-$201   &75   &7.9  &1.8  &C &93.8 &A     \\
G 24-25 &$-$307.2   &191  &$-$225   &61   &11.2 &1.9  &C &201  &D \\
G 26-12 &$-$239.3 &299 &$-$164 &$-$39 &19.9 &10.87 &V, Sc, Bk05 &306.7 &A \\
G 59-24 &$-$57      &217  &$-$234  &$-$109  &13.8 &3.2  &C &243.2 &A,R     \\
G 63-46 &$-$24.8   &$-$77   &$-$24   &$-$44   &9.4  &0.6  &C &215.1 &D     \\
G 64-12 &442.5  &$-$77   &$-$272   &398  &44.7 &25.0  &C &408.7 &A,R       \\
G 64-37 &91     &$-$165  &$-$252  &$-$86   &10.6 &1.7  &C &188.8 &A,R     \\
G 74-5  &24.2    &46   &$-$45   &$-$32   &8.4  &0.3  &C &183.8 &D     \\
G 75-56 &23.8   &$-$62   &$-$246  &$-$47   &8.0  &1.5  &C &82.0 &A,R     \\
G 88-10 &83      &56   &$-$218  &$-$69   &8.0  &2.3  &C,Be &88.9 &A     \\
G 90-3  &29.1    &19   &$-$86   &$-$13   &8.1  &0.1  &C &136.0 &D     \\
G 92-6  &49.2   &$-$173  &$-$97    &65   &11.6  &11.6  &C &222.0 &D     \\
G 108-58 &143.0   &8    &$-$209  &$-$61   &7.9  &1.2  &C &62.5 &A    \\
G 113-9 &191   &$-$31   &$-$291  &$-$2    &\nodata &\nodata &RN &77.5 &A,R \\
G 113-22 &53.8   &$-$29   &$-$63    &59   &8.2  &0.9 &C &170.2 &D       \\
G 115-49 &$-$51    &$-$81   &$-$308  &$-$36   &8.7  &0.4  &C &124.9 &A,R     \\
G 126-52 &$-$241.9  &$-$79   &$-$290  &$-$8    &8.5  &0.1  &C &105.9 &A,R   \\
G 130-65 &$-$271.0  &$-$143  &$-$300   &14   &10.1 &0.2  &C &164.5 &A,R     \\
G 153-21 &$-$64.3    &72   &$-$28    &21   &9.1  &0.2  &C &206.1 &D     \\
G 180-24 &$-$157    &$-$64   &$-$192  &$-$46   &7.9  &1.3  &C &83.6 &A     \\
G 181-28 &$-$169.8 &12 &$-$263 &60 &8.0 &0.5 &C &74.8 &A,R \\
G 188-22 &$-$94.9   &$-$112  &$-$86    &63   &9.6  &1.0  &C &185.7 &D     \\
\\
\\
\\
\\
\\
G 191-55 &$-$258.2  &$-$224  &$-$168   &16   &13.4 &0.2  &C &230.5 &D     \\
G 192-43 &189     &265  &$-$128   &29   &17.5  &0.3  &C &282.0 &A     \\
G 201-5 &$-$35.7   &$-$262  &$-$51    &28   &20.6  &0.9  &C &313.0 &A      \\
G 206-34 &$-$72     &$-$140  &$-$144  &$-$89   &9.9  &2.0  &C &182.5 &D      \\
G 268-32 &80.8    &194  &$-$241   &156  &\nodata &\nodata &Sc &249.8 &A,R   \\
G 275-4 &135   &114  &$-$293  &$-$261  &\nodata &\nodata  &RN,Sc &294.0 &A,R \\
LP 553-62 &177  &$-$1   &$-$314   &35   &\nodata &\nodata &RN,Sc &100.3 &A,R \\
LP 635-14 &$-$106   &$-$83   &$-$211   &41   &\nodata &\nodata &RN &93.0 &A \\
LP 651-4 &34  &$-$157  &$-$211  &$-$138  &\nodata &\nodata &RN,Sc &209.2 &A \\
LP 752-17 &$-$225  &73  &$-$382   &25   &\nodata &\nodata &RN,Sc &179.4 &A,R \\
LP 815-43 &15  &$-$128  &$-$195   &13    &\nodata &\nodata &RN,Sc &131.1 &A \\
LP 831-70 &$-$33  &$-$16    &$-$175   &100   &6.5 &\nodata &RN,Sc &110.8 &D \\
LTT 1566 &154     &13   &$-$336  &$-$123  &\nodata &\nodata &RN &169.6 &A,R \\
Ross 390 &89     &$-$152  &$-$228   &52   &\nodata &\nodata  &RN &160.8 &A,R \\
Ross 797 &12   &$-$260  &$-$262  &$-$81  &\nodata &\nodata   &RN &275.5 &A,R \\
\enddata

\tablenotetext{1}{
APS = Allen et al.~(1991),
Bk04 = Borkova et al.~(2004),
Bk05 = Borkova et al.~(2005),
Bo = Bobylev et al.~(2006),
C = Carney et al.~(1994),
Ch = Chen et al.~(2000),
Eg = Eggen (1996),
F = Fulbright (2002),
GC = Geneva-Copemhagen (Nordstrom et al.~(2004) as recalibrated Holmberg et al.~(2007), 
Gr03 = Gratton et al.~(2003),
M = Mishenina et al.~(2004),
N = Nissen et al.~(2000),
R06 = Reddy et al.~(2006),
RN = Ryan \& Norris (1991),
Sc = Schuster et al.~(2006),
So = Soubiran \& Girard (2005),
V = Venn et al.~(2004),
}

\end{deluxetable}

\clearpage
\begin{figure}
\plotone{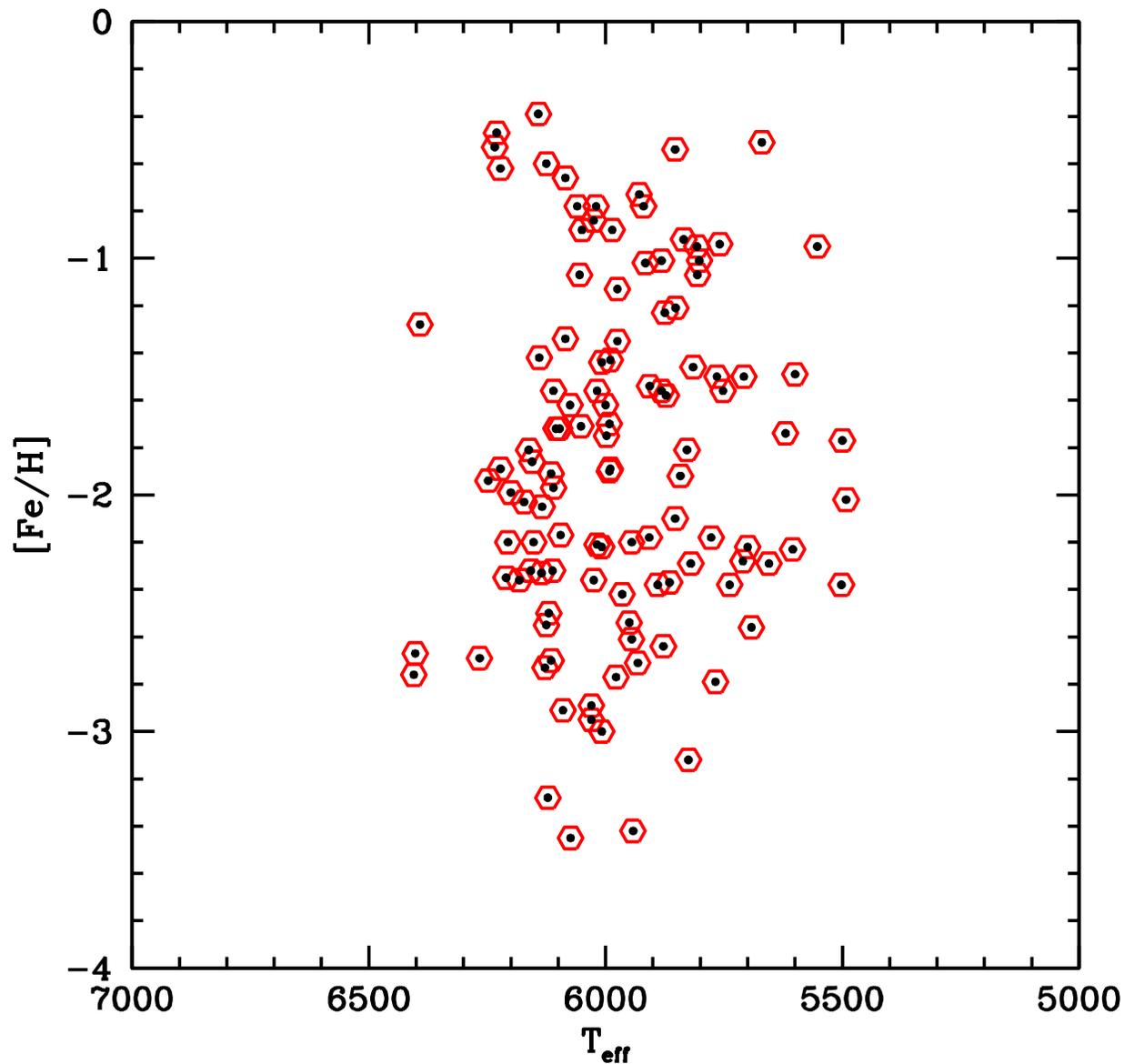}
\caption{Distribution of our sample of stars in the $T_{\rm eff}$ -- [Fe/H]
plane.  Our stars cover more than 3 orders of magnitude in [Fe/H].  We set a
limit on $T_{\rm eff}$ of $>$ 5500 K due to the growing weakness of the Be II
lines and increasing strengths of the blending features below that
temperature.  There are six stars with [Fe/H] $\leq$ $-$3.0.  The parameters
plotted are those derived in section 3.}
\end{figure} 

\begin{figure}
\plotone{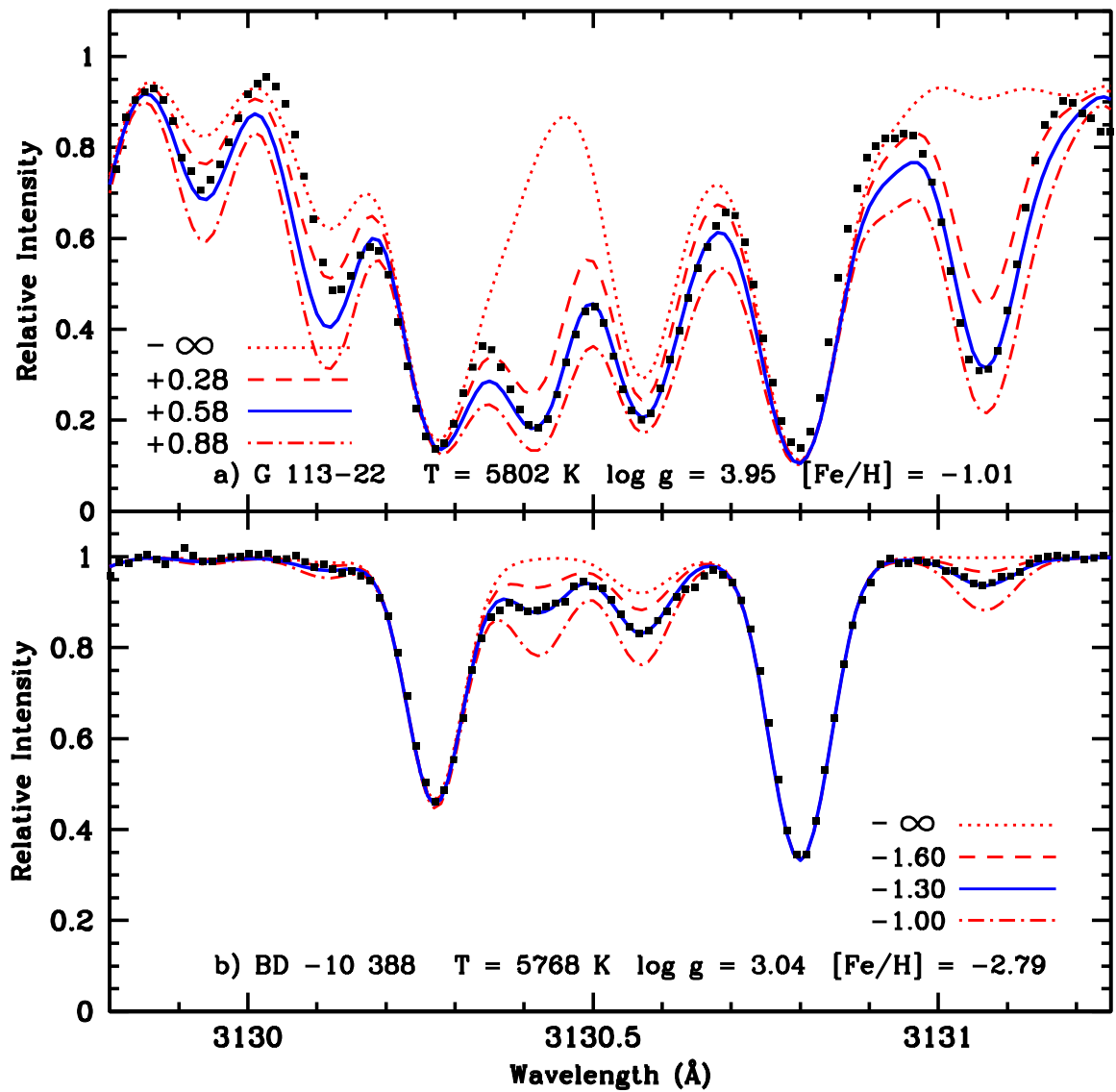}
\caption{Spectrum synthesis in the Be region of stars with similar $T_{\rm
eff}$ but very different metallicities.  The Be II lines are at 3130.421 and
3131.065 \AA.  The filled squares are the observational data points and the
solid line is the best fit.  The dotted line contains no Be.  The dashed and
dot-dash lines are a factor of two lower and higher in Be abundance.  For G
113-22, with a value for [Fe/H] of $-$1.01, the spectrum is full of line
blends and has strong lines.  For BD $-$10$\arcdeg$ 388 at [Fe/H] = $-$2.79
the synthesis is much more straight-forward and both the Be II lines and the
OH features are well-fit.  In this example the solid line is also for the best
fit for OH and the other syntheses differ in [O/H] by 0.20 dex.  The best fit
for this OH feature gives [O/H] of $-$0.53 in G 113-22 and $-$2.00 in BD
$-$10$\arcdeg$ 388.  It is clearly easier to derive good Be abundances in
metal-poor stars.}
\end{figure} 

\begin{figure}
\plotone{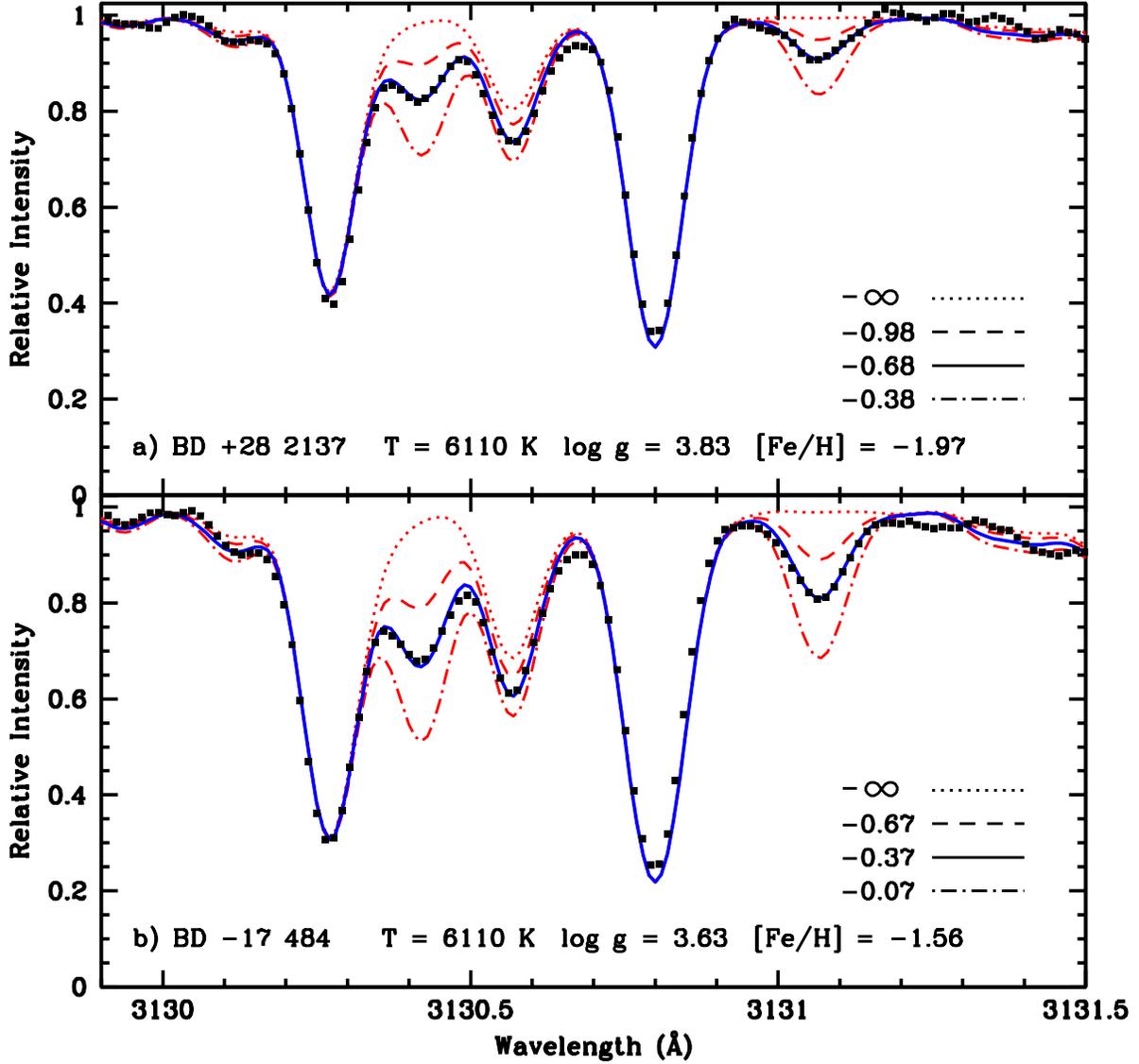}
\caption{Spectrum synthesis in the Be region of two stars with the same
$T_{\rm eff}$.  The metallicities differ by a factor of 2.6 and the Be
abundances by a factor of 2.0.  The lower metallicity star has the lower Be
abundance.  The points and the lines are as in Figure 2, but the O abundances
differ by 0.10 dex in this figure.  The best fit for this OH feature gives
[O/H] of $-$1.37 in BD +28$\arcdeg$ 2137 and $-$0.98 in BD $-$17$\arcdeg$
484.}
\end{figure} 

\begin{figure}
\plotone{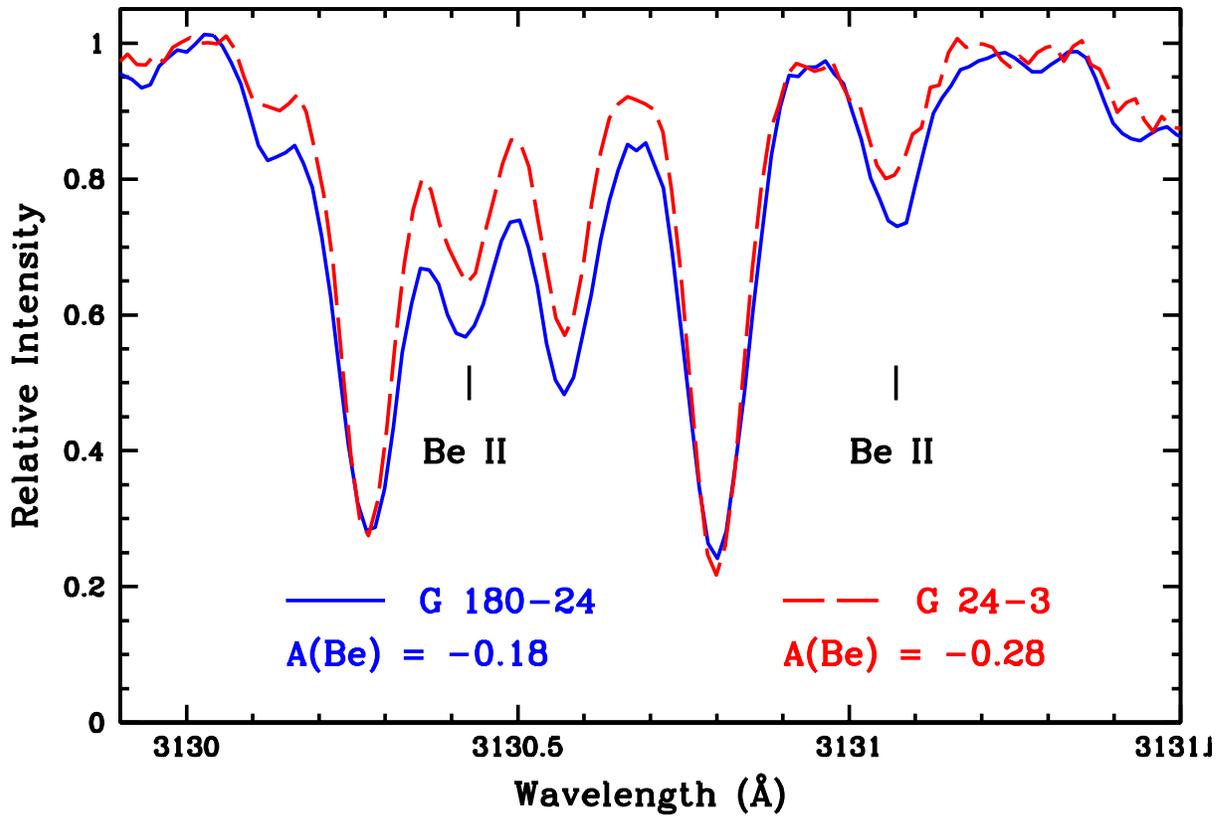}
\caption{The observed spectra of two similar stars.  The star with the broader
lines and the stronger Be lines, G 180-24, has a lower log g by 0.21 dex.
This comparison shows the effect of log g on A(Be): the Be II lines are
stronger for a given Be abundance in subgiants than in dwarf stars with
similar parameters.}
\end{figure} 

\begin{figure}
\plotone{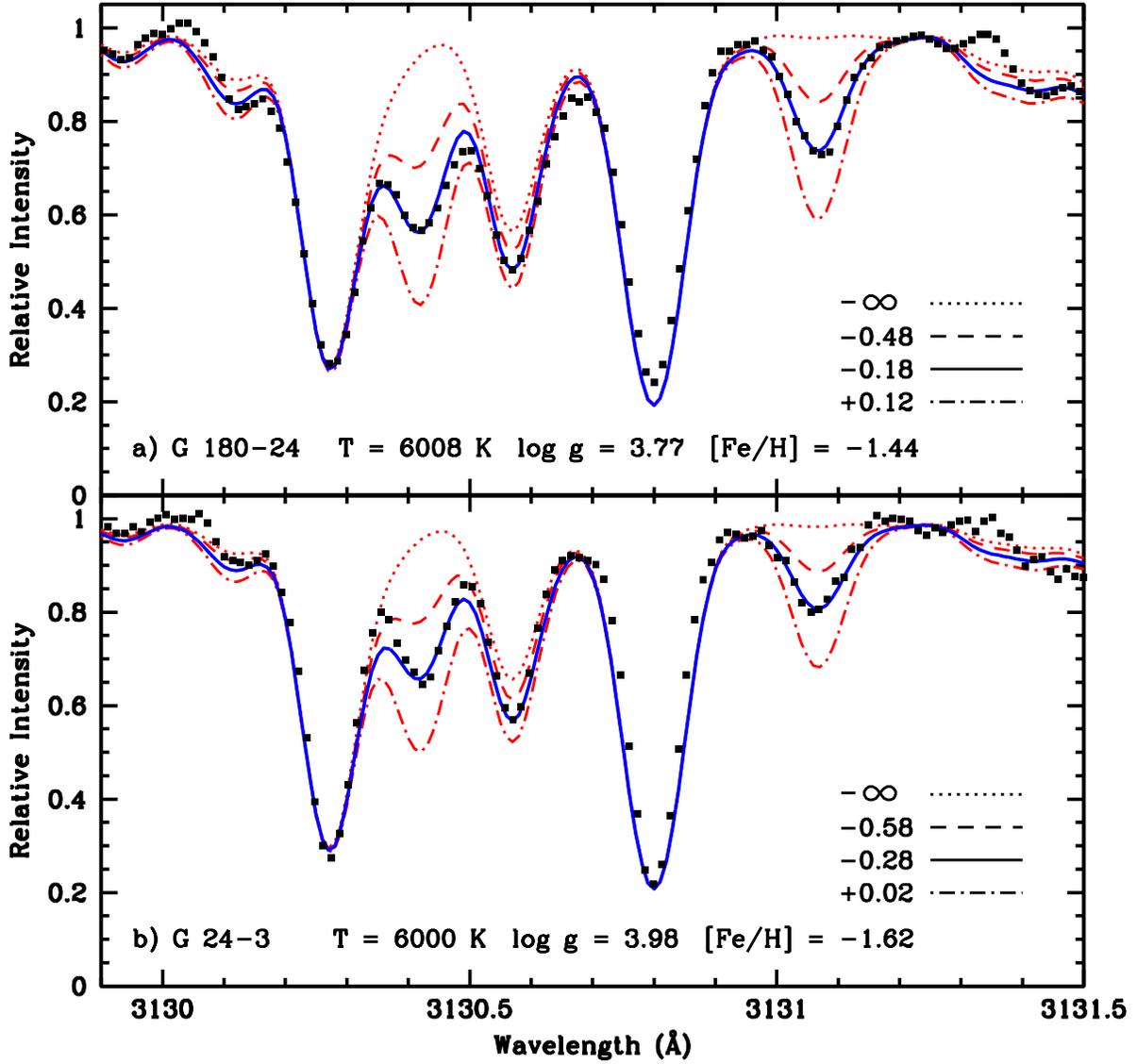}
\caption{The spectrum syntheses for the two stars in Figure 4.  The best fit
to the data points is the solid line. The dotted line contains no Be.  The
dashed and dot-dash lines are a factor of two lower and higher in Be
abundance.  The solid line is also for the best fit for OH and the other
syntheses differ in [O/H] by 0.10 dex, with the best fit for this OH feature
of [O/H] = $-$0.92 in G 180-24 and $-$1.20 in G 24-3.}
\end{figure} 

\begin{figure}
\plotone{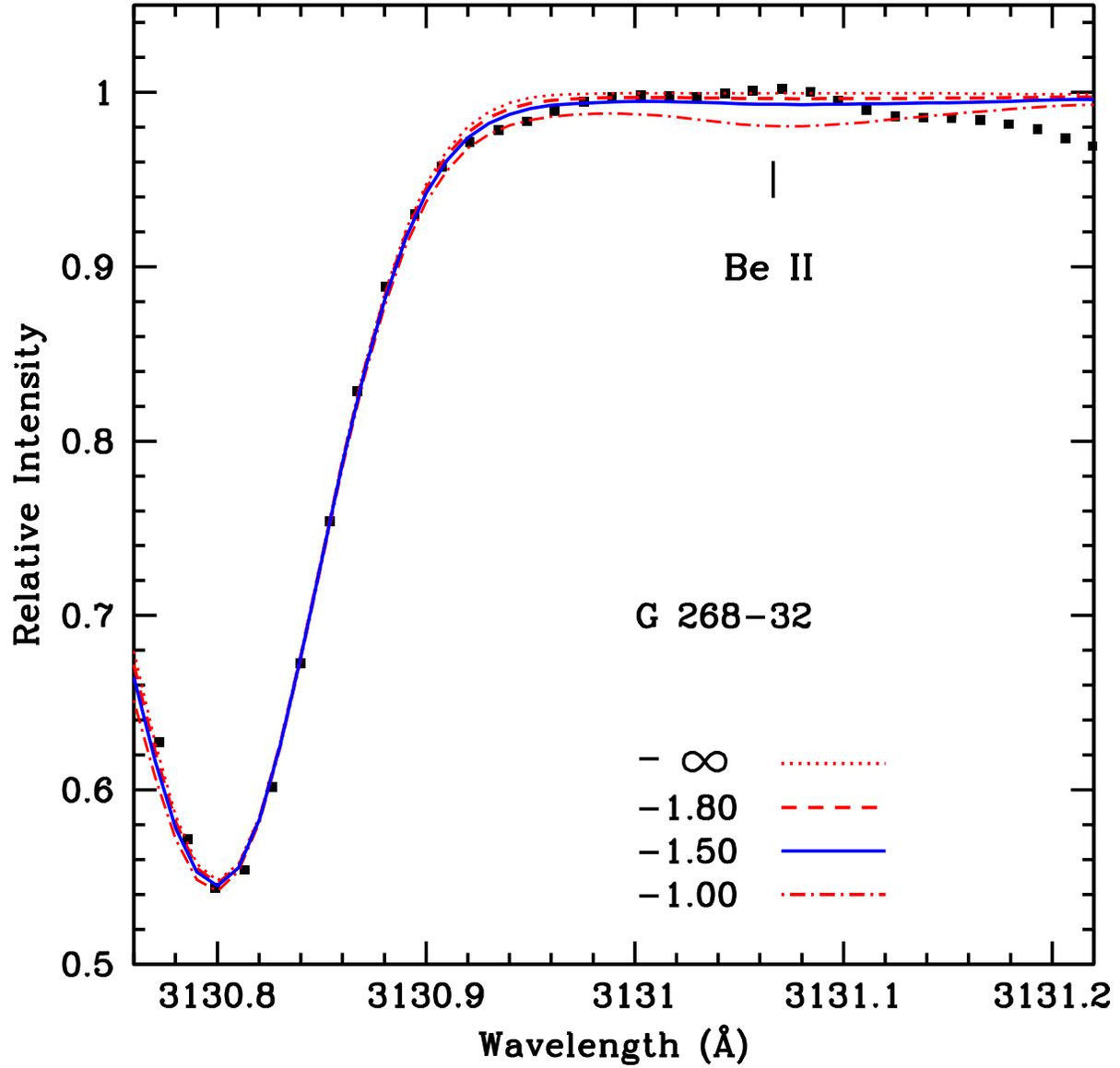}
\caption{An expanded view of the 3131 \AA{} line of the Be II in the CEMP
star, G 268-32.  The CH lines were removed from the line list in order to get
an upper limit for A(Be).  The feature at 3130.8, mostly Ti II, is well-fit.}
\end{figure} 

\begin{figure}
\plotone{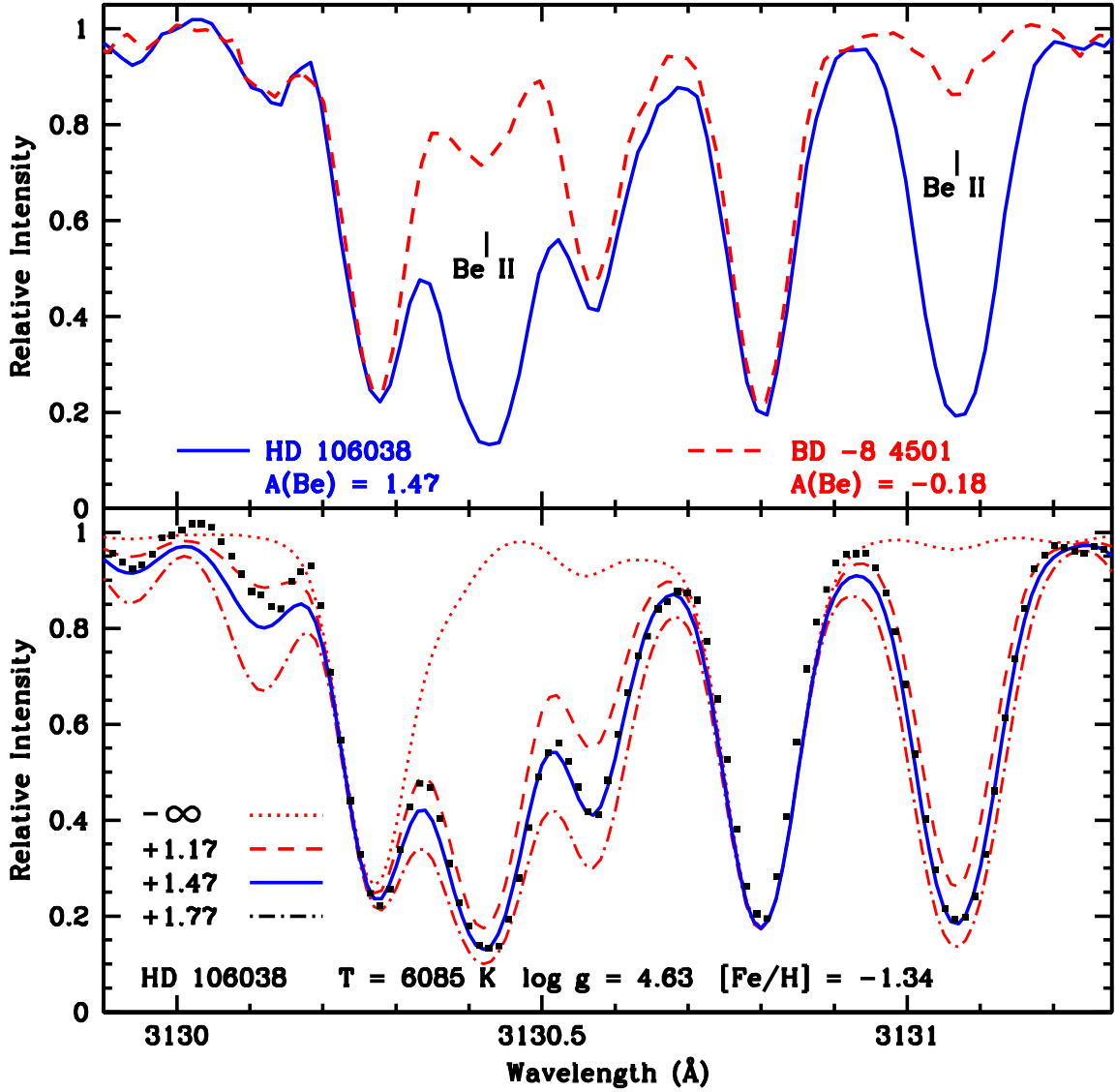}
\caption{Top panel: Our Keck spectrum of the Be-rich star, HD 106038
discovered by Smiljanic et al (2008) compared to BD $-$8$\arcdeg$ 4801, a star
of similar metallicity ([Fe/H] = $-$1.23) and log g (4.39).  Lower panel: The
spectrum synthesis for HD 106038 with our spectroscopically determined
parameters.  The solid line is the best fit with A(Be) = 1.47 and [O/H] =
$-$0.95. }
\end{figure} 

\begin{figure}
\plotone{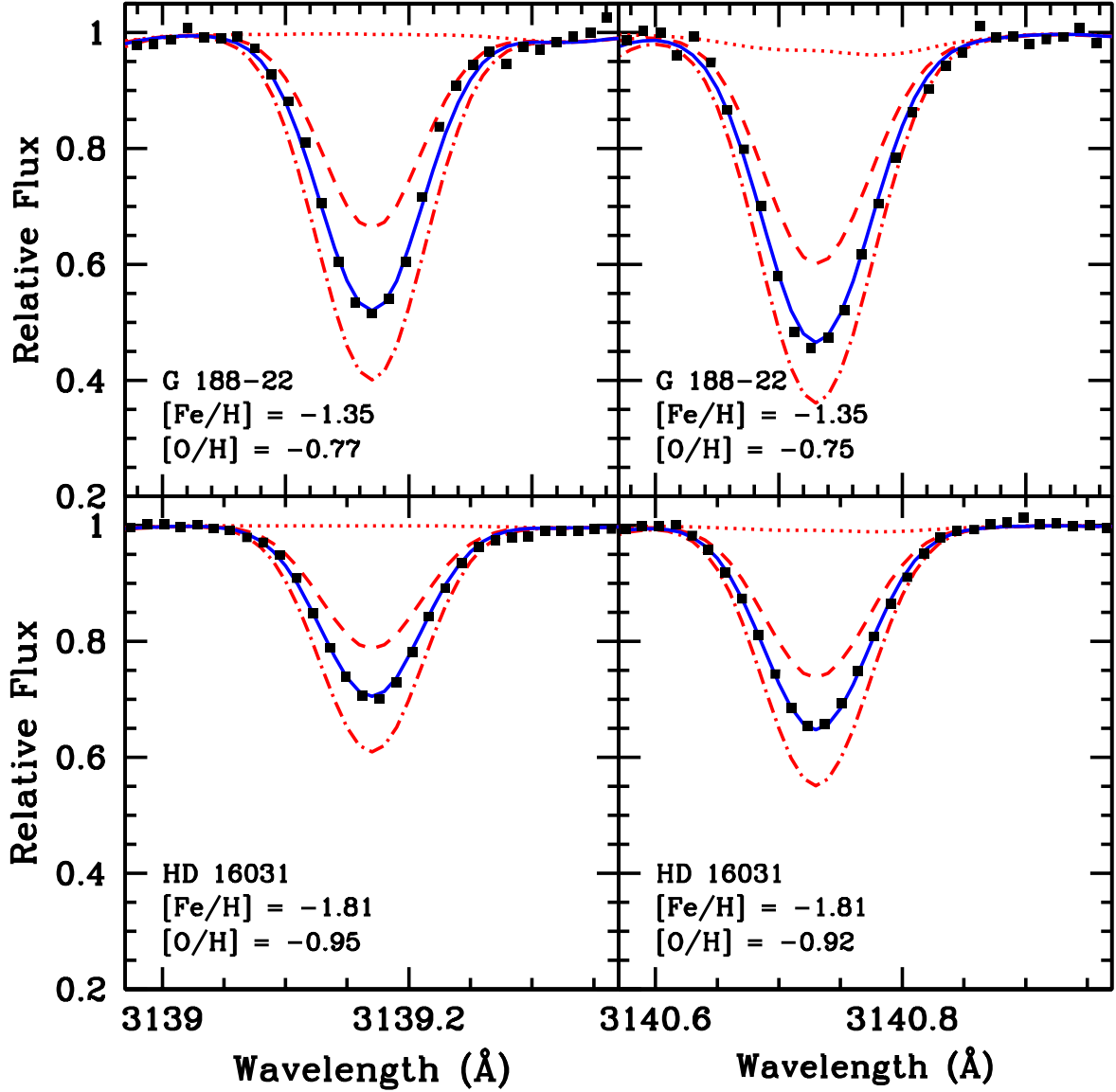}
\caption{Spectrum syntheses for the other two OH features used to find O
abundances in two of our stars.  The observations are the solid squares and
the solid line is the best fit O abundance.  The dotted line corresponds to no
oxygen.  The dashed and dot-dash lines are 0.2 dex less and 0.2 dex more O.
The final [O/H] abundance for G 188-22 is $-$0.78 and for HD 16031 is $-$0.97,
showing good agreement among the three features.}
\end{figure} 

\begin{figure}
\plotone{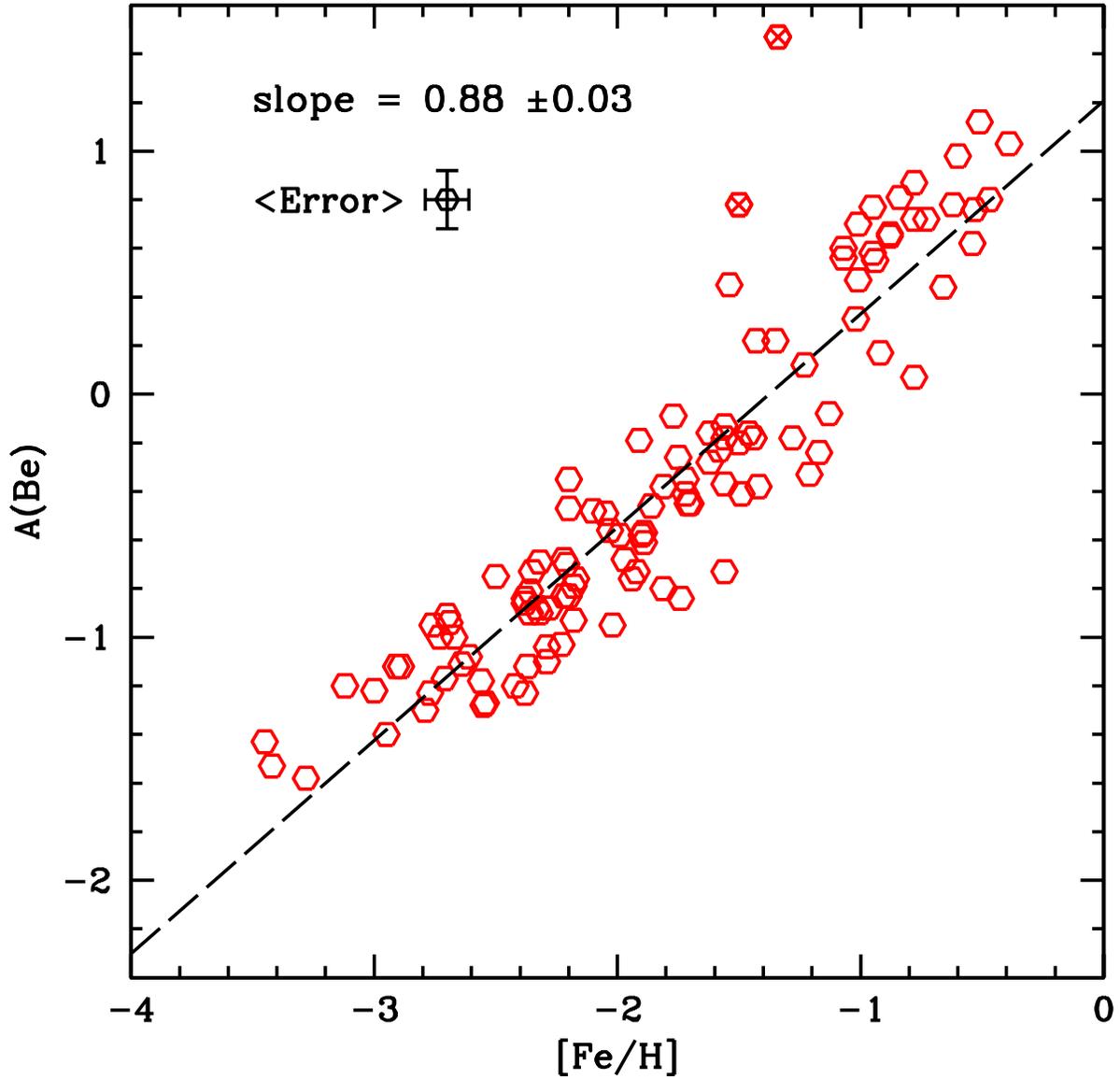}
\caption{Our Be abundances for the stars in Table 2 plotted against our
derived Fe abundances.  The slope of +0.89 was calculated excluding the two
points that are indicated by a cross within the hexagon.  Those two high
points are the Be-rich stars, HD 106038 and HD 132475.  The mean 1$\sigma$
error bar is shown in the upper left in this figure and in subsequent figures.}
\end{figure}

\begin{figure}
\plotone{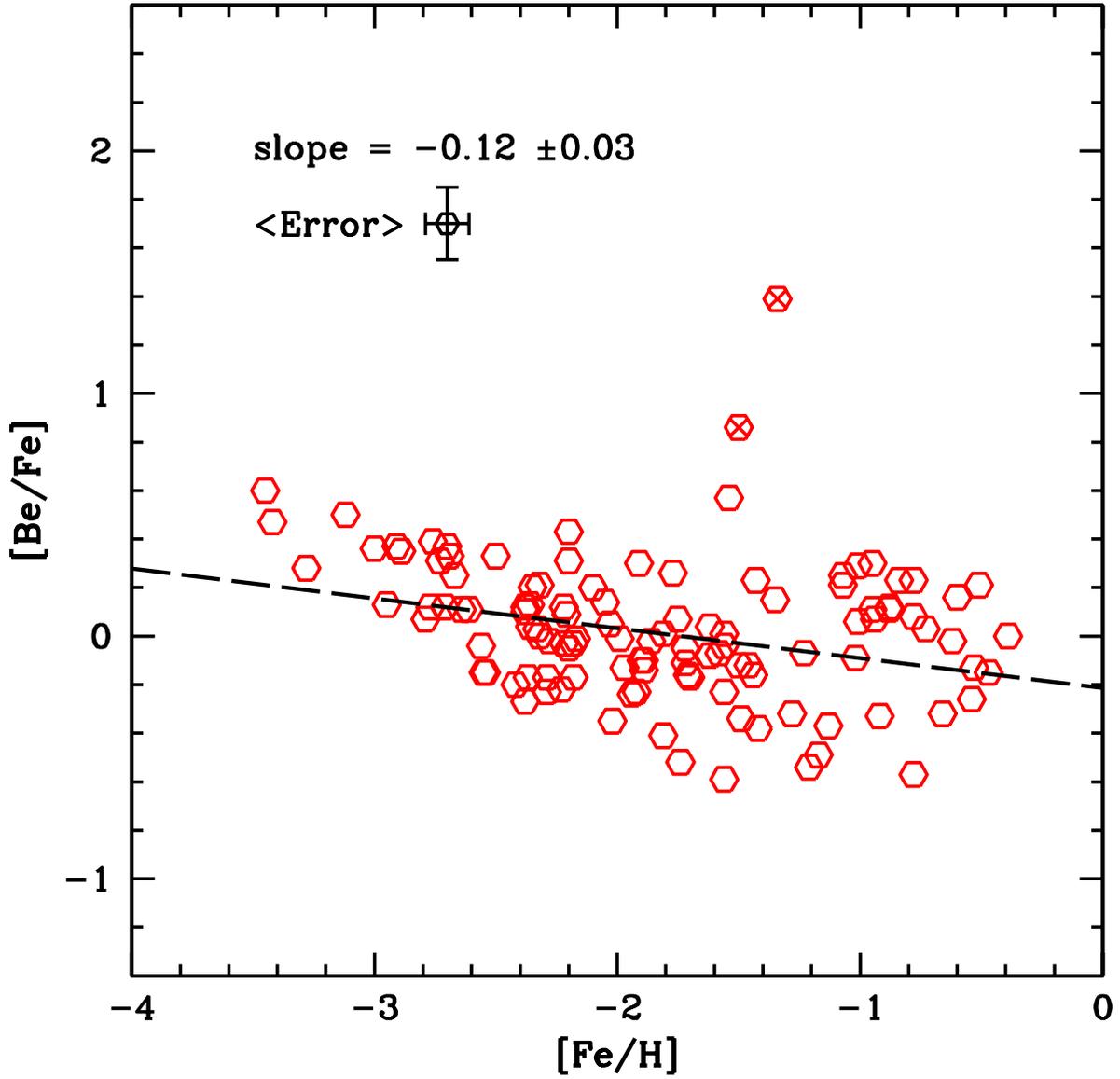}
\caption{This plot shows Be values as normalized to the Fe values compared to
Fe.  Again the two Be-rich stars were not used in the calculation of the slope
and are indicated by a cross within the hexagon.}
\end{figure}

\clearpage

\begin{figure}
\plotone{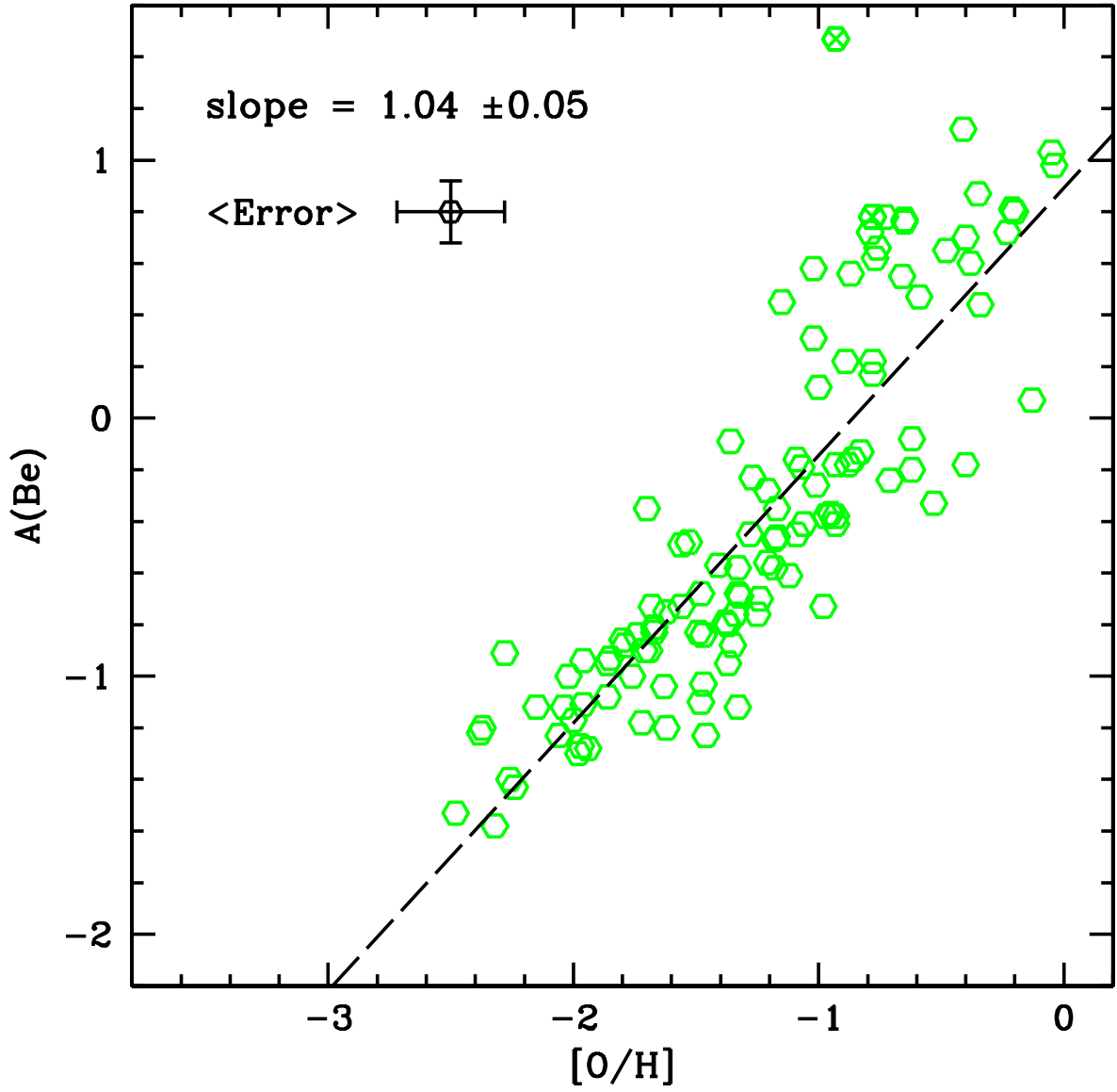}
\caption{Our Be abundances for the stars in Table 2 plotted against our
derived O abundances.  Both the Be-rich and the Be-poor stars lie above the
best fitting straight line indicating that a polynomial fit might be better. }
\end{figure}

\begin{figure}
\plotone{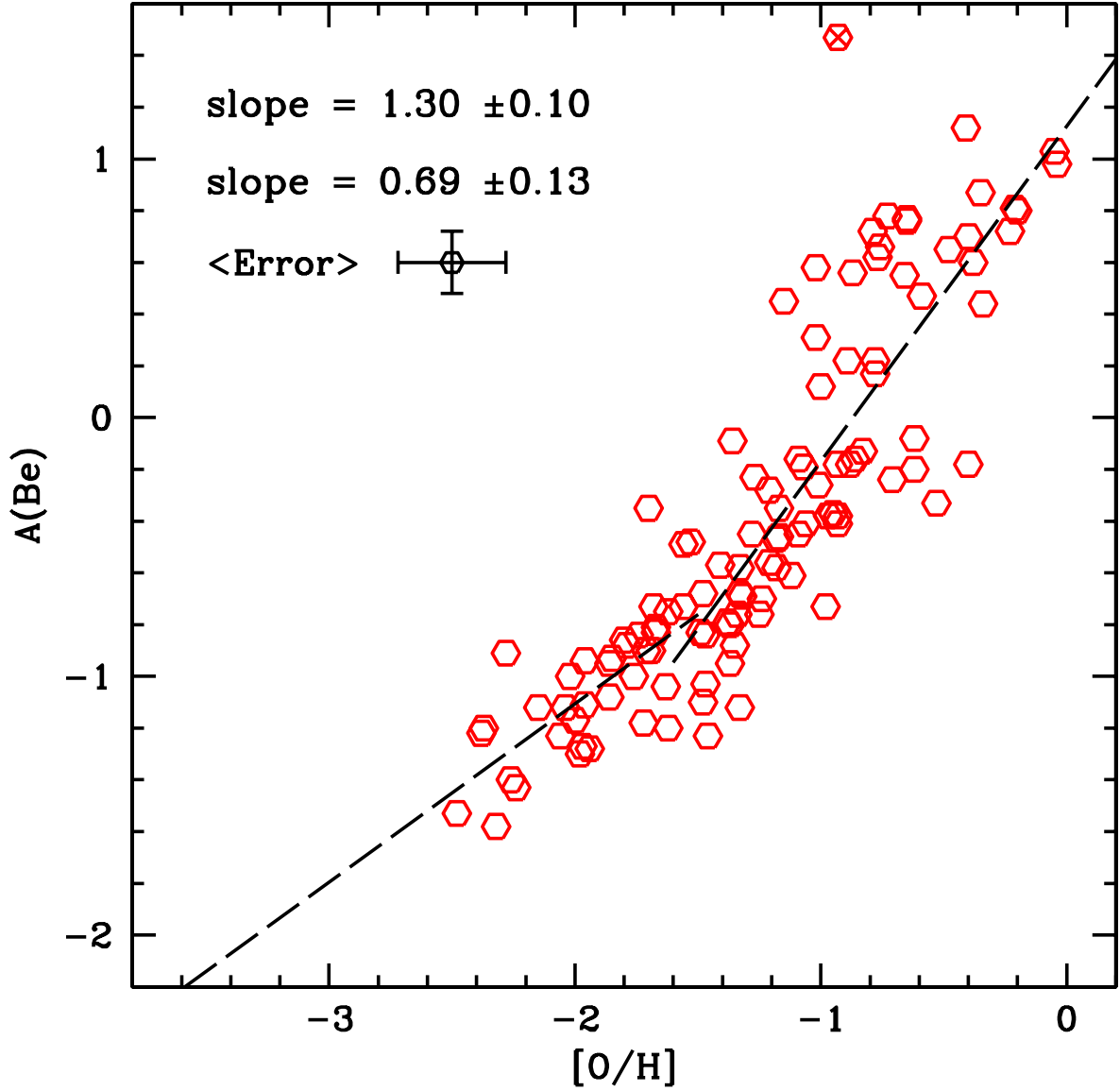}
\caption{A fit for the A(Be) with [O/H] with 2 lines.  We have separated the
stars into high-O and low-O groups.  See the text in $\S$4.2 for discussion.}
\end{figure}

\begin{figure}
\plotone{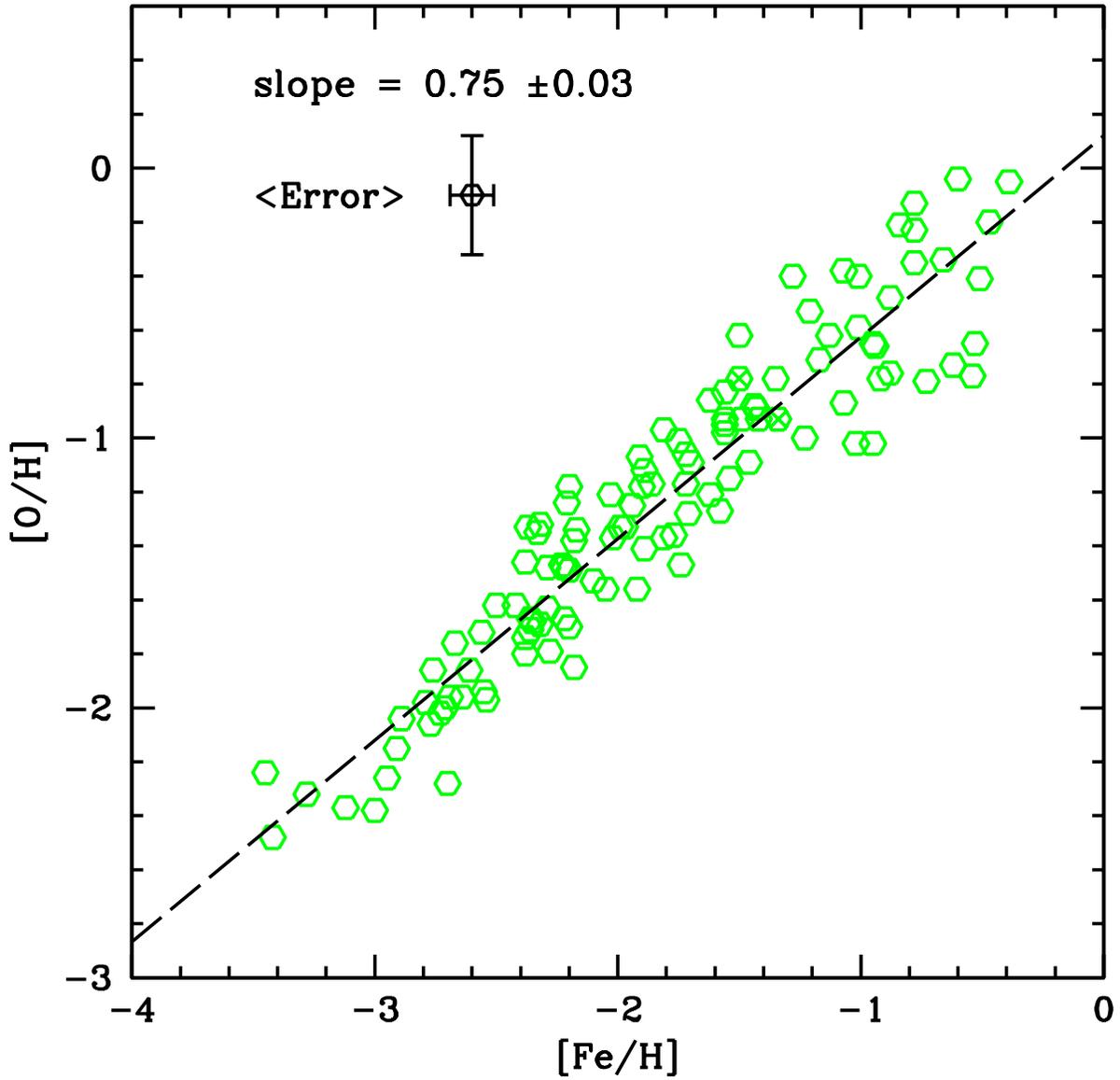}
\caption{The relationship between [O/H] and [Fe/H].  This shows a good linear
fit with smaller scatter than in the A(Be) vs.~[Fe/H] plot in Figure 11.  This
in turn implies that the slope change in Figure 12 is due to Be, not O. even
though the 1$\sigma$ error bar is larger for [O/H] than for A(Be).}
\end{figure}

\begin{figure}
\plotone{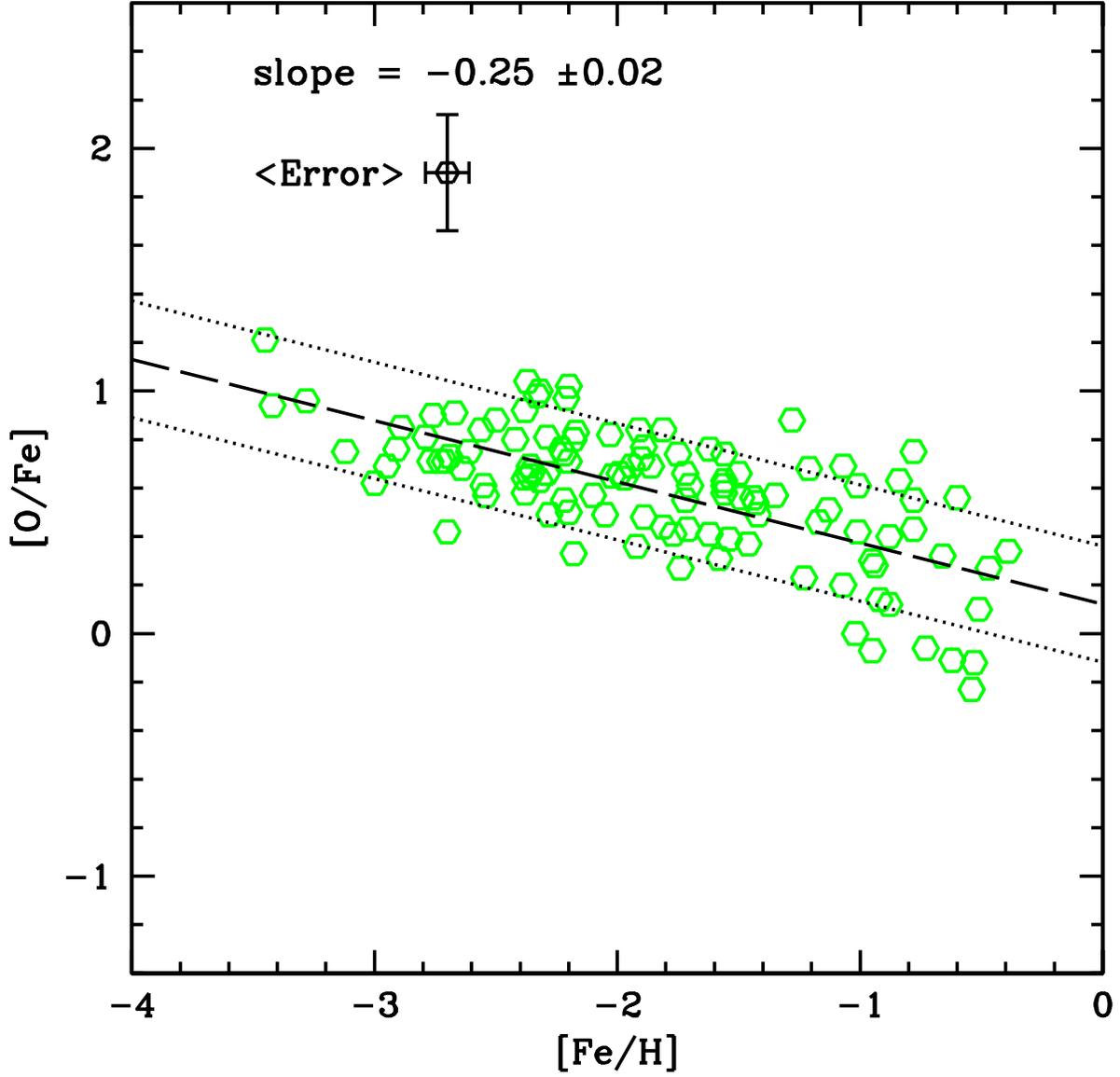}
\caption{ Oxygen as normalized to Fe vs.~Fe. The 1$\sigma$ error bars due to
[O/Fe] are drawn parallel to the best fit.  There is more scatter in [O/Fe]
for values of [Fe/H] $>$ $-$1.4.}
\end{figure}

\begin{figure}
\plottwo{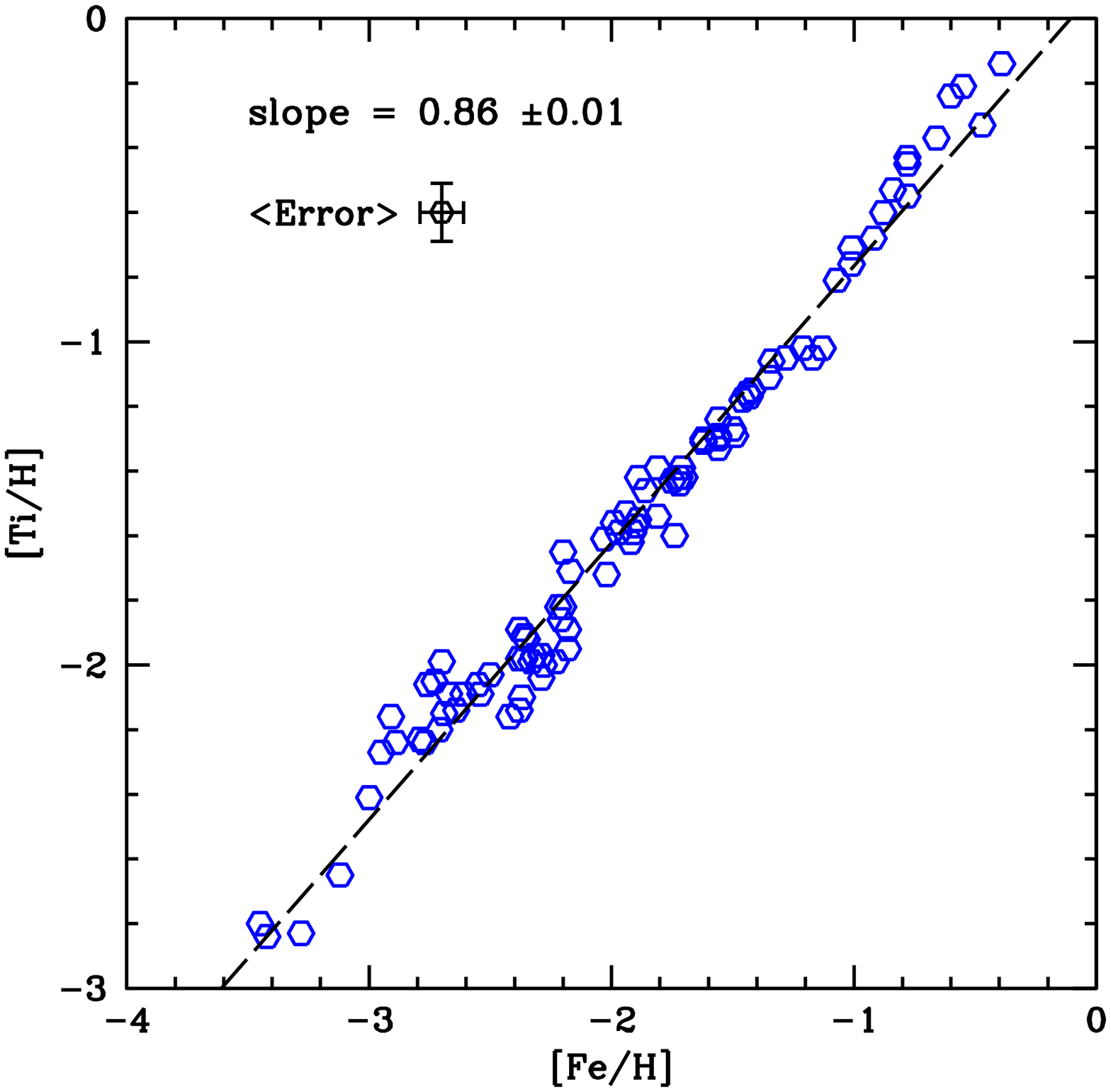}{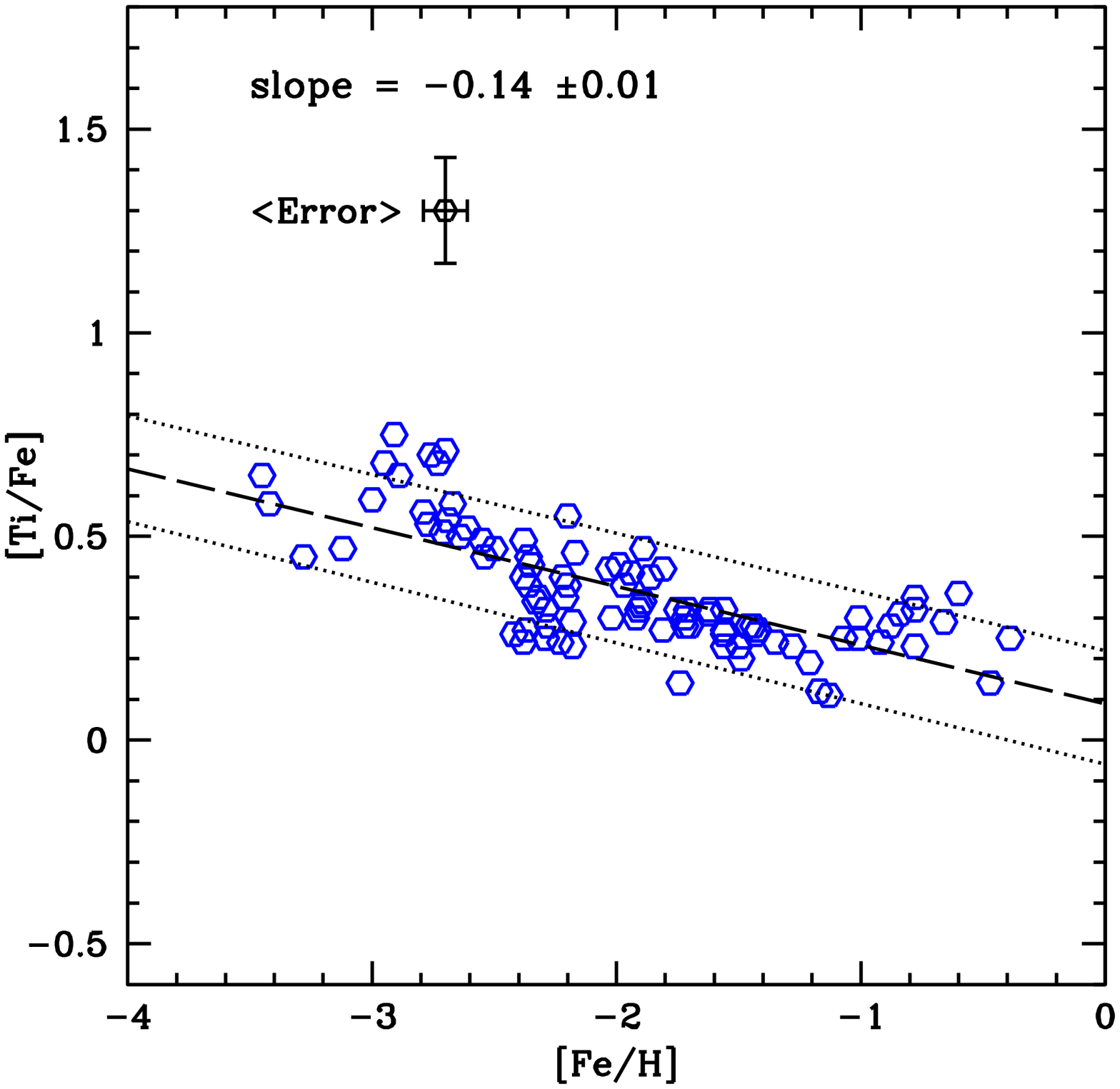}
\caption{These plots are the analogs of Figures 13 and 14 for Ti instead of O.
The correlation between [Ti/H] and [Fe/H] is remarkably tight.}
\end{figure}

\clearpage

\begin{figure}
\plotone{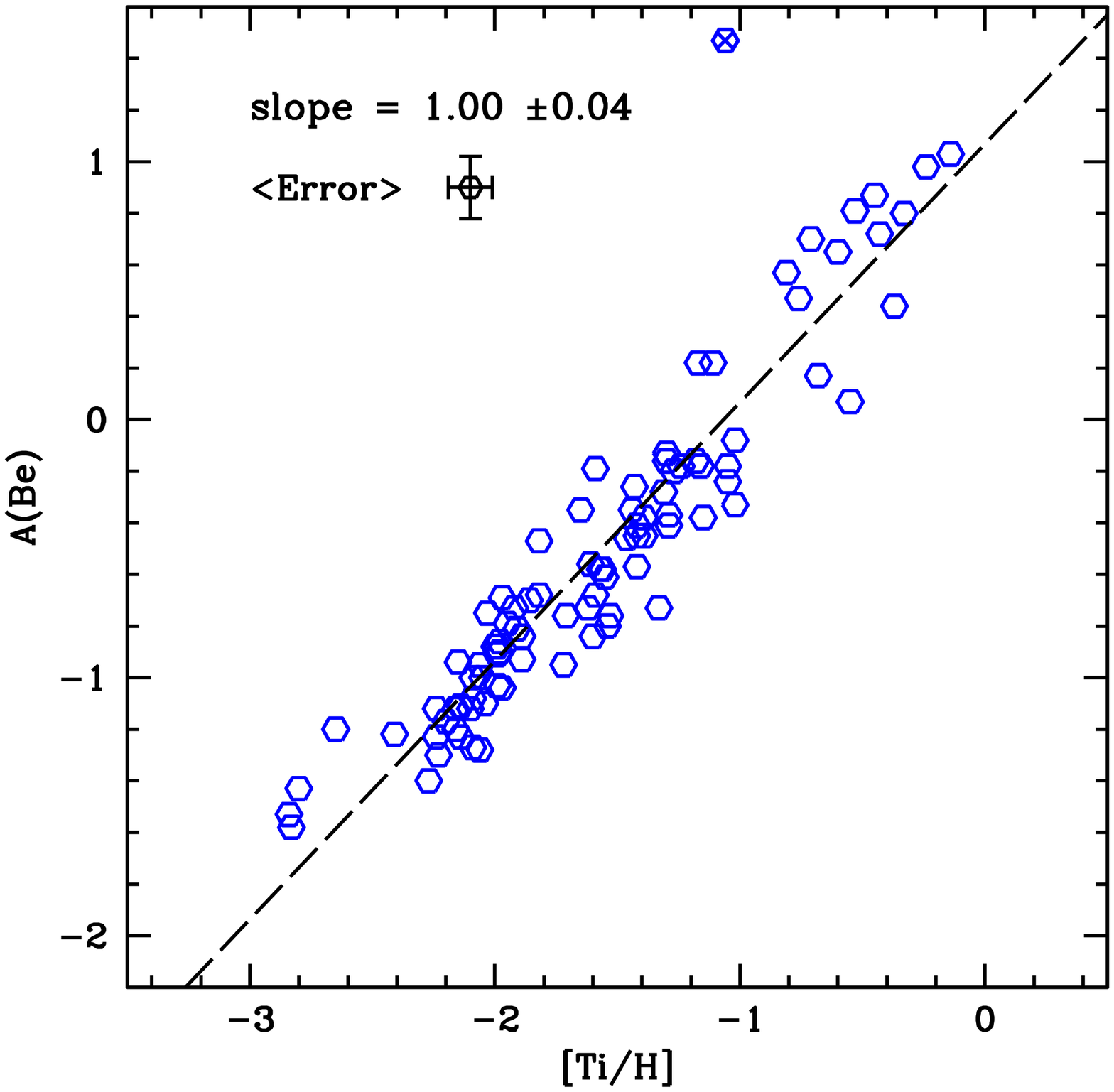}
\caption{The relationship between the alpha-element, Ti, with Be.  This
correlation is considerably tighter than Be and O in Figure 11.  The Be-rich
star, HD 106038 is the hexagon with the cross in it and was not used to find
the slope.  (We do not have a Ti abundance for the other Be-rich star.)}
\end{figure}

\begin{figure}
\plottwo{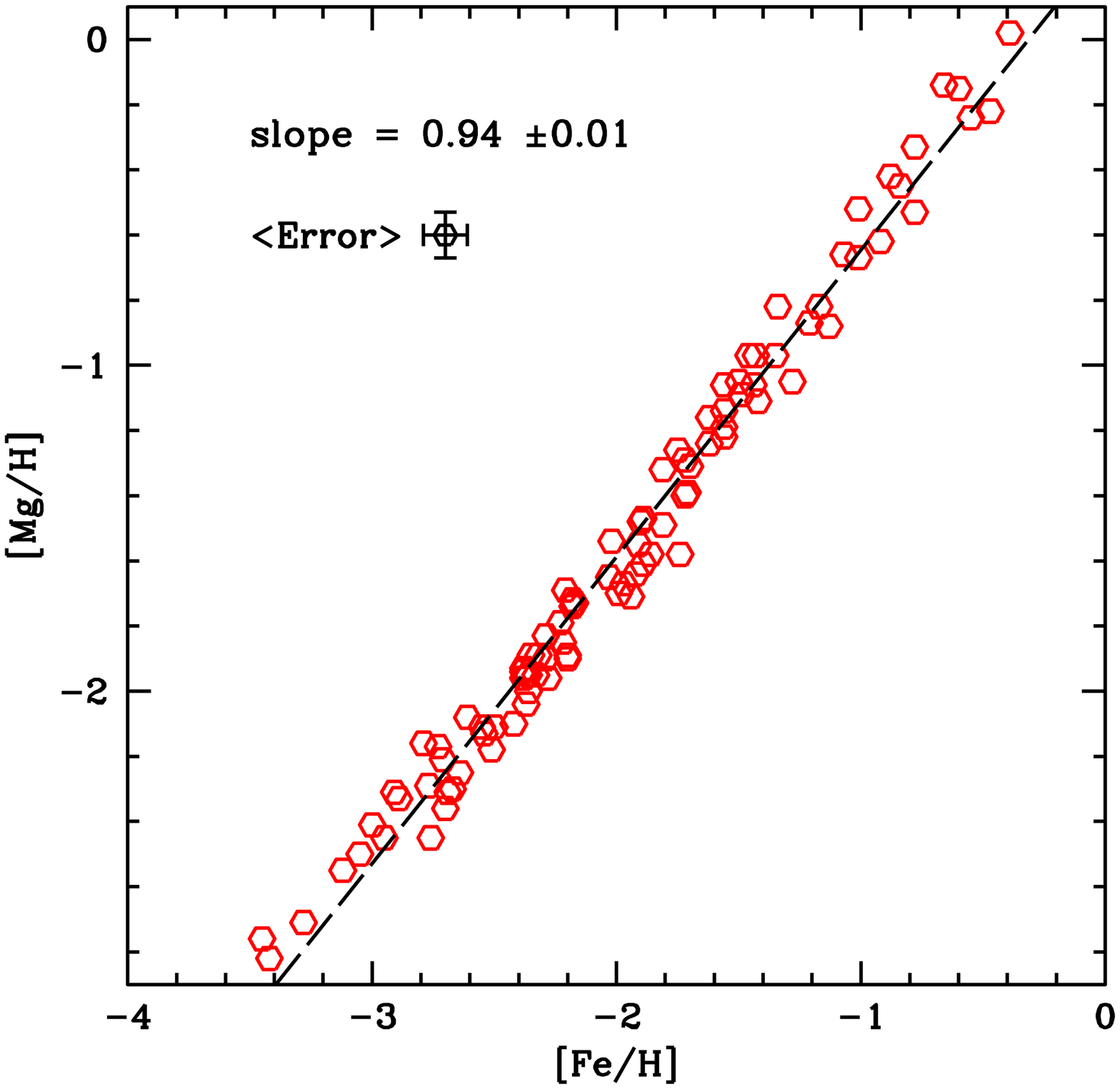}{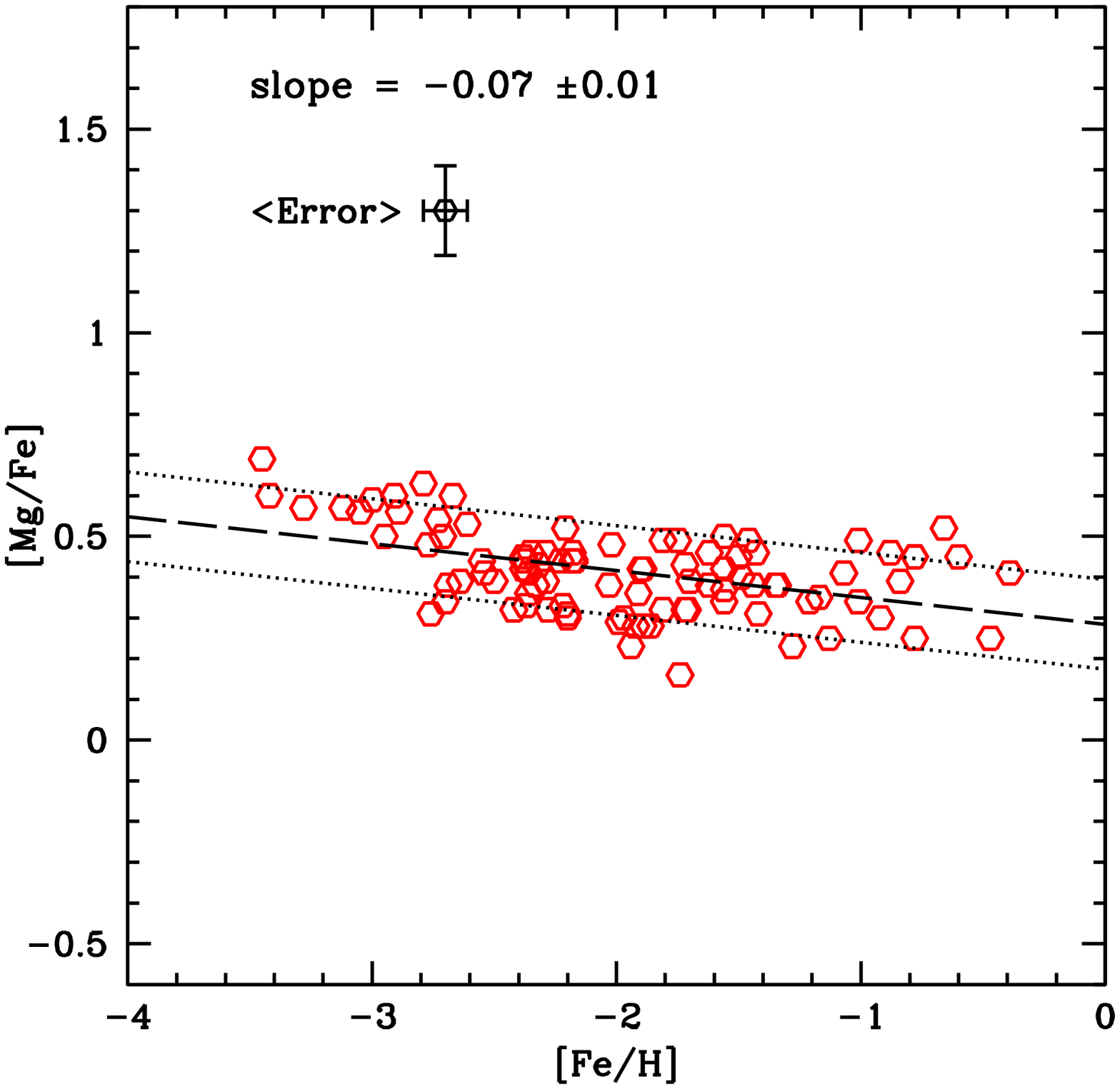}
\caption{These plots are the analogs of Figures 13 and 14 for Mg instead of O.
There is a very tight correlation between [Mg/H] and [Fe/H].}
\end{figure}

\begin{figure}
\plotone{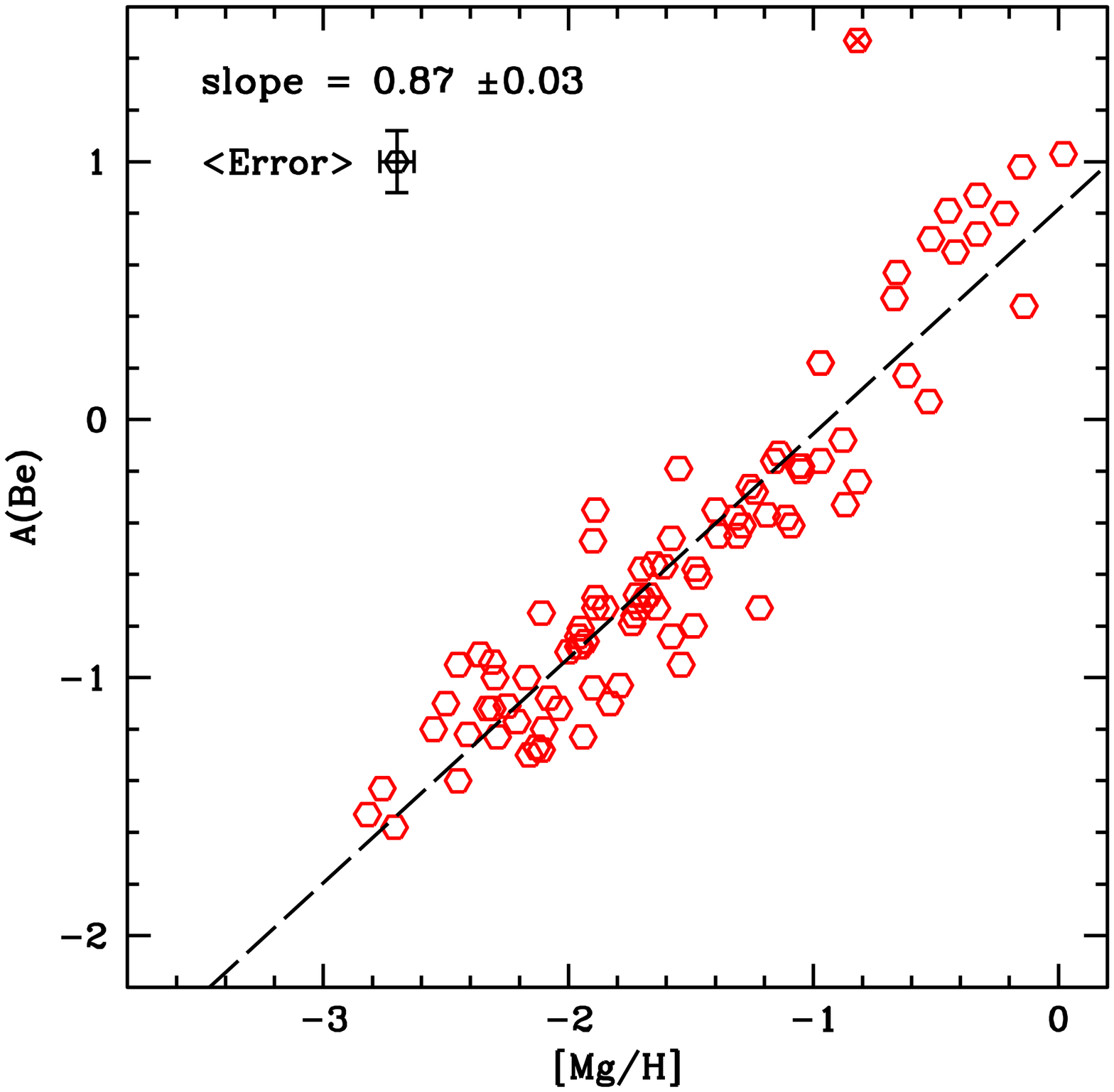}
\caption{he relationship between the alpha-element, Mg, with Be.  This
correlation is considerably tighter than Be and O in Figure 11.  The Be-rich
star, HD 106038 is the hexagon with the cross in it and was not used to find
the slope.  (We do not have a Mg abundance for the other Be-rich star.)}
\end{figure}

\begin{figure}
\plotone{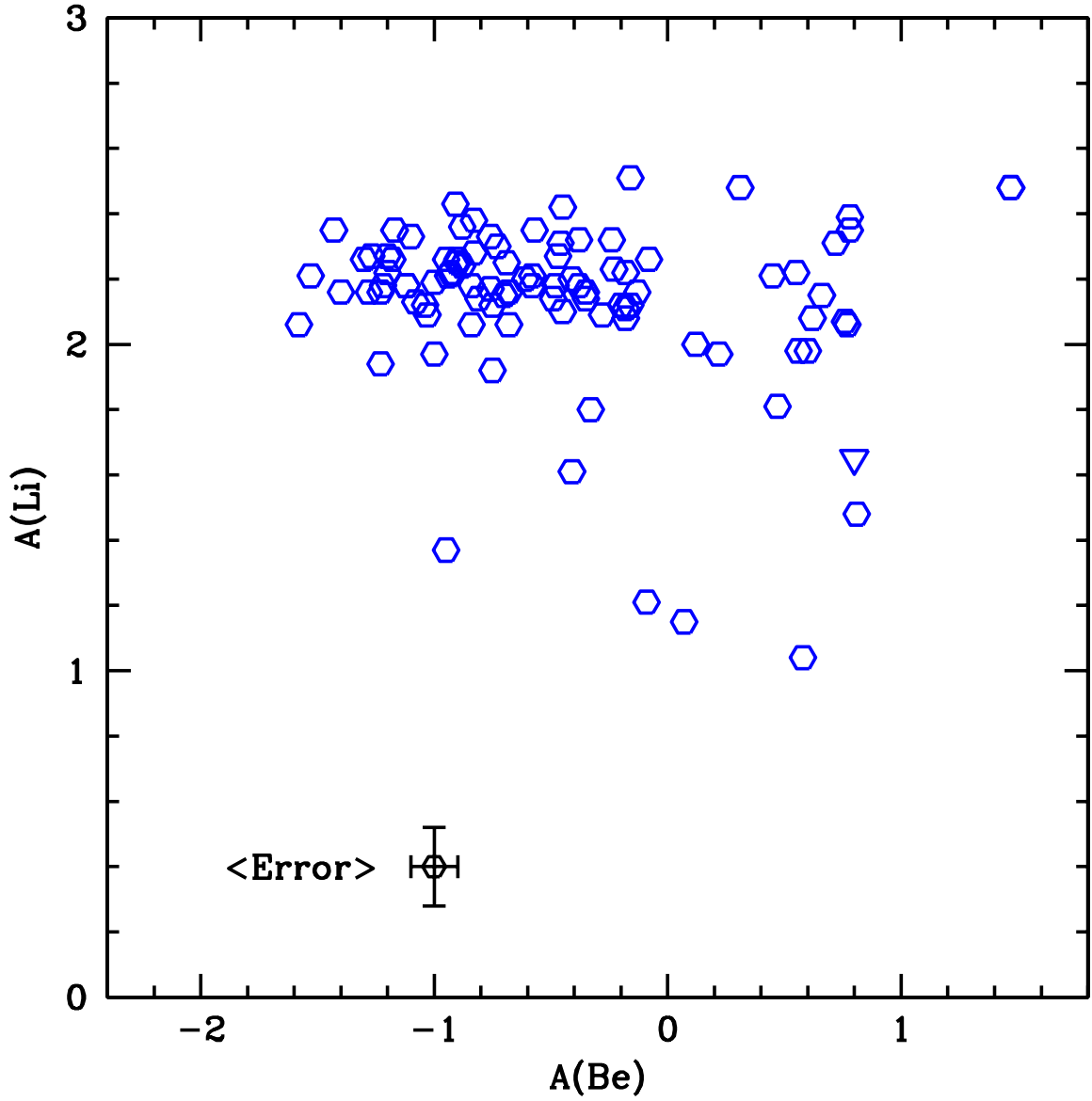}
\caption{Our Be abundances compared with Li abundances in the literature.
There are seven stars that are Li-depleted, but are apparently normal in Be.}
\end{figure}

\begin{figure}
\plotone{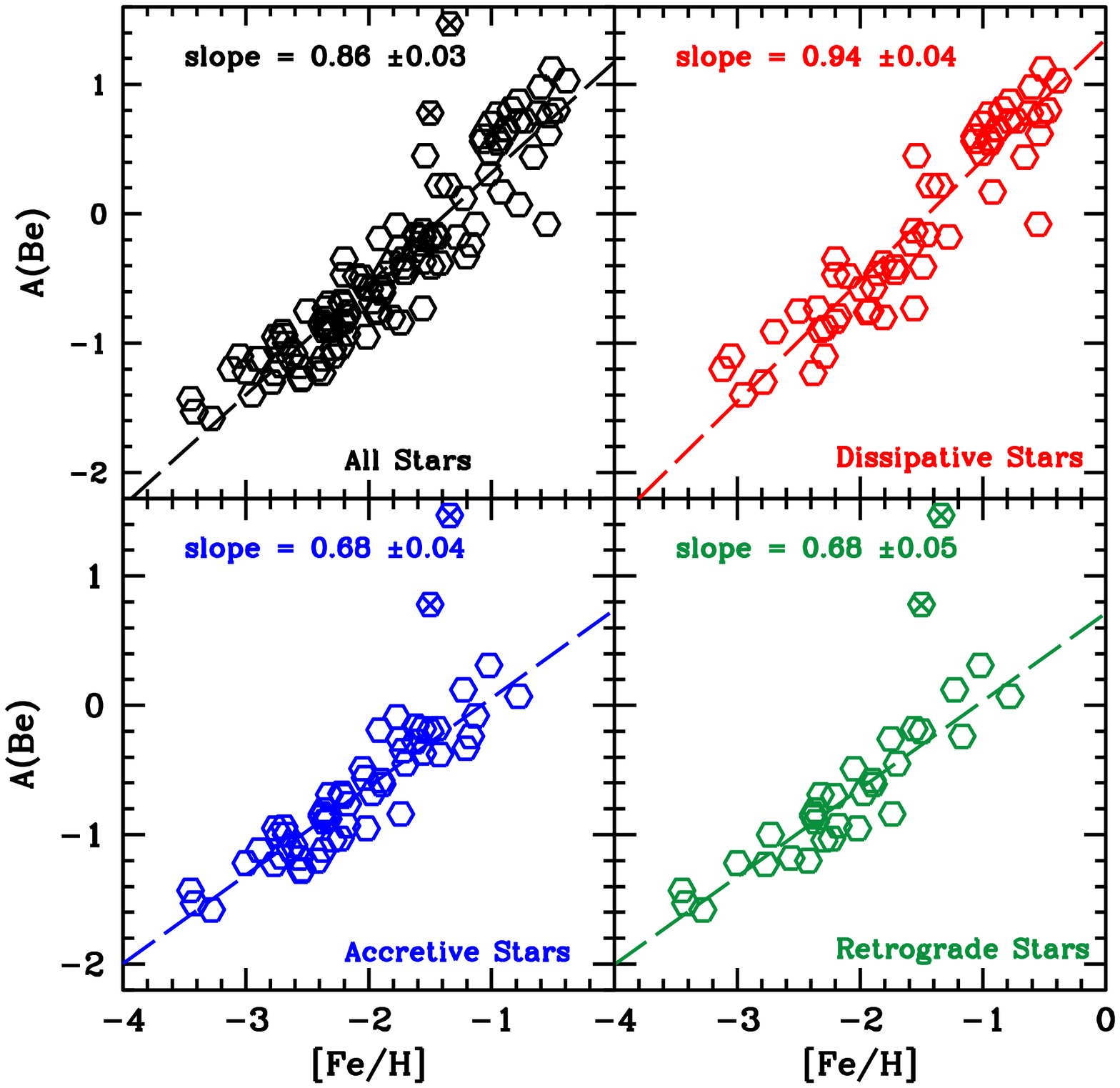}
\caption{The distribution of A(Be) with [Fe/H] for our total sample, for the
dissipative stars, for the accretive stars, and for retrograde subset of the
accretive stars.  There is a steeper slope, 0.94, for the dissipative stars
than for the accretive stars, 0.68.  The hexagons with the crosses are the
Be-rich stars which are both accretive and retrograde.}
\end{figure}

\begin{figure}
\plotone{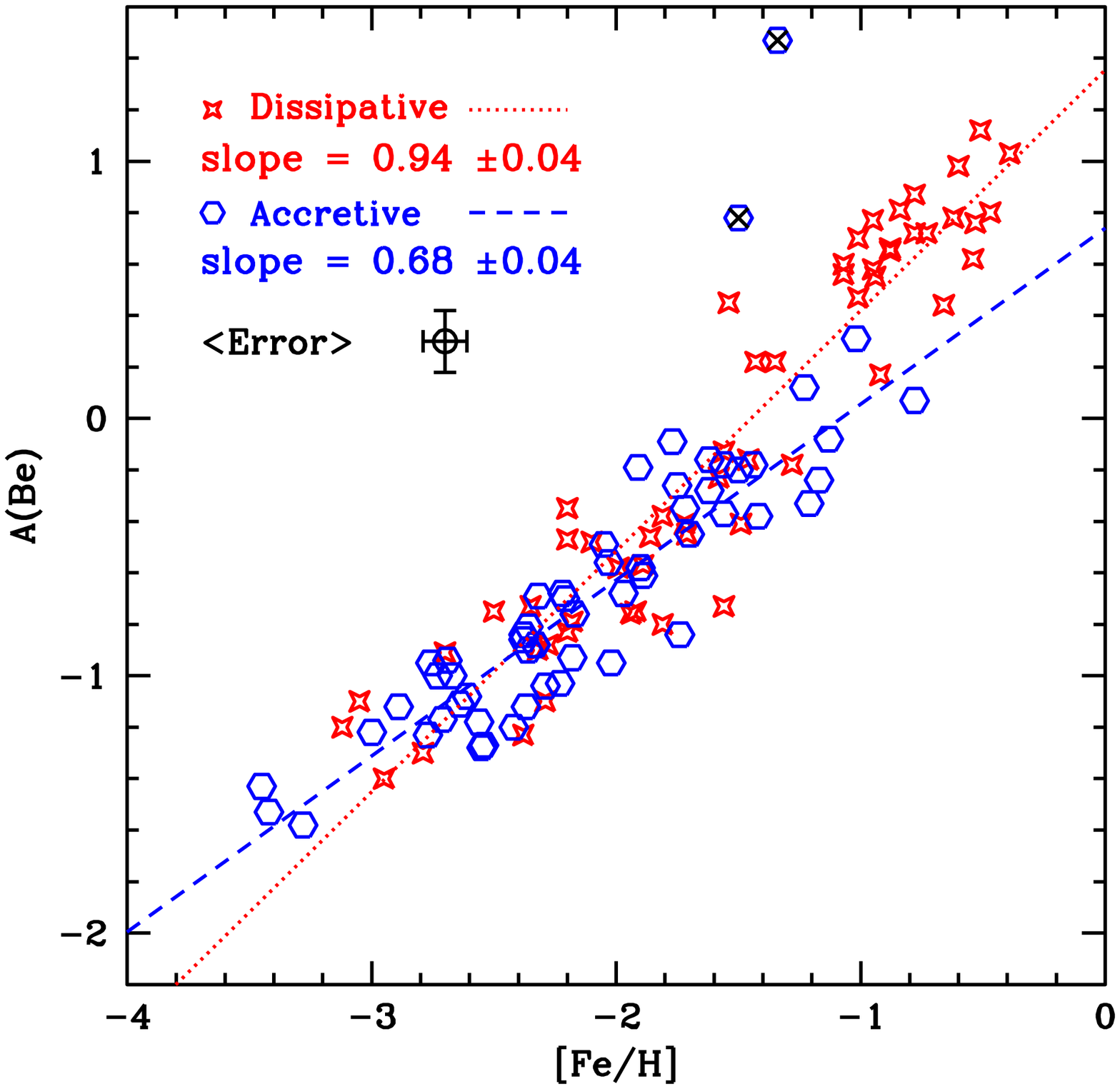}
\caption{The dissipative stars (starred crosses) and their slope (dotted line)
and the accretive stars (hexagons) and their slope (dashed line) are shown
together. }
\end{figure}

\clearpage

\begin{figure}
\plotone{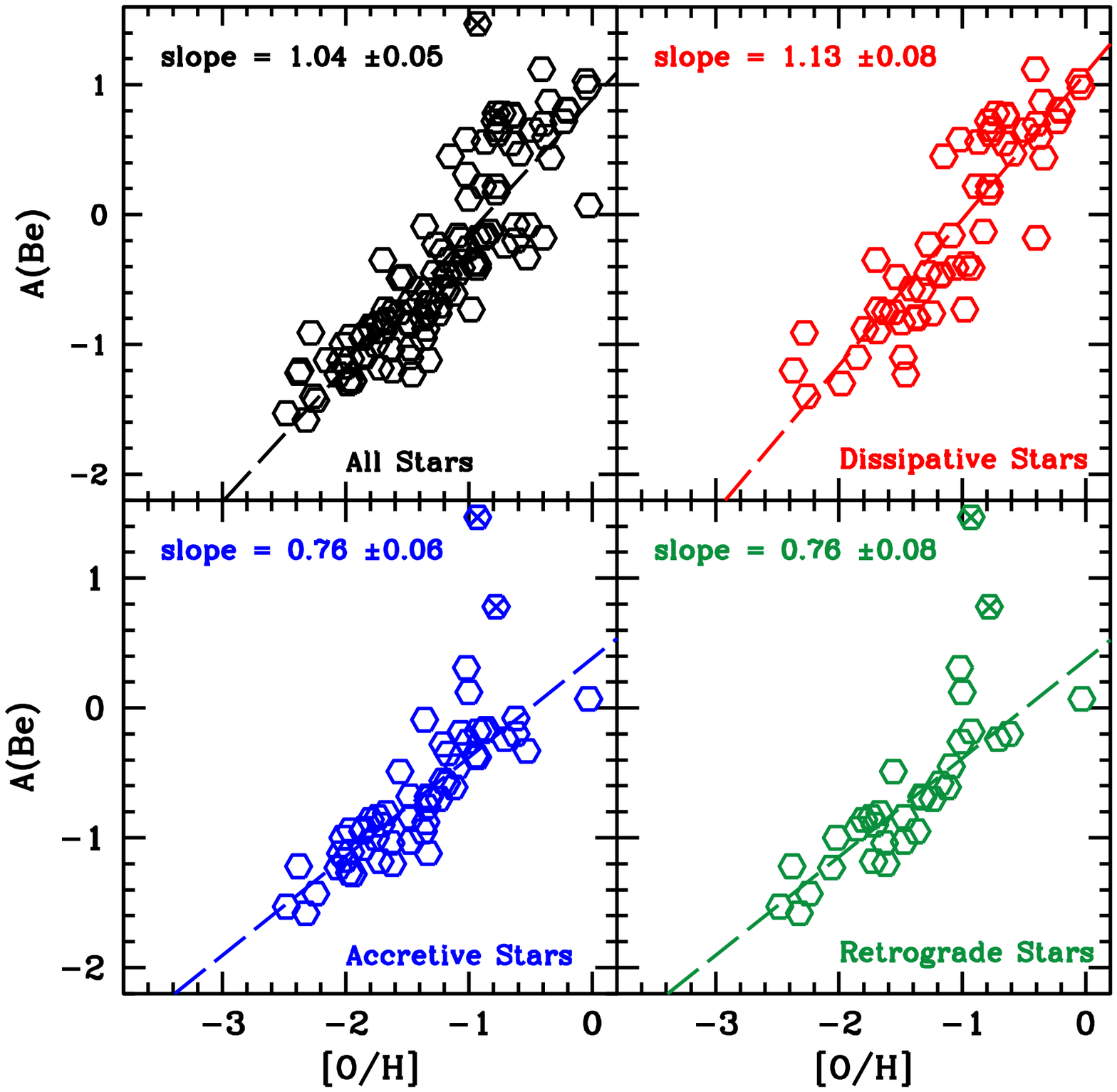}
\caption{The distribution of A(Be) with [O/H] for our total sample, for the
dissipative stars, for the accretive stars, and for retrograde subset of the
accretive stars.  There is a steeper slope, 1.13, for the dissipative stars
than for the accretive stars, 0.76.  The hexagons with the crosses are the
Be-rich stars which are both accretive and retrograde.}
\end{figure}

\begin{figure}
\plotone{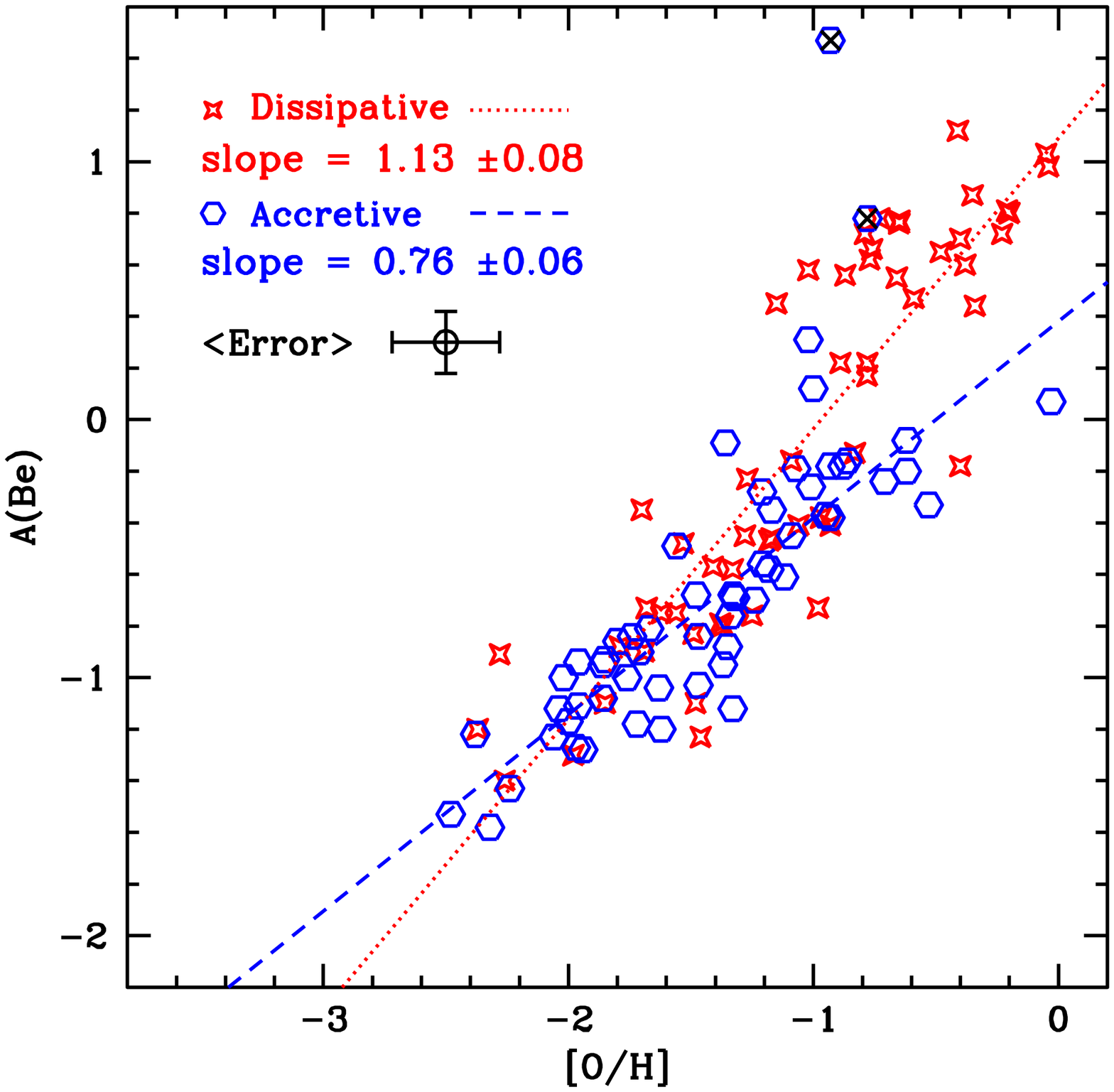}
\caption{The dissipative stars (starred crosses) and their slope (dotted line)
and the accretive stars (hexagons) and their slope (dashed line) are shown
together.}
\end{figure}

\begin{figure}
\plotone{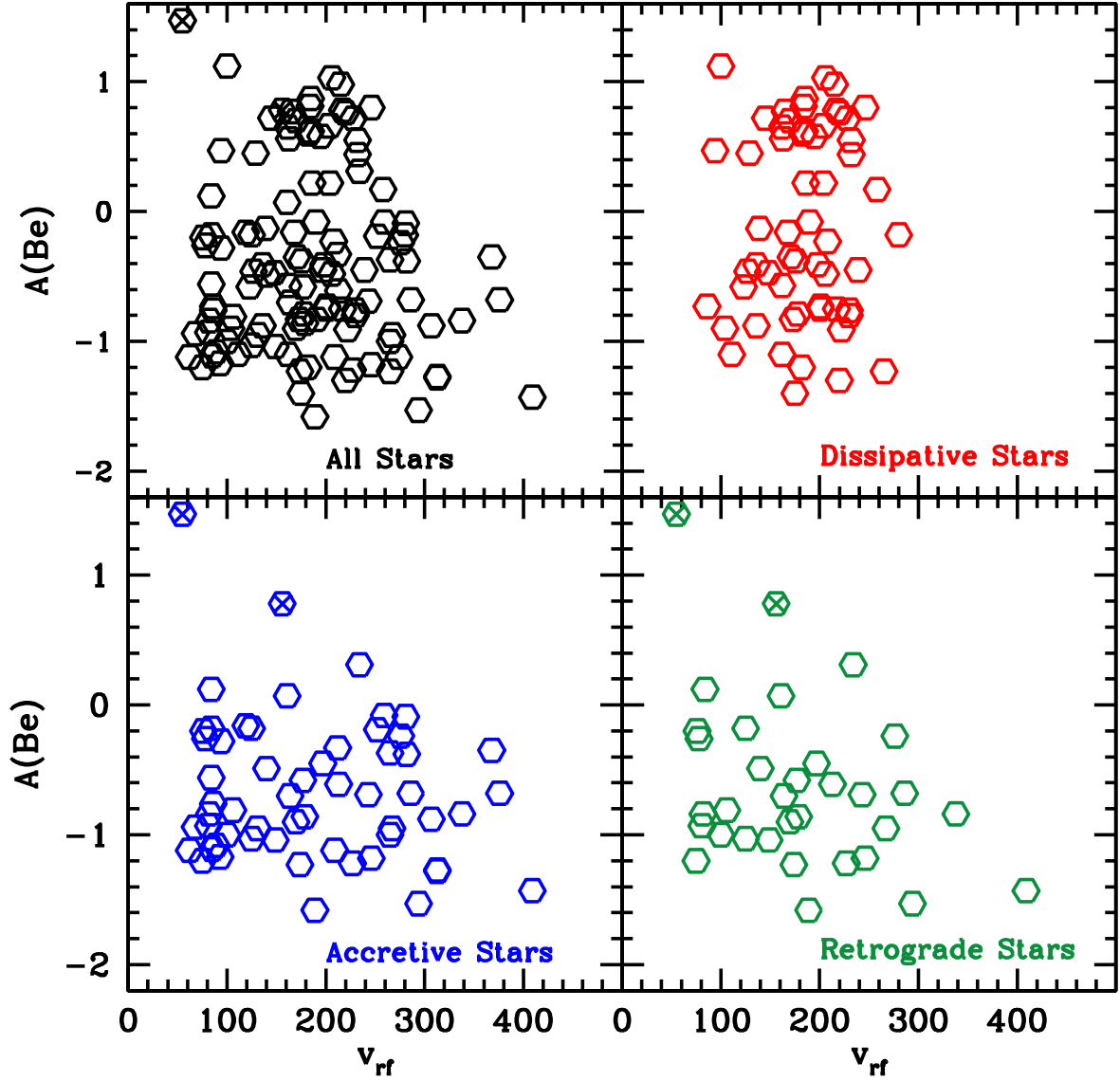}
\caption{The values of A(Be) shown as a function of the rest-frame velocity
for the different groupings of stars.}
\end{figure}

\begin{figure}
\plotone{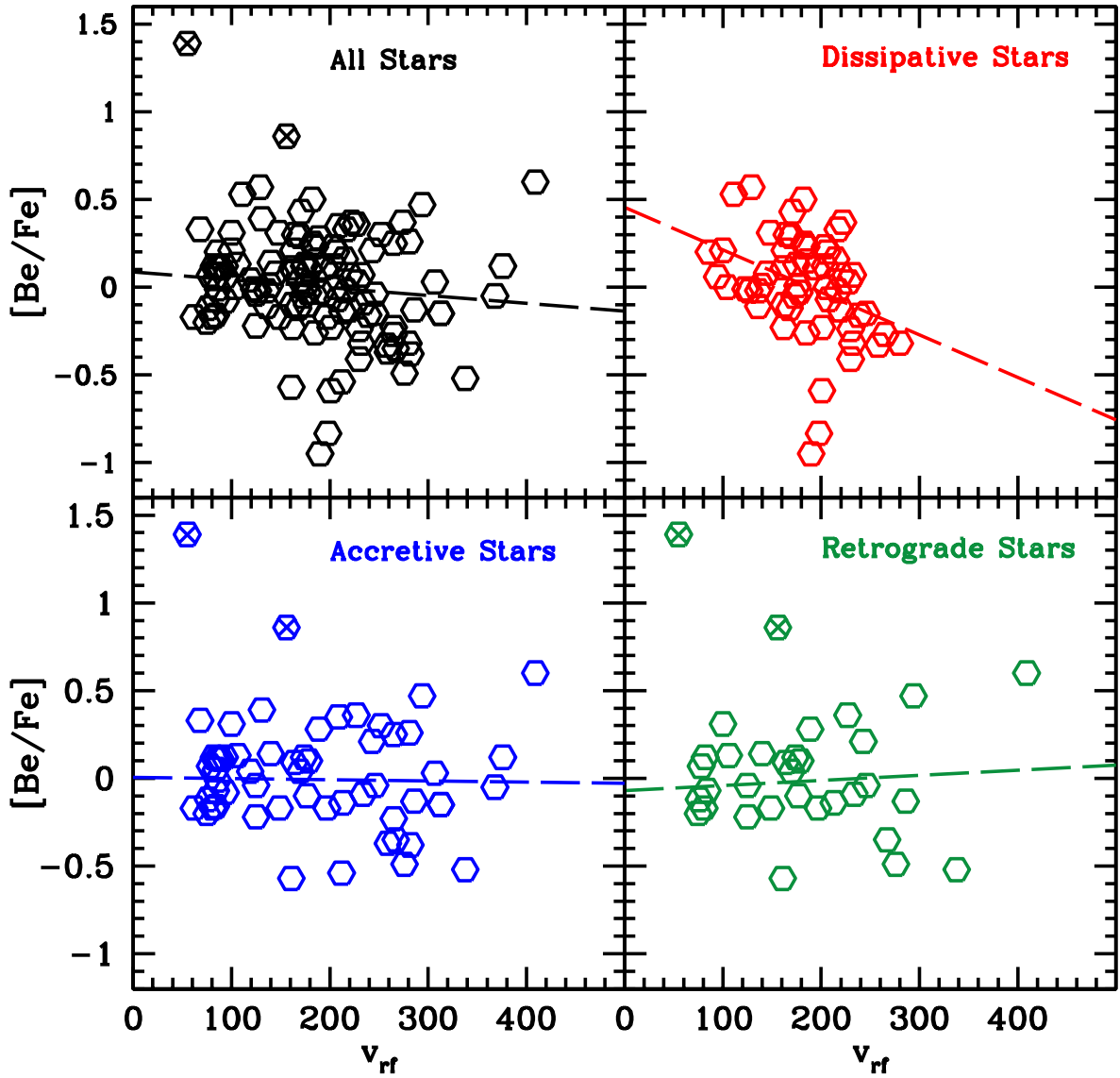}
\caption{The Be abundances normalized to Fe as a function of the rest-frame
velocity for the different groupings of stars.  In the bottom two panels the
highest-velocity point is G 64-12.}
\end{figure}

\begin{figure}
\plotone{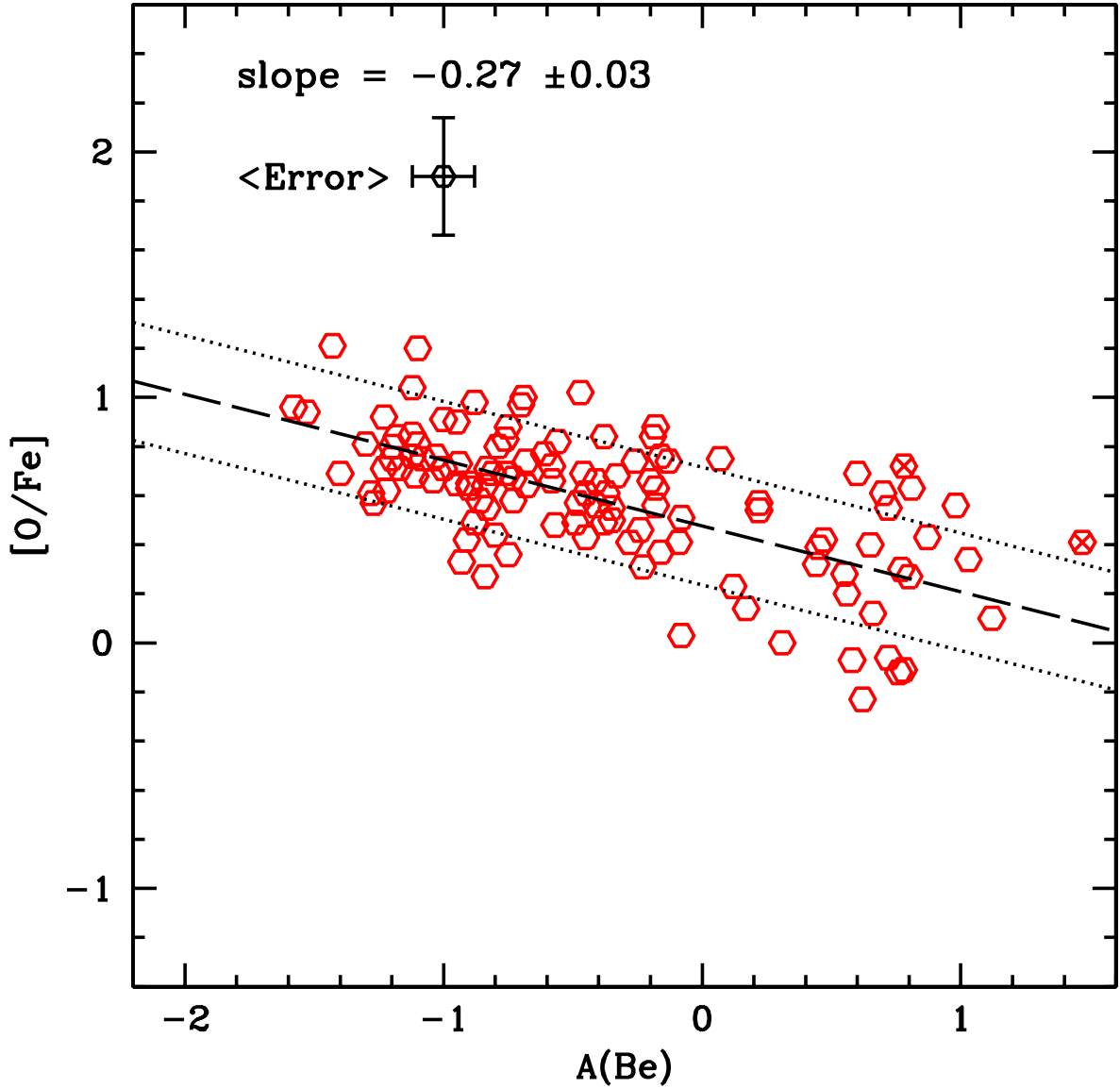}
\caption{[O/Fe] as a function of the Be abundance, A(Be).  The dashed line is
the linear fit and the dotted lines are plus and minus the error in [O/Fe].
There is more scatter at the higher values of A(Be) where more of the disk
stars reside.  Comparison of this with Figure 14, where [Fe/H] is the
abscissa, shows that [Fe/H] is a better chronometer than A(Be) overall.}
\end{figure}

\begin{figure}
\vskip -1.5truein
\plotone{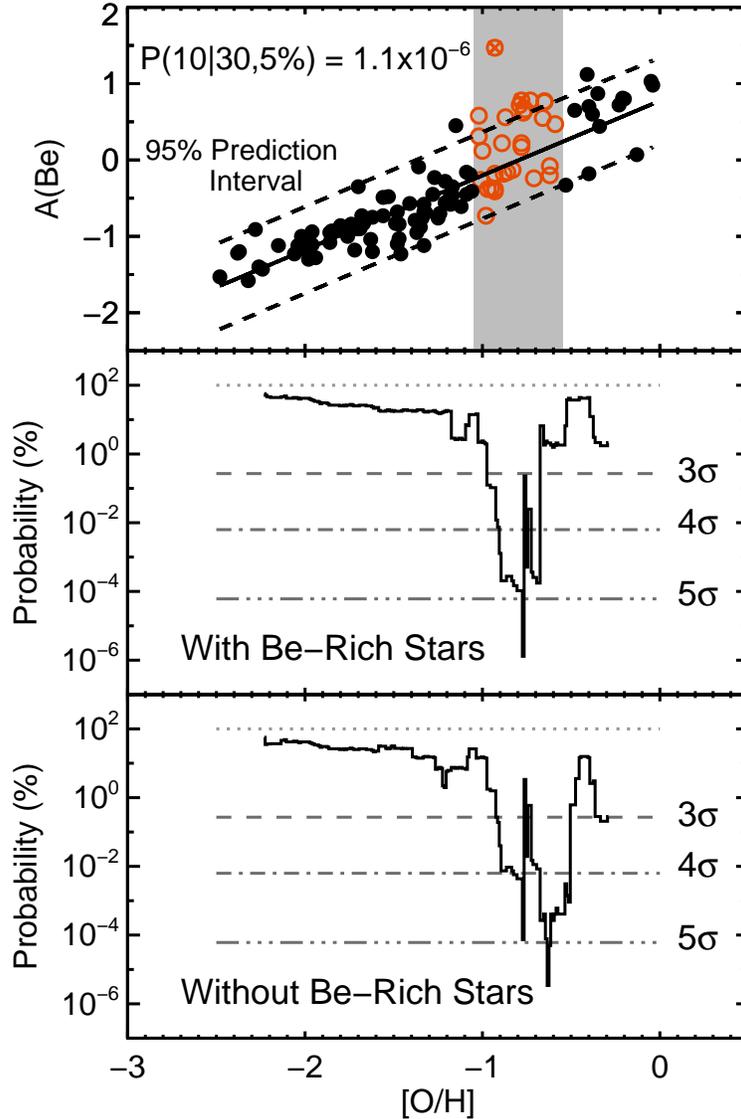}
\vskip -2.5truein
\caption{Example of our A(Be) spread analysis for [O/H].  The top panel shows
the test for the interval centered at [O/H] = $-$0.8 with a width of 0.5 dex
(gray shaded region).  We use the data outside the interval (black filled
circles) to derive the best-fit line and prediction interval we expect the
data to follow within the interval (red open circles).  In this example 10
out of 30 points lie outside the 95\% prediction interval.  Based on the
binomial theorem, the probability of this occurring by chance is
1.1$\times$10$^{-6}$.  This process is repeated in running steps of 0.01
dex, producing the probability curve shown in the middle panel.  For central
values of [O/H] between $\sim-$0.75 to $-$0.95 (or a range from $\sim$0.5
to $-$1.2), there appears to be a spread in Be at $>$4-$\sigma$ for a given
[O/H].  In the bottom panel we show the results ignoring the two Be-rich stars
HD~106038 and HD~132475 (marked with crosses in the top panel); even in this
case the spread appears to be significant.
}
\end{figure}

\end{document}